\documentclass{article}

\usepackage{arxiv}

\usepackage[utf8]{inputenc} 
\usepackage[T1]{fontenc}    
\usepackage{hyperref}       
\usepackage{url}            
\usepackage{booktabs}       
\usepackage{amsfonts}       
\usepackage{nicefrac}       
\usepackage{microtype}      
\usepackage{lipsum}		
\usepackage{graphicx}
\usepackage{natbib}
\usepackage{doi}

\usepackage{amsmath}
\usepackage{bm}
\usepackage{graphicx,psfrag,epsf}
\usepackage[title]{appendix}

\usepackage{float}

\usepackage{tikz}
\usetikzlibrary{positioning}

\newcommand{\widesim}[2][1.5]{
	\mathrel{\overset{#2}{\scalebox{#1}[1]{$\sim$}}}
}

\newcommand{\btheta}{{\mbox{\boldmath $\theta$}}}
\newcommand{\hbtheta}{{\mbox{\boldmath $\hat{\theta}$} }}
\newcommand{\bdeta}{{\mbox{\boldmath $\eta$}}}
\newcommand{\bbeta}{{\mbox{\boldmath $\beta$}}}

\newcommand{\bSigma}{{\mbox{\boldmath $\Sigma$}}}

\newcommand\tab[1][1.7cm]{\hspace*{#1}}

\title{The scale transformed power prior for use with historical data from a different outcome model}

\author{ Brady Nifong\\
	Department of Biostatistics\\
	University of North Carolina\\
	Chapel Hill, NC 27599 \\
	\texttt{brady\_nifong@unc.edu} \\
	\And
	Matthew A.~Psioda \\
	Department of Biostatistics\\
	University of North Carolina\\
	Chapel Hill, NC 27599 \\
	\texttt{matt\_psioda@unc.edu} \\
	\And
	Joseph G.~Ibrahim \\
	Department of Biostatistics\\
	University of North Carolina\\
	Chapel Hill, NC 27599 \\
	\texttt{ibrahim@bios.unc.edu} \\
}

\renewcommand{\shorttitle}{The Scale Transformed Power Prior}

\hypersetup{
pdftitle={A template for the arxiv style},
pdfsubject={q-bio.NC, q-bio.QM},
pdfauthor={David S.~Hippocampus, Elias D.~Striatum},
pdfkeywords={First keyword, Second keyword, More},
}

\begin{document}
\maketitle

\begin{abstract}
	We develop the scale transformed power prior for settings where historical and current data involve different data types, such as binary and continuous data, respectively. This situation arises often in clinical trials, for example, when historical data involve binary responses and the current data involve time-to-event or some other type of continuous or discrete outcome.  The power prior proposed by Ibrahim and Chen (2000) does not address the issue of different data types. Herein, we develop a new type of power prior, which we call the scale transformed power prior (straPP). The straPP is constructed by transforming the power prior for the historical data by rescaling the parameter using a function of the Fisher information matrices for the historical and current data models, thereby shifting the scale of the parameter vector from that of the historical to that of the current data.  Examples are presented to motivate the need for a scale transformation and simulation studies are presented to illustrate the performance advantages of the straPP over the power prior and other informative and non-informative priors. A real dataset from a clinical trial undertaken to study a novel transitional care model for stroke survivors is used to illustrate the methodology. 
\end{abstract}


\section{Introduction}
The availability and use of historical data has become increasingly common in the design and analysis of clinical trials 
and observational studies covering a wide array of biomedical and non-biomedical research applications. The incorporation of 
historical data into a design or analysis of a study has been a widely discussed topic over the past 25 years with a vast 
literature including the use of hierarchical models and various informative priors of many styles and types. One such tool 
for incorporating historical data into a Bayesian design or analysis is the power prior proposed by \cite{IbrahimEtAl00}. The power 
prior has been established to have several important uses and attractive properties in the design and analysis of clinical 
trials and observational studies when historical data are available. The power prior has been used by many researchers in several applications in design and analysis of clinical trials and observational studies including 
\cite{DeSantis06}, \cite{DeSantis07}, and \cite{HobbsEtAl11}. 
Books illustrating the use of the power prior in epidemiological studies and clinical trials contexts
include \cite{SpiegEtAl04}, and \cite{BerryEtAl10}. \cite{IbrahimEtAl15} gives a thorough review of the power prior 
and its variants, presents a detailed discussion on the theory, and provides numerous applications of the power prior 
for both study design and analysis. 

The construction of the power prior inherently assumes that the data types of the historical and current data are the same 
(e.g., if the historical data are binary, so are the current data). A challenging current issue that arises in a variety of study
designs or analyses is that the historical and current data have different data types. For example, such a scenario arises 
often in cancer clinical trials in which the historical data may be binary, such as objective response data from a Phase 2 trial, 
but the current data may be continuous such as a normally distributed data or time-to-event data in a Phase 3 trial. There are many 
other settings other than cancer clinical trials for which this phenomenon is common. 

The power prior \citep{IbrahimEtAl00} is not theoretically equipped to handle this setting since the scales of the response variables 
in the historical and current data may be quite different and, as a result, the regression coefficients for the corresponding models
may have non-comparable magnitudes. To solve this dilemma, the regression coefficients from the power prior based on the historical 
data need to be rescaled in such a way so as to result in a reasonable prior for the regression coefficients 
in the current data likelihood. We achieve such a rescaling by transforming the parameter for the power prior 
using a function of the Fisher information matrices for the historical and current data models. We call this newly scaled power prior the scale transformed power prior (straPP). The straPP and its generalization, called the generalized straPP, are a broad class of priors in which the power prior and the commensurate prior are closely related. A key aspect of the power prior is that the historical and current data share a common parameter, whereas the commensurate prior, developed by \cite{HobbsEtAl11}, assume the parameter for the current data is normally distributed around the historical data parameter.
We show in simulation studies that the straPP often gives better performance than the 
power prior and other priors in terms of bias and mean square error. Thus, the straPP is a potentially powerful tool for 
the design and analysis of studies in which historical data are available that does not match the data type of a current dataset but where
the outcomes in the two datasets are related. 

The rest of this paper is organized as follows. In Section~\ref{motivate}, we give a detailed motivating example illustrating the need of the straPP for a Bayesian analysis. Section~\ref{existing priors} gives a brief review of the power prior and several variants. Section~\ref{methods: straPP} presents the proposed methodology for the straPP in detail, develops the generalized straPP, and discusses connections with the commensurate prior, while focusing on the straPP for generalized linear models (GLMs). Section~\ref{sims} presents detailed simulation studies using the straPP and its generalizations under various settings and data types within the class of GLMs. Section~\ref{compass} presents a real data analysis using the straPP and its generalizations to demonstrate the advantages of the straPP over other priors. We close the article with some discussion in Section~\ref{disc}.  

\section{\label{motivate} Motivating Example}

The Comprehensive Post-Acute Stroke Services (COMPASS) study (\cite{compass}) was a two arm, cluster-randomized pragmatic trial 
designed to evaluate the effectiveness of a novel transitional care model (COMPASS care model) compared to usual care in 
mild-to-moderate stroke and transient ischemic attack (TIA) patients across a diverse set of hospitals within North Carolina, 
USA. The study consisted of two phases. In Phase 1 of the COMPASS study, 40 hospital units were randomized in a 1:1 
allocation scheme to either implement the COMPASS care model (i.e., the intervention) or maintain their usual care practices. 
Phase 1 intervention hospitals that continued into Phase 2 attempted to sustain the intervention with minimal resources 
provided by the study team. Phase 1 usual care hospitals that continued into Phase 2 were transitioned to provide the intervention 
as their standard of care.

One of the exploratory objectives of the COMPASS study was to assess whether intervention arm patients who received a specialized 
electronic care plan (eCare plan) had better health outcomes than patients who did not after adjustment for key covariates 
to account for potential selection bias regarding which patients choose to attend the clinic visit at which they received a 
customized eCare plan. Of note, Phase 2 of the COMPASS study included a continuous measure of physical health (the PROMIS Global 
Health Scale) that was not collected in Phase 1. We consider 
the analysis of Phase 2 patient outcomes based on the PROMIS measure from one large hospital that provided the COMPASS care 
model during both phases of the study. This relatively large hospital was selected due to having provided the intervention with consistency and high fidelity in both phases of the study.
We use data from Phase 1 as historical data to inform the straPP. 

Since the PROMIS outcome was not collected for Phase 1 patients, we consider the incidence of one or more falls as the Phase 1 outcome.This variable is an indicator of whether the participant had at least one fall between hospital discharge and 90 days post-stroke (no falls versus at least one fall).
As the historical and current outcomes measure related concepts (global disability versus global health) but are different scales (e.g., one binary, one continuous), these data make an ideal case study for comparing performance of the straPP to other commonly used informative prior distributions. 
In fact, we were able to investigate this relationship empirically as the incidence of falls outcome was collected during Phase 2 of the COMPASS study. Using a simple logistic regression model with incidence of falls as the outcome and the continuous PROMIS measure as the covariate, we estimated the area under the Receiver Operating Characteristic (ROC) curve to be 0.64 indicating 
fair predictive ability of the incidence of falls for the PROMIS measure. Accordingly, the Phase 1 dataset based on incidence of falls may be useful for inference on covariate effects for the PROMIS Global Health as measured in Phase 2. 

The covariates of interest considered for our analyses were an indicator for receipt of the eCare plan within 30 days of hospital discharge, an indicator for having a history of stroke or TIA, an indicator for having non-white race, and categorized NIH stroke scale score (abbreviated as NIHSS, 0 = no stroke symptoms, 1-4 = minor stroke symptoms, and $\ge5= $ moderate-to-severe 
stroke symptoms).

The analyses presented in this paper are for illustration purposes only as they make use of data only from complete cases from the aforementioned hospital that participated in the COMPASS study. A more sophisticated analysis that incorporates information from patients with partially missing covariates and/or missing outcomes is beyond the scope of this paper.

\section{Existing Priors for Incorporation of Historical Data \label{existing priors}}

\subsection{The Power Prior \label{intro: pp}}

The power prior was developed by \cite{IbrahimEtAl00} as an informative prior derived from historical data that contain information on the same response and covariates as measured in a current study. The power prior, denoted $\pi_{pp}(\cdot)$, creates a meaningful semi-automatic informative prior for the $p \times 1$ parameter of interest $\bm{\theta}$ as 
\begin{eqnarray}
\pi_{pp}(\bm{\theta}\mid D_0) \propto \mathcal{L}(\bm{\theta}\mid D_0)^{a_0}\pi_0(\bm{\theta}),\label{PP}
\end{eqnarray}
where $\mathcal{L}(\bm{\theta}\mid D_0)$ denotes the historical data likelihood, $D_0 = (n_0, \bm{Y}_0, \bm{X}_0)$ denotes the historical data, $n_0$ denotes the sample size, $\bm{Y}_0$ denotes the $n_0\times 1$ response vector, and $\bm{X}_0$ denotes the $n_0\times p$ covariate matrix. The distribution $\pi_0(\bm{\theta})$ is called the initial prior and is often taken to be non-informative. The scalar $0 \le a_0 \le 1$ is called the discounting parameter and its value
controls the weight given to the historical data. For example, a value of $a_0=0$ discards the historical data altogether resulting in a prior equal to the initial prior (complete discounting) and a value of $a_0=1$ weights the historical and current data equally. The power prior is robust under many settings (\cite{IbrahimEtAl00}) but, does not account for scale differences in the response variable for the historical and current data.

\subsubsection{Variations of the Power Prior}

Several variations and generalizations of the power prior have been developed including several that take $a_0$ to be a random variable. For a complete discussion of the power prior and several variations and generalizations, including variations that take $a_0$ to be a random variable, see \cite{IbrahimEtAl15}. Additional information on how to choose $a_0$ may be found in  \cite{NeuenEtAl09} and \cite{DeSantis06}. In this section, we briefly discuss variations of the power prior where $a_0$ is treated as fixed. 

One variation of the power prior is the asymptotic power prior. The asymptotic 
power prior can be developed using the asymptotic normality properties of the power prior as described by \cite{IbrahimEtAl15}. It is given as
\begin{equation}
\pi(\btheta, a_0\mid D_0) \propto N(\hbtheta_0, a_0^{-1}H^{-1}(\hbtheta_0)),
\label{app}
\end{equation}
where $\hbtheta_0$ denotes the mode of the power prior and $H(\btheta) = - d^2/(d\btheta d\btheta^T)\log\{\pi(\btheta\mid D_0,a_0)\}$ 
is its associated Hessian matrix evaluated at $\hbtheta_0$. 

The partial-borrowing power prior, developed by \cite{IbrahimEtAl2012}, provides a useful generalization of the power prior described above. This power prior borrows information on a subset of parameters common to both  the historical and current models. Let $\bm{\theta} = (\bm{\theta}_1, \bm{\theta}_2)$, where $\bm{\theta}_1$ is a set of parameters common to both models and $\bm{\theta}_2$ is a set of parameters relevant only to the current data model. The partial-borrowing power prior is given as 
\begin{equation}
\pi(\bm{\theta}\mid D_0) \propto \mathcal{L}(\bm{\theta}_1\mid D_0)^{a_0}\pi_0(\bm{\theta}),\label{pb}
\end{equation}
where $\pi_0(\bm{\theta})$ is an initial prior for all components of $\bm{\theta}$. This is a useful and flexible extension 
of the power prior since there are many cases where historical information may only be available for certain parameters (e.g., parameters 
associated with a control group).

\subsection{Commensurate Prior}

The commensurate prior was proposed by \cite{HobbsEtAl11} to utilize historical information when there is strong evidence of agreement between the historical and current data. Unlike the power prior, the commensurate prior assumes that though the historical and current data covariates comprise the same set, the model parameters are different, denoted $\bdeta$ and $\btheta$, respectively. The influence of the historical data is then controlled by the commensurability parameter, $\tau$, which measures the degree to which the historical and current data are commensurate \citep{HobbsEtAl11}.

The commensurate prior specifies a prior on $\btheta$ centered at $\bdeta$, which is often taken to be normally distributed. The commensurate prior by \cite{HobbsEtAl11} is then
\begin{equation}
\pi_c(\btheta | D_0, \bdeta, \tau) \propto \mathcal{L}(\bdeta|D_0)N_p\left(\btheta | \bdeta, \tau^{-1}\bm{I}_p\right)\pi_0(\btheta),\label{com}
\end{equation}
where $\pi_0(\cdot)$ and $\mathcal{L}(\bm{\theta}\mid D_0)$ are as defined by the power prior, and $\bm{I}_p$ is the $p \times p$ identity matrix.

\section{\label{methods: straPP} The Scale Transformed Power Prior}

In this section, we develop the scale transformed power prior (straPP). The straPP is a generalization of the power prior (\cite{IbrahimEtAl00}) from Section~\ref{intro: pp} that is designed to modify the scale of $\btheta$ to better align with that of the current data. The derivation for the straPP is based on the assumption that the standardized parameter values
are approximately equal for the historical and current data models. 

Denote the asymptotic covariance matrix of $\btheta$, i.e. the inverse of the Fisher information matrix, 
by $I^{-1}(\btheta)$ and $I^{-1/2}(\btheta)$ as the square root of $I^{-1}(\btheta)$ obtained via a spectral decomposition. The quantity $I^{1/2}(\btheta) \btheta$ can be viewed as a standardized or
scaled $\btheta$. The resulting quantity is unitless and the scaling action can be interpreted as converting the parameter from the original scale into standard deviation units based on the asymptotic covariance matrix. Let $\bdeta$ and $\btheta$ denote the parameters for the historical and current data models, respectively. Formally, the 
assumption of equal standardized parameter values can be expressed as follows
\begin{equation}
I_0^{1/2}(\bdeta) \bdeta = I_1^{1/2}(\btheta) \btheta, \label{transf1} 
\end{equation} 
where $I_0(\cdot)$ and $I_1(\cdot)$ denote the Fisher information matrix based on the historical data model and the current data model, respectively. In order to rescale the power prior in \eqref{PP} by applying the transformation implied by \eqref{transf1}, one would have to solve for the historical or current parameter vector. When the historical data information matrix does not depend on the regression parameters, such as when the historical data model is a linear model, one can solve algebraically for $\bdeta = g(\btheta)$. When the historical data information matrix depends on the regression parameters but the current data information matrix does not, one can solve for $\btheta = g^{-1}(\bdeta)$ and perform analysis using the straPP by sampling from what we refer to as the complementary posterior distribution. More details on the complementary posterior distribution can be found in Appendix~\ref{comp_straPP}.

In general, \eqref{transf1} cannot be solved algebraically for either parameter (i.e., solved to obtain $\bdeta = g(\btheta)$ or $\btheta = g^{-1}(\bdeta)$ in closed form), as the information matrices generally depend on the parameters of interest.
For the case where neither information matrix is free of the parameter, e.g., when the historical data follow a Poisson distribution 
and the current data follow a Bernoulli distribution, it is still possible to analyze data using the straPP by considering a posterior distribution representation involving both $\bdeta$ and $\btheta$. Then a Metropolis-Hastings sampling algorithm can be used in which proposed values of $\bdeta$ and $\btheta$ satisfy the constraint 
in \eqref{transf1}. This is discussed more in Appendix~\ref{alg}.

In what follows, without loss of generality, we present mathematical developments based on the general transformation
$\bdeta = g(\btheta)$ while acknowledging that actually fitting the model using the straPP may require alternative techniques such as sampling from the complementary posterior as described in Appendix.
Using \eqref{transf1}, one can take the power prior,
indexed by parameter $\bdeta$ and apply the transformation to obtain a prior 
for $\btheta$ that is rescaled to match the current data. The scale transformed power prior, denoted $\pi_s(\cdot)$, is then
\begin{eqnarray*}
	\pi_{s}(\bm{\theta}\mid D_0) \propto \mathcal{L}(g(\bm{\theta})|D_0)^{a_0}\pi_0(g(\bm{\theta}))\left|\frac{dg(\bm{\theta})}{d\bm{\theta}}\right|,
\end{eqnarray*} 
where $a_0$ and $\pi_0(\cdot)$ are as described in Section~\ref{intro: pp} and $\left|dg(\bm{\theta})/d\bm{\theta}\right|$ is the determinant of the 
Jacobian of the transformation. The expression of the Jacobian can be found in Section~\ref{methods: straPP: jacobian}. 

The fact that the transformation in \eqref{transf1} is locally one-to-one follows directly from the implicit function 
theorem (\cite{taylor1983advanced}). As a result, propriety of the power prior implies propriety of the 
straPP. Establishing propriety of the power prior for GLMs and survival models has received significant treatment in the literature, including 
works by \cite{ChenEtAl99},
\cite{ChenEtAl00}, among others.

When $a_0 = 0$ and $\pi_0(\cdot)$ is a uniform improper 
prior, the straPP will be an improper prior and its use may result in an improper posterior since the kernel of the straPP is 
simply the determinant of the Jacobian. To avoid this complexity, we simply define the straPP to be equal to the initial prior 
when $a_0 = 0$. The intended purpose of the straPP is for cases where one would want to incorporate historical data to some degree 
in the analysis of the current data. Thus, the case where $a_0 = 0$ is of no practical relevance. 

\subsection{\label{methods: straPP: jacobian} The Expression for the Jacobian}
In general, the transformation implied by \eqref{transf1} cannot be calculated algebraically. However, the derivative inside the Jacobian can be calculated via implicit derivation as shown below.

\begin{eqnarray*}
	\left[\left\{\frac{d}{d\bdeta}I_0^{1/2}(\bdeta)\right\}\bdeta + I_0^{1/2}(\bdeta)\right]\frac{d\bdeta}{d\btheta} & = & \left\{\frac{d}{d\btheta}I_1^{1/2}(\btheta)\right\}\btheta + I_1^{1/2}(\btheta)\\
	\Rightarrow \frac{d\bdeta}{d\btheta} & = & \left[\left\{\frac{d}{d\bdeta}I_0^{1/2}(\bdeta)\right\}\bdeta + I_0^{1/2}(\bdeta)\right]^{-1}\left[\left\{\frac{d}{d\btheta}I_1^{1/2}(\btheta)\right\}\btheta + I_1^{1/2}(\btheta)\right].
\end{eqnarray*}

The matrix $\left\{dI_1^{1/2}(\btheta)/d\btheta\right\}\btheta$ can be written as $ (\{dI_1^{1/2}(\btheta)/d\theta_0\}\btheta , \ldots, \{dI_1^{1/2}(\btheta)/d\theta_{p-1}\}\btheta ).$
The derivative can then be decomposed using a direct application of the product rule,  for $j = 0, \ldots, p-1$,
as
\begin{eqnarray}
dI_1(\bm{\theta})/d\theta_j &= &
I_1^{1/2}(\bm{\theta})\{dI_1^{1/2}(\bm{\theta})/d\theta_j\} + 
\{dI_1^{1/2}(\bm{\theta})/d\theta_j\}I_1^{1/2}(\bm{\theta}). \label{eq:decomp1}  																			
\end{eqnarray} 
Equation \eqref{eq:decomp1} can be expressed in the form of the Sylvester equation (\cite{Sylvester}) which allows 
for vectorized representations of the needed derivatives using the approach suggested by \cite{Laub}.
Let $\bm{I}_p$ denote the $p\times p$ identity matrix. 
Then, following \cite{Laub}, the required derivatives may be represented as
\begin{eqnarray}
\text{vec}\left(\frac{dI_{1}^{1/2}(\bm{\theta})}{d\theta_j}\right)
&=&\left\{I_1^{1/2}(\bm{\theta})\otimes \bm{I}_p +\bm{I}_p\otimes I_1^{1/2}(\bm{\theta})\right\}^{-1}\
\text{vec}\left(\frac{dI_{1}(\bm{\theta})}{d\theta_j}\right)\label{i1_sqrt},
\end{eqnarray}
where $\text{vec}(\cdot)$ denotes the vectorization of a matrix in which columns are stacked to convert a $n \times p$ matrix into a $np \times 1$ vector. The derivative of $I_0^{1/2}(\bdeta)$ is calculated analogously.

\subsection{\label{methods: straPP: subset} The Partial-Borrowing straPP}

One may wish to utilize the straPP for only a subset of parameters. 
Let the ${p\times 1}$ vector $\bm{\theta}$ be partitioned into two vectors such that $\bm{\theta} = (\bm{\theta}_1, \bm{\theta}_2)$, where 
$\bm{\theta}_1$ is $r\times 1$ and $\bm{\theta}_2$ is $(p-r)\times 1$. 
Suppose we would like to specify a straPP that rescales and borrows information on $\bm{\theta}_1$ but 
does not borrow information on $\btheta_2$.
We construct a similar partition of the ${s\times 1}$ vector $\bdeta$, in which $\bdeta_1$ is $r\times 1$ and $\bdeta_2$ is $(s-r)\times 1$.
In this more general setting, we must modify the partial-borrowing power prior in \eqref{pb} to account for the fact that the historical data contain parameters that are not in the current data model. In particular, the partial-borrowing power prior can be revised as
$\pi_{pp}(\bdeta\mid D_0) \propto \mathcal{L}(\bdeta\mid D_0)^{a_0}\pi_0(\bdeta)\pi_0(\btheta_2)$,
where we now use $\bdeta$ to represent the historical data model parameters as has been our convention
for exposition related to the straPP. The only material difference between the power prior and the partial-borrowing power prior is the inclusion of an initial prior on $\btheta_2$ denoted by $\pi_0(\btheta_2)$.

The partial-borrowing straPP is then derived according to the transformation 
\begin{equation}
I_{0}^{-1/2}(\bdeta)\bdeta_1 = I_{1}^{1/2}(\btheta)\btheta_1, \label{pbTran}
\end{equation}
where here $I_{0}(\cdot)$ and $I_{1}(\cdot)$ denote the $r\times r$ submatrices 
of the Fisher information matrices for the current and historical data models, respectively. In general, $\bdeta_1$ cannot be solved for algebraically, so we will refer to the transformation as $g_1(\btheta, \bdeta_2)$. The transformation yields 
\begin{align}
\pi(\bm{\theta},\bdeta_2\mid D_0) &= \pi_{s}(\bm{\theta}_1, \bdeta_2\mid  \btheta_2, D_0, )\pi_0(\bm{\theta}_2) \nonumber \\ &\propto 
\mathcal{L}(g_1(\btheta, \bdeta_2),\bdeta_2\mid \btheta_2, D_0)^{a_0}\pi_0(g_1(\btheta,\bdeta_2),\bdeta_2 \mid  \btheta_2)
\left|\frac{dg_1(\bm{\theta}, \bdeta_2)}{d\bm{\theta}_1}\right|\pi_0(\bm{\theta}_2). \label{sub}
\end{align}
Technically speaking, the transformation to obtain the partial-borrowing straPP is a combination of \eqref{pbTran} and 
identity transformations for $\bdeta_2$ and $\btheta_2$.
It is straightforward to show the determinant of that full-rank transformation is equal to the determinant shown 
in \eqref{sub}.

The partial-borrowing straPP in \eqref{sub} enables use of the straPP in a variety of settings. Perhaps the most intuitive example
is for situations where it is not appropriate to borrow information on the intercept parameter for the current data model. For example, the partial-borrowing straPP may be desirable when the historical response data is binary and the current response data is normal. In this case, the intercept terms will often have no logical connection even though covariate effects will when outcomes are related. Another instance where the partial-borrowing straPP may be preferred is the case where it is of paramount interest to borrow information on a treatment effect parameter but  borrowing information on nuisance parameters in the regression model is avoided simply to add a degree of robustness provided the current data alone are sufficient to estimate the other covariate effects.

\subsection{\label{methods: straPP: gen} The Generalized Scale Transformed Power Prior (Gen-straPP)}

As discussed in Section~\ref{methods: straPP}, the straPP is derived under the assumption that the standardized parameter values
for the historical and current data models are equal. Such an assumption leads to a reasonable transformation to rescale the parameter in
a power prior when the historical data model is not the same as that of the current data. Nonetheless it is important to be able to conduct sensitivity analyses of the core assumption of the straPP.
Thus, it is useful to develop a generalization of the straPP that provides a degree of robustness 
when the assumption of equal standardized parameter values does not hold.  

Towards this goal, we develop a generalized scale transformed power prior (Gen-straPP), in which we specify 
\begin{equation}
I_0^{1/2}(\bdeta) \bdeta = I_1^{1/2}(\btheta) \btheta + \bm{c_0}, \label{transf-gen} 
\end{equation} 
where $\bm{c}_0$ is a $p \times 1$ vector that allows component-specific deviations from the assumption of equal standardized parameter values for $\bdeta$ and $\btheta$. We denote the transformation as $\bdeta  = g_{\bm{c}_0}(\btheta)$. We note that $\bm{c}_0 = \bm{0}$ corresponds to the straPP. 

In practice, one could take $\bm{c}_0$ to be fixed or a random vector. 
If one takes $\bm{c}_0$ to be fixed, one possible choice is $\bm{c}_0 = d_0 \mathbf{J}$, where $\mathbf{J}$ is a $p\times1$ vector of ones and $d_0$ is a scalar.
One could then conduct sensitivity analyses by varying $d_0$.
There are innumerable choices for $\bm{c}_0$ with the aforementioned choice best equipped to identify violations in the assumption of the straPP when the violations occur across all covariates but poorly equipped to identify violations which present in
only one.
Therefore, we suggest taking $\bm{c}_0$ to be a random vector and specify its prior to be $\bm{c_0} \sim N_p(\bm{0}, \omega_0\bm{I}_p)$ 
where $\omega_0$ is a positive scalar.

The Gen-straPP can be derived from the power prior in~\eqref{PP} using the transformation in~\eqref{transf-gen}.
The Gen-straPP is
\begin{equation}
\pi_{g}(\btheta,\bm{c}_0\mid D_0) = \pi_{s}(\btheta|\bm{c}_0,D_0)\pi_0(\bm{c}_0) \propto \mathcal{L}(g_{\bm{c}_0}(\btheta)| \bm{c}_0,D_0)^{a_0}\pi_0(g_{\bm{c}_0}(\btheta)| \bm{c}_0)\left|\frac{dg_{\bm{c}_0}(\bm{\theta})}{d\bm{\theta}}\right|\pi_0(\bm{c}_0),\label{gen-straPP}
\end{equation}
where $\pi_0(\bm{c}_0)$ denotes the prior for $\bm{c}_0$. 
With this development, more broad sensitivity analyses can be conducted by varying $\omega_0$.  For example, one could investigate the results for various values of $\omega_0$, or use an empirical estimate for $\omega_0$, which we denoted $\omega_{0,E}$. Under the empirical estimate, we calculate the maximum likelihood estimate (MLE), denoted as $\hat{\bdeta}$ and $\hat{\bbeta}$ for the historical and current data, respectively, and then take $\omega_{0,E} = \max|I_0^{1/2}(\hat{\bdeta})\hat{\bdeta} - I_1^{1/2}(\hat{\bbeta})\hat{\bbeta}|.$
Similar to the development of the partial-borrowing straPP, the transformation to obtain the Gen-straPP 
is a combination of \eqref{transf-gen} and an identity transformation for $\bm{c}_0$.
It is straightforward to show the determinant of that full-rank transformation is equal to the determinant shown 
in \eqref{gen-straPP}.

The Gen-straPP can accommodate partial-borrowing as well. Following an approach identical to that from Section~\ref{methods: straPP: subset}, 
we define the transformation
\begin{equation}
I_{0}^{1/2}(\bdeta)\bdeta_1 = I_{1}^{1/2}(\btheta)\btheta_1 + \bm{c}_0, \label{gTranSub}
\end{equation}
where here $\bm{c}_0$ is an $r\times 1$ vector. We denote the transformation implied by \eqref{gTranSub} as $g_{1, \bm{c}_0}(\btheta, \bdeta_2)$.
The partial-borrowing Gen-straPP is
\begin{align}
\pi(\bm{\theta},\bdeta_2,\bm{c}_0\mid D_0) &= \pi_{s}(\bm{\theta}_1, \bdeta_2\mid  \bm{c}_0,\btheta_2, D_0, ) 
\pi_0(\bm{c}_0) \pi_0(\bm{\theta}_2)&&\label{gen-sub}\\
&\propto \mathcal{L}(g_{1,\bm{c}_0}(\btheta,\bdeta_2),\bdeta_2| \bm{c}_0,\btheta_2, D_0)^{a_0} 
\pi_0(g_{1,\bm{c}_0}(\btheta,\bdeta_2),\bdeta_2 \mid  \bm{c}_0, \btheta_2)
\left|\frac{dg_{1, \bm{c}_0}(\bm{\theta},\bdeta_2)}{d\bm{\theta}_1}\right|\pi_0(\bm{c}_0) \pi_0(\bm{\theta}_2).&& \nonumber
\end{align}
Similar to the development of the partial-borrowing straPP and Gen-straPP, the transformation to obtain the partial-borrowing 
Gen-straPP is a combination of \eqref{gTranSub} and an identity transformation for $\bdeta_2$, $\btheta_2$, and $\bm{c}_0$.
It is straightforward to show the determinant of that full-rank transformation is equal to the determinant shown 
in \eqref{gen-sub}.

\subsubsection{Relationship between the Gen-straPP and the Commensurate Prior \label{relation}}

To derive the relationship between the Gen-straPP and the commensurate prior, we derive the Gen-straPP in an alternate manner. The Gen-straPP transformation in \eqref{transf-gen}, can be re-written as
\begin{equation}
I_1^{1/2}(\btheta) \btheta = I_0^{1/2}(\bdeta) \bdeta - \bm{c_0}. \label{transf-com} 
\end{equation}
When $\bm{c_0} \sim N_p(\bm{0}, \omega_0\bm{I}_p)$, independently of $\bdeta, \btheta$, the standardized current parameter $\btheta^* = I_1^{1/2}(\btheta)\btheta$, conditional on $\bdeta$, is distributed normally about the standardized historical parameter as $\btheta^* ~|~ \bdeta \sim N_p(I_0^{1/2}(\bdeta)\bdeta, ~\omega_0\bm{I}_p).$
To complete the specification of the joint prior for the historical and standardized current parameters, we specify a power prior on $\bdeta$. The joint prior is
$\pi(\btheta^*, \bdeta) \propto  N_p(\theta^* | I_0^{1/2}(\bdeta)\bdeta, ~\omega_0\bm{I}_p)\mathcal{L}(\bdeta | D_0)^{a_0}\pi_0(\bdeta).$

To derive the Gen-straPP, one must calculate the joint distribution of the untransformed current and historical parameters. When $a_0 = 1$, we can write the Gen-straPP as 
\begin{equation}
\pi(\btheta, \bdeta) \propto  \mathcal{L}(\bdeta | D_0)N_p(I_1^{1/2}(\btheta)\btheta ~|~I_0^{1/2}(\bdeta)\bdeta, ~\omega_0\bm{I}_p)\pi_0(\bdeta)\left|\frac{dI_1^{1/2}(\btheta)\btheta}{d\btheta}\right|. \label{gen-com}
\end{equation}
The transformation to obtain this equation is a combination of $\theta^* = I_1^{1/2}(\btheta)\btheta$ and an identity transformation for $\bdeta$. It is straightforward to show the determinant of that full-rank transformation is equal to the determinant shown in \eqref{gen-com}. This modified Gen-straPP can be thought of as a type of commensurate prior in which the standardized current parameter is normally distributed about the standardized historical parameter when the commensurability parameter is equal to the inverse of $\omega_0$. 

\subsection{\label{method: glm} The straPP for Generalized Linear Models}
For the remainder of the paper, we assume that outcomes for the historical and current data arise from the class of 
generalized linear models (GLMs). 
Without loss of generality, we focus on development of the partial-borrowing straPP and note that with minor modifications one can similarly develop
the straPP, Gen-straPP, or partial-borrowing Gen-straPP.
When discussing GLMs, in a departure from previous notation, we will represent the parameters for the historical data model as $\bm{\xi}_0 = (\bbeta_0, \phi_0)$ and current data model as $\bm{\xi}_1 = (\bbeta_1, \phi_1)$, where $\bbeta_0$ and $\bbeta_1$ are  the regression parameter vectors and $\phi_0$ and $\phi_1$ are the scalar dispersion parameters.

Let $k$ index the historical ($k=0$) and current ($k=1$) data models, $\bm{Y}_{k}^T = (y_{k1}, \ldots, y_{kn_k})$ be the $n_k \times 1$
response vector, $\bm{X}_k$ be the $n_k \times p$ covariate matrix (with intercept) with $\bm{x}^T_{ki}$ denoting 
the covariate vector for the $i$th observation, and $\bm{\xi}_k = (\bbeta_k, \phi_k)$ be the GLM
parameters. 
The likelihood contribution for the $i$th case for dataset $k$ can be written as
\begin{equation}
f(y_{ki}\mid \bm{\xi}_k) 
= \exp\left[  
\phi_k\{y_{ki}h_k(\bm{x}_{ki}^T\bbeta_k) - b_k(h_k(\bm{x}_{ki}^T\bbeta_k)) - c_k(y_{ki})\} - \frac{1}{2}s_k(y_{ki},\phi_k) \right],
\label{glm}
\end{equation}
where $h_k(\cdot)$ is the link function, and  $b_k(\cdot)$,  $c_k(\cdot)$, and $s_k(\cdot)$ are known functions based on the particular
GLM family member. For the canonical link function, the  $p \times p$ Fisher information matrix for the regression parameters based on the likelihood associated with \eqref{glm} is given as
\begin{equation}
I_k(\bm{\xi}_k\mid \bm{X}_0) = \phi_k \bm{X}_0^TV_k(\bbeta_k)\bm{X}_0, \label{eq:regim}
\end{equation}
where  $V_k(\bbeta_k)$ = $\text{diag}\left\{ v_{ki}(\bbeta_k) \right\}$, with  $v_{ki}(\bbeta_k) = \ddot{b}_k(h_k(\bm{x}_{0i}^T\bbeta_k))$ 
for $i = 1,...,n_k$. Here $\text{diag}\left\{ v_{ki}(\bbeta_k) \right\}$ denotes a diagonal matrix with the $(i,i)$ element as $\left( v_{ki}(\bbeta_k) \right)$ and $\ddot{b}_k(\cdot)$ represents the second derivative of the function $b_k(\cdot)$ taken with respect to its scalar argument.

Consider a situation in which we only wish to borrow information on the regression parameters. Based on the general form of \eqref{eq:regim}, the transformation leading to the partial-borrowing straPP for GLMs is given by
\begin{equation}
I_0^{1/2}\left( \bm{\xi}_0 \right)\bbeta_0 = I_1^{1/2}\left( \bm{\xi}_1 \big| \bm{X}_0 \right)\bbeta_1. \label{gReg}
\end{equation}
We denote the transformation implied by \eqref{gReg} as $g_{\text{reg}}\left( \bm{\xi}_1, \phi_0 \big| \bm{X}_0 \right)$.
Note that in \eqref{gReg}, we develop the straPP by computing the information matrices for the current and historical data
models using the covariate matrix $\bm{X}_0$ associated with the historical data.
This is done for several reasons. 
First, it ensures the effective sample size for the straPP is equal to that of
the power prior from which it is derived. 
Second, it allows the straPP to be constructed using information entirely derived from the historical data
(e.g., outcome and covariates) which is appealing.

The formulation of the straPP leads to a conditional distribution for $\bbeta_1$ and $\phi_0$ given $\phi_1$ and thus
the straPP is completed by specifying an initial prior for $\phi_1$. Upon doing so, we obtain
$\pi_{g_{reg}}\left(\bbeta_1,\phi_0,\phi_1\big| D_0\right) 
= \pi_{s}\left( \bbeta_1, \phi_0 \big|\phi_1,D_0 \right)  \pi_0(\phi_1)$, 
where $\pi_{s}(\bbeta_1, \phi_0\mid \phi_1,D_0)$ is proportional to the expression
\begin{eqnarray}
\mathcal{L}\left( g_{\text{reg}}\left( \bm{\xi}_1, \phi_0 \big| \bm{X}_0 \right),\phi_0 \big|\phi_1,D_0\right)^{a_0}
\pi_0(g_{\text{reg}}\left( \bm{\xi}_1, \phi_0 \big| \bm{X}_0 \right),\phi_0\big|\phi_1)
\left|\frac{dg_{\text{reg}}\left( \bm{\xi}_1, \phi_0 \big| \bm{X}_0 \right)}{d\bbeta_1}\right|.  \label{gen-Regsub2}
\end{eqnarray}
Similar to above developments, the transformation to obtain the partial-borrowing straPP for GLMs is a 
combination of \eqref{gReg} and identity transformations for $\phi_0$ and $\phi_1$.
It is straightforward to show the determinant of that full-rank transformation is equal to the determinant 
in \eqref{gen-Regsub2}. The expression inside the Jacobian in which both the historical and current data models have the canonical link can be found in Appendix~\ref{jac_canon}.

\subsection{\label{linear model}The Linear Model Special Case}

For illustrative purposes, we consider a special case in which both the historical and current 
data arise from linear regression models with known variances (henceforth referred to as the normal-normal case). 
When variances are known, $\bm{\xi}_k = \bbeta_k$ and so, for ease of exposition, we simply write $\bbeta_0$ and $\bbeta_1$ to represent 
the complete parameter vector for the historical and current data models, respectively.
For this simple case, an elegant closed-form can be derived for the straPP which provides insight into its rescaling 
properties. For the linear model, the information matrix in \eqref{eq:regim} reduces to
$I_k(\bbeta_k\mid \bm{X}_0) =  \sigma_k^{-2} \bm{X}_0^T\bm{X}_0$, 
where $\phi_k = \sigma_k^{-2}$ is the inverse variance. Based on this, the transformation leading to the straPP in \eqref{transf1} reduces
to $\bbeta_0/\sigma_0 = \bbeta_1/\sigma_1$,
which nicely illustrates the equality of parameter values once scaled by the standard deviations for 
the associated outcomes.
In this simple setting, both the power prior and the straPP can be shown to be normal distributions, with  power prior as
\begin{eqnarray}
\bbeta_1  \widesim{PP} N\left((\bm{X}_0^T\bm{X}_0)^{-1}\bm{X}_0^T\bm{Y}_0,\text{ }\left(\frac{\sigma_0^2}{a_0}\right)(\bm{X}_0^T\bm{X}_0)^{-1}\right). \label{pp-norm}
\end{eqnarray}

For deriving the straPP, one can understand the scale transformation as a variable transformation on the regression parameter of the power prior for the historical data. We can write the power prior for the historical parameter as
$\bbeta_0  \sim N\Big((\bm{X}_0^T\bm{X}_0)^{-1}\bm{X}_0^T\bm{Y}_0,$ \\$\left(\sigma_0^2/a_0\right)(\bm{X}_0^T\bm{X}_0)^{-1}\Big)$.
The straPP transformation $\bbeta_1=(\sigma_1/\sigma_0)\bbeta_0$ is a function of the historical parameter, thus the current regression parameter is also distributed normally with mean $(\sigma_1/\sigma_0)\text{E}(\bbeta_0)$ and variance $(\sigma^2_1/\sigma^2_0)\text{Var}(\bbeta_0)$. We have
\begin{eqnarray}
\bbeta_1 \widesim{straPP} N\left(\frac{\sigma_1}{\sigma_0}(\bm{X}_0^T\bm{X}_0)^{-1}\bm{X}_0^T\bm{Y}_0,\text{ }\left(\frac{\sigma_1^2}{a_0}\right)(\bm{X}_0^T\bm{X}_0)^{-1}\right) \label{straPP-norm}.
\end{eqnarray}
One can see that the mean for the power prior in \eqref{pp-norm} is equal to the maximum likelihood estimate of $\bbeta$ based on the historical 
data, which we denote as $\hat{\bbeta}_0$. Therefore, the mean for the straPP is equal to $(\sigma_1/\sigma_0)\hat{\bbeta}_0$. 
Of equal importance, the variance of the straPP is equal to the variance of the power prior apart from the former
being a function of $\sigma_1^2$ and the latter $\sigma_0^2$. 
Thus, the information contained in the straPP is essentially recalibrated to be a function of the variance associated with 
the current data model instead of the historical data model.

\subsection{\label{prop} Properties of the straPP for the Linear Model}

For the normal-normal case when the assumption of the straPP transformation hold, the posterior mean based on an analysis with the straPP can be shown to be unbiased as a point estimator in the frequentist sense (see Section~\ref{proof}). It follows that the posterior mean based on an analysis with the (unscaled) power prior is biased. However, the rescaling property of the straPP can result in a prior with less precise information about the parameter (e.g., when $\sigma_1>\sigma_0$) and thus the variance of the posterior mean from an analysis with the straPP can exceed that of the power prior. This implies a trade-off between the bias and variance of the posterior mean point estimators, which becomes apparent when comparing their mean-squared error (MSE). Theorem \ref{thm} below gives conditions under which the posterior mean based on the straPP has a smaller MSE than the posterior mean based on the power prior.

\newtheorem{theorem}{Theorem}
\begin{theorem}
	Let $\beta_{1j}$ denote the $(j+1)$th element of $\bm{\beta}_1$ $(j = 0, \ldots, p-1)$.
	Further let $\hat{\beta}_{s,1j}$ and $\hat{\beta}_{p,1j}$ denote the posterior mean point estimators for the straPP and
	power prior, respectively.
	For the normal-normal case, when $\bbeta_0 = g(\bbeta_1)$ (i.e., the relationship 
	of the 	straPP transformation holds), the straPP estimator $\hat{\beta}_{s,1j}$ has a lower MSE than the power prior 
	estimator $\hat{\beta}_{p,1j}$ under the following condition:
	\begin{eqnarray}
	\tfrac{\emph{Var}(\hat{\beta}_{s,1j}) - \emph{Var}(\hat{\beta}_{p,1j})}{\left[\emph{Percent Bias}(\hat{\beta}_{p,1j})\right]^2} \le \beta_{1j}^2\label{mse thresh}.
	\end{eqnarray}
	\label{thm}
\end{theorem} 
In general, the percent bias of $\hat{\beta}_{p,1j}$ depends on $\beta_{1j}$. 

\subsubsection{Proof of Theorem \ref{thm}\label{proof}}

Let $\bm{Y}_{0} \sim N_{n_0}(\bm{X}_{0}g(\bm{\beta}_1), ~\sigma_0^2\bm{I}_{n_0})$ be the $n_0 \times 1$ vector of responses for the historical data and 
$\bm{Y}_{1} \sim N_{n_1}(\bm{X}_{1}\bm{\beta}_1, ~\sigma_1^2\bm{I}_{n_1})$ be the $n_1 \times 1$ vector of responses for the current data, with known historical and current data 
variance. 
Then the posterior distribution associated with an analysis based on the straPP and power prior are given by \eqref{straPP post} and \eqref{pp post}, respectively.
\begin{eqnarray}
\bm{\beta}_1 \mid  \bm{Y}_1, \bm{Y}_0  \widesim{straPP}N_p(\widehat{\bbeta}_{s}, ~\bSigma_{s})\label{straPP post} \\
\bm{\beta}_1 \mid  \bm{Y}_1, \bm{Y}_0  \widesim{PP} N_p(\widehat{\bbeta}_{p}, ~\bSigma_{p})\label{pp post},
\end{eqnarray}
where $\widehat{\bbeta}_{s} = \bSigma_{s}\left\{(1/\sigma_1^2)\bm{X}_1^T\bm{Y}_1 + a_0/(\sigma_0\sigma_1)\bm{X}_0^T\bm{Y}_0\right\}$, 
$\bSigma_{s} = \sigma_1^2(\bm{X}_1^T\bm{X}_1+ a_0\bm{X}_0^T\bm{X}_0)^{-1}$, \\
$\widehat{\bbeta}_{p} = \bSigma_{p}\left\{(1/\sigma_1^2)\bm{X}_1^T\bm{Y}_1 + (a_0/\sigma_0^2)\bm{X}_0^T\bm{Y}_0\right\}$, and 
$\bSigma_{p} = \{(1/\sigma_1^2)\bm{X}_1^T\bm{X}_1 + (a_0/\sigma_0^2)\bm{X}_0^T\bm{X}_0\}^{-1}$.

Now let $\widehat{\beta}_{s,1j}$ and $\widehat{\beta}_{p,1j}$ denote the straPP and power prior estimator for $\beta_{1j}$, respectively, where $\beta_{1j}$ is the $(j+1)$th element of $\bm{\beta}_1$ $(j = 0,\ldots, p-1)$. 
We wish to find the smallest $\beta_{1j}$ such that MSE$(\widehat{\beta}_{s,1j}) \le $ MSE$(\widehat{\beta}_{p,1j})$. Note that the MSE can be decomposed into the sum of the
point estimator's variance and squared bias (i.e., MSE = Var + $\left[\text{Bias}\right]^2$). 

First, we show that Bias($\widehat{\bbeta}_{s}$) = $\bm{0}$, and thus Bias($\widehat{\beta}_{s,1j}$) = 0. For the normal-normal case, the assumed relationship between the historical and current data model parameters is $\bbeta_0 = g(\bbeta_1) = (\sigma_0/\sigma_1)\bbeta_1$. Then E$(\bm{Y}_0)/(\sigma_0\sigma_1)$ = $ \bm{X}_0\bbeta_0/(\sigma_0\sigma_1) = \bm{X}_0 \bbeta_1/\sigma_1^2$. We can calculate the expectation of the straPP estimator as
\begin{flalign*}
\text{Bias}(\widehat{\bbeta}_{s}) = E(\widehat{\bbeta}_{s}) -\bbeta_1 & = \bSigma_{s}\left\{\frac{1}{\sigma_1^2}\bm{X}_1^TE(\bm{Y}_1) + \frac{a_0}{\sigma_0\sigma_1}\bm{X}_0^TE(\bm{Y}_0)\right\} -\bbeta_1\\
&= \bSigma_{s}\left\{\frac{1}{\sigma_1^2}\left(\bm{X}_1^T\bm{X}_1 + a_0\bm{X}_0^T\bm{X}_0\right)\right\}\bm{\beta}_1 -\bbeta_1\\
&= \bSigma_{s}\left(\bSigma_{s}\right)^{-1}\bm{\beta}_1 -\bbeta_1
=  \bm{0}.
\end{flalign*}

Then, for non-zero $\beta_{1j}$, it follows that 
\begin{eqnarray*}
	\text{Var}(\widehat{\beta}_{s,1j}) \le \text{Var}(\widehat{\beta}_{p,1j}) + \left\{\text{Bias}(\widehat{\beta}_{p,1j})\right\}^2 & \Leftrightarrow &
	\text{Var}(\widehat{\beta}_{s,1j}) \le  \text{Var}(\widehat{\beta}_{p,1j}) + \beta_{1j}^2\left\{\text{Percent Bias}(\widehat{\beta}_{p,1j})\right\}^2\\
	& \Leftrightarrow &
	\frac{\text{Var}(\widehat{\beta}_{s,1j}) - \text{Var}(\widehat{\beta}_{p,1j})}{\left\{\text{Percent Bias}(\widehat{\beta}_{p,1j})\right\}^2} \le \beta_{1j}^2.
\end{eqnarray*}

\section{Simulation Studies \label{sims}}

In this section, we present and discuss results from a collection of simulation studies designed to evaluate the performance of the straPP and Gen-straPP compared to each other as well as to the power prior, asymptotic power prior, commensurate prior, and uniform improper prior. For the commensurate prior, the prior for the commensurability parameter was taken to be gamma with shape hyperparameter equal to 2, and with multiple choices for the inverse scale parameter, which we denoted $b_0$.
In Section~\ref{sim:NN}, we present simulation studies for the normal-normal case described above where both the historical and current data models are linear regression models with known variances. 
In Section~\ref{sim:BP}, we present simulation studies for a case where the historical data follow a logistic regression model and the current data follow a loglinear regression model, which we call the binary-Poisson case. Lastly, in Section~\ref{sim:BN violate}, we present simulation studies for a case in which the straPP assumption does not hold and the historical data follow a logistic regression model and the current data follow a linear regression model with known variance, which we call the binary-normal case. 

In all of the simulations, we simulated the historical and current data to have an intercept and a treatment indicator
such that 50\% of simulated patients were treated. 
For each simulation, we generated 5,000 historical and current data sets for each unique parameter combination, and 
used Metropolis-Hastings Markov chain Monte Carlo (MCMC) methods to obtain 25,000 samples from the posterior using 
various priors to analyze each dataset. Additional simulations can be found in Appendix~\ref{add_sims}.

\subsection{Simulation Studies for the Normal-Normal Case \label{sim:NN}}

For the normal-normal case, we performed simulation studies using parameter values that obey the assumption of the 
straPP transformation (e.g., $\bbeta_0 = g(\bbeta_1)$). We then simulated the historical data and 
current data based on the corresponding linear regression models. Thus, for this simulation, we are effectively 
evaluating the performance of the straPP compared to the alternative priors for the case where
the straPP transformation assumption holds. 

For the normal-normal case, the percent bias of the power prior posterior mean estimator for the treatment effect 
(i.e., $\hat{\beta}_{p,11}$) does not depend on the actual treatment effect (i.e., $\beta_{11}$). As a 
result, one can calculate the exact threshold, denoted as $\beta_{11}^*$, where the MSE for the posterior mean estimators
based on the straPP and power prior are equal, as
\begin{align}
\beta_{11}^* &= \pm \frac{\sqrt{\text{Var}(\hat{\beta}_{s,11}) - \text{Var}(\hat{\beta}_{p,11}) }}{\text{Percent Bias}(\hat{\beta}_{p,11})}\nonumber
= \pm  \frac{2(n_1\sigma_0^2+a_0n_0\sigma_1^2)}{a_0n_0(\sigma_0 - \sigma_1)}\sqrt{\frac{n_1 + a_0^2n_0}{(n_1+a_0n_0)^2} - \sigma_0^2\frac{n_1\sigma_0^2 + a_0^2n_0\sigma_1^2}{(n_1\sigma_0^2 + a_0n_0\sigma_1^2)^2}}.&&
\end{align}

We performed two sets of simulation studies for the normal-normal case. In both sets of simulations, the following were considered: $n_0$ = 50, $n_1$ =  100, $a_0 = 0.5$, $\beta_{10} = 1$, and $\beta_{11} \in [0.0,1.8]$. 
For the first set of simulation studies, we considered a case where the historical data variance exceeded that of the current data (i.e., $\sigma_0 = 3 > \sigma_1 = 1$). For the second set of simulation studies, we reversed the relationship between the variances. 
The values of the historical data model parameters were then identified by solving $\bbeta_0 = (\sigma_0/\sigma_1)\bbeta_1$.
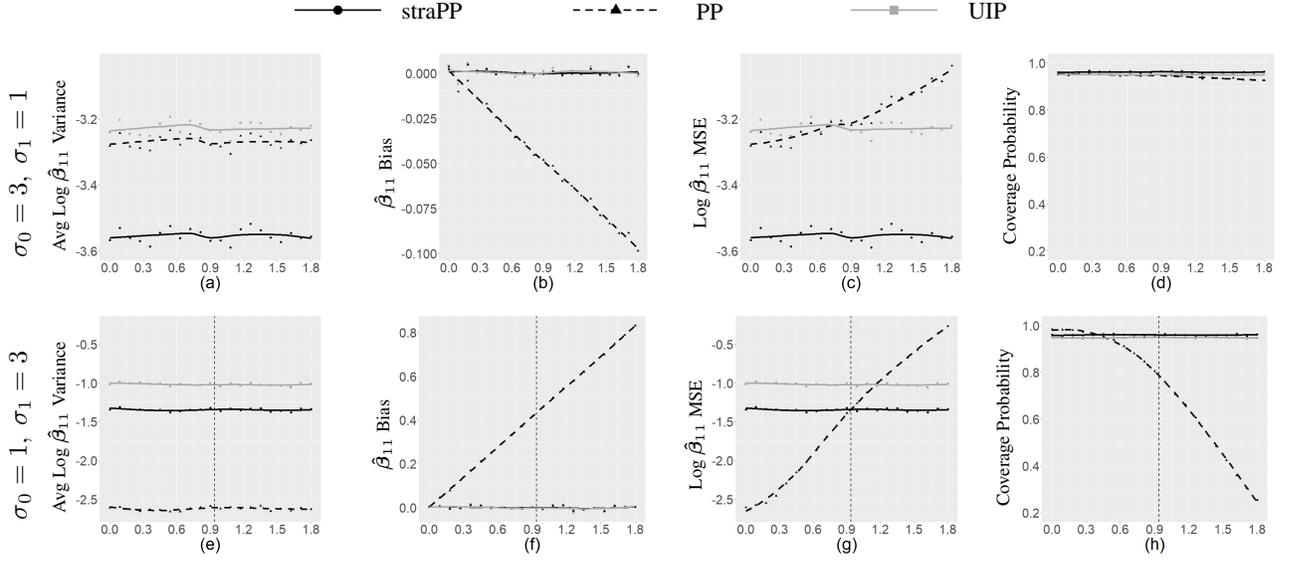
\begin{figure}
	\centering
	\begin{minipage}[t]{0.1\textwidth}
		\text{ }
	\end{minipage}
	\hfill
	\begin{minipage}[t]{0.1\textwidth}
		\begin{tikzpicture}
		\node (ls)  {	\includegraphics[scale=0.05]{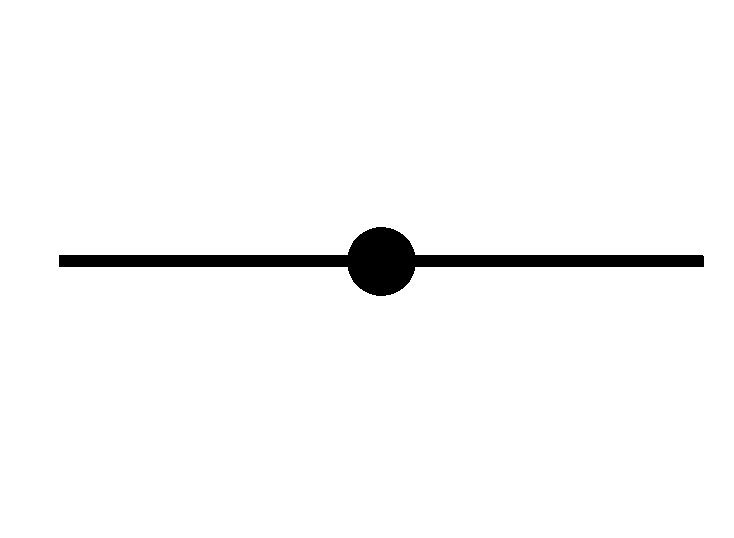} };
		\node[right=of ls, node distance=0cm, anchor=center, xshift = -0.5cm, font=\color{black}] {\footnotesize straPP};
		\end{tikzpicture}
	\end{minipage}
	\hfill
	\begin{minipage}[t]{0.1\textwidth}
		\begin{tikzpicture}
		\node (lp)  {	\includegraphics[scale=0.05]{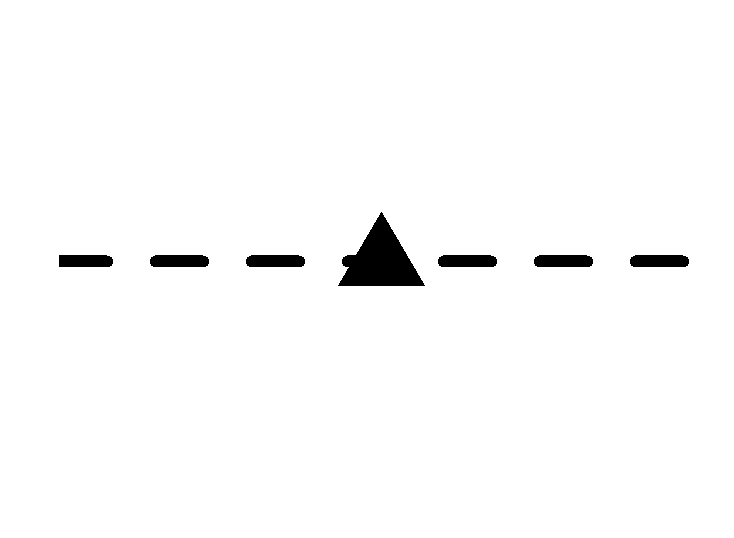} };
		\node[right=of lp, node distance=0cm, anchor=center, xshift = -0.5cm, font=\color{black}] {\footnotesize PP};
		\end{tikzpicture}
	\end{minipage}
	\hfill
	\begin{minipage}[t]{0.1\textwidth}
		\begin{tikzpicture}
		\node (lu)  {	\includegraphics[scale=0.05]{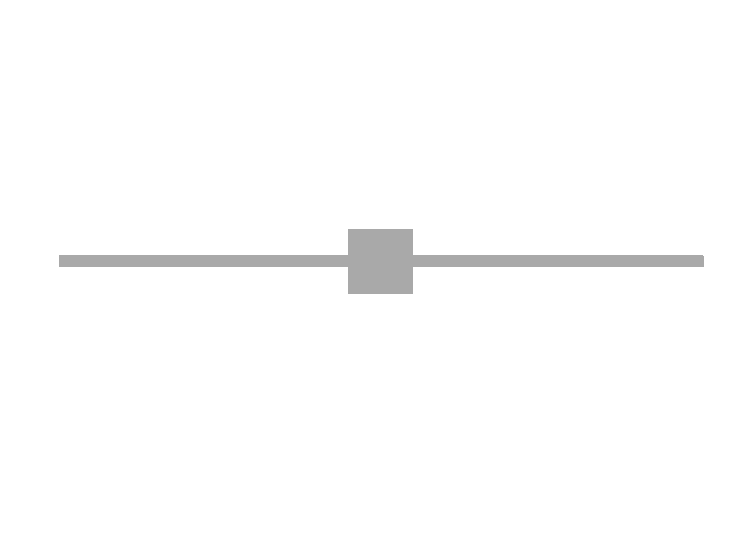} };
		\node[right=of lu, node distance=0cm, anchor=center, xshift = -0.5cm, font=\color{black}] {\footnotesize UIP};
		\end{tikzpicture}
	\end{minipage}
	\hfill
	\begin{minipage}[t]{0.1\textwidth}
		\text{ }
	\end{minipage}
	\vspace{-0.2cm}
	
	\begin{minipage}[t]{0.24\textwidth}
		\begin{tikzpicture}
		\node (img5)  {\includegraphics[scale=0.16]{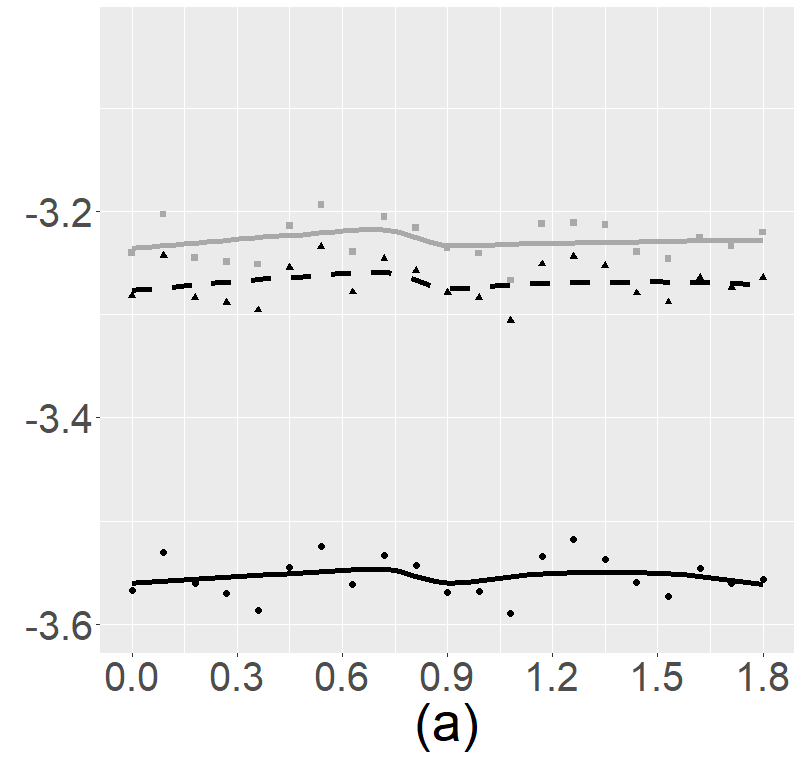}};
		\node[left=of img5, node distance=0cm, rotate=90, anchor=center,yshift=-1cm, xshift = 0.1cm, font=\color{black}] {\scriptsize Avg Log $\bm{\hat{\beta}}_{11}$ Variance};
		\node[left=of img5, node distance=0cm, rotate=90, anchor=center,yshift=-.5cm,font=\color{black}] { $\sigma_0 = 3, \text{ } \sigma_1 = 1$};
		\end{tikzpicture}
	\end{minipage}
	\hfill
	\begin{minipage}[t]{0.2\textwidth}
		\begin{tikzpicture}
		\node (img6)  {\includegraphics[scale=0.16]{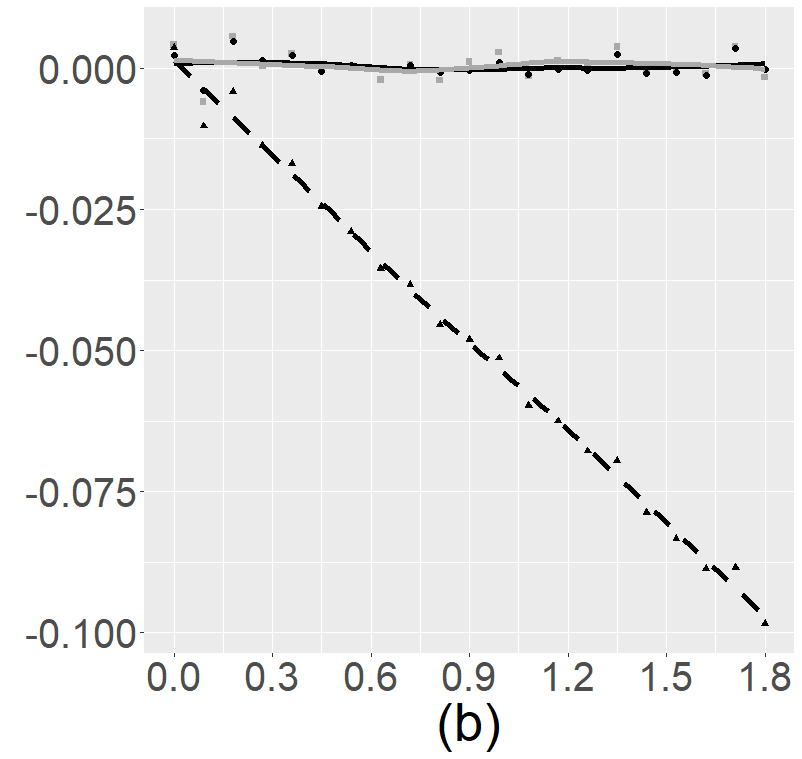}};
		\node[left=of img6, node distance=0cm, rotate=90, anchor=center,yshift=-1cm] {\scriptsize $\bm{\hat{\beta}}_{11}$ Bias};
		\end{tikzpicture}
	\end{minipage}
	\hfill
	\begin{minipage}[t]{0.2\textwidth}
		\begin{tikzpicture}
		\node (img7)  {\includegraphics[scale=0.16]{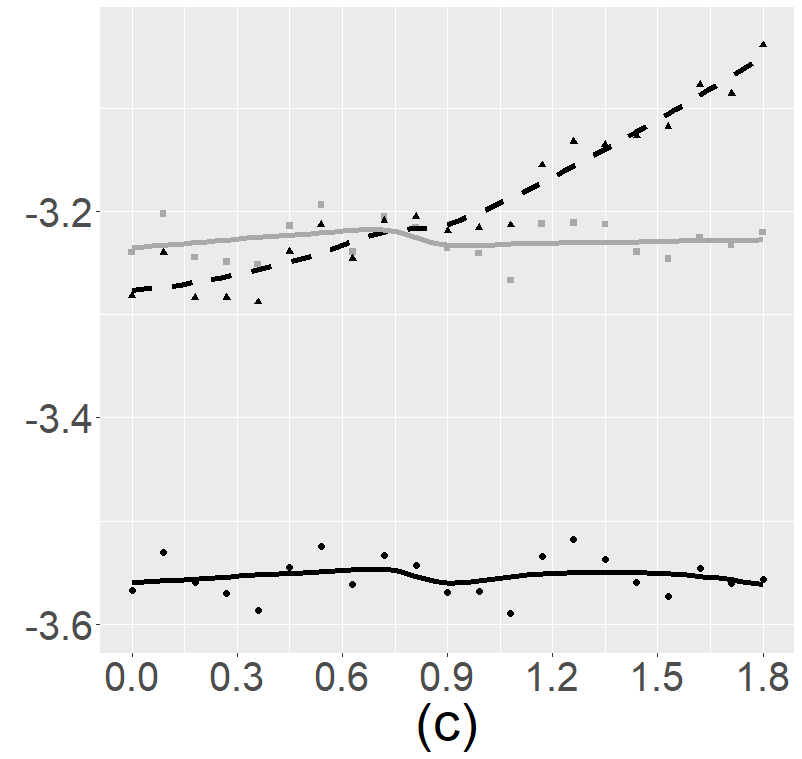}};
		\node[left=of img7, node distance=0cm, rotate=90, anchor=center,yshift=-1cm] {\scriptsize Log $\bm{\hat{\beta}}_{11}$ MSE};
		\end{tikzpicture}
	\end{minipage}
	\hfill
	\begin{minipage}[t]{0.2\textwidth}
		\begin{tikzpicture}
		\node (img8)  {\includegraphics[scale=0.16]{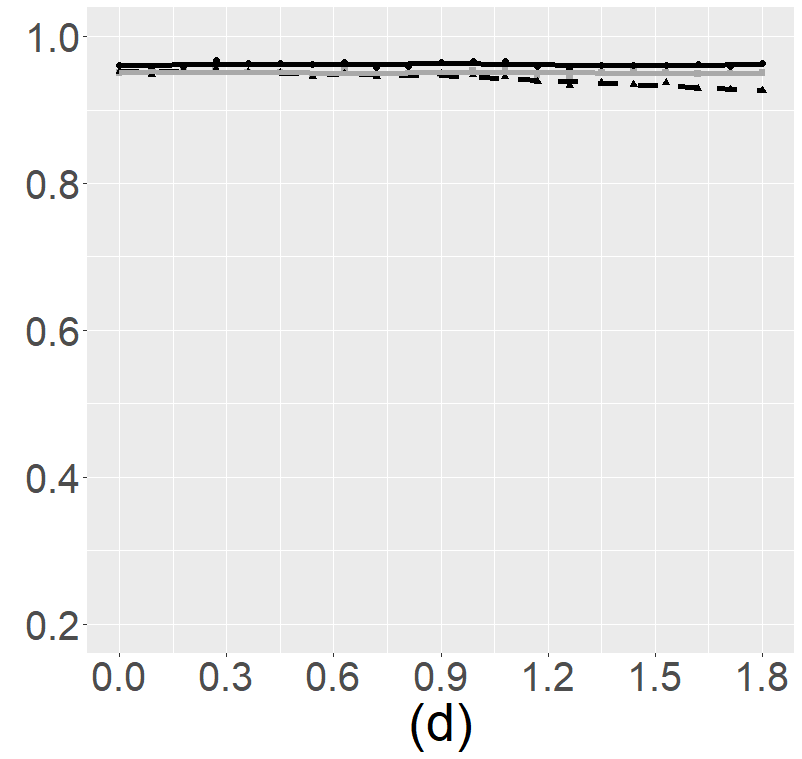}};
		\node[left=of img8, node distance=0cm, rotate=90, anchor=center,yshift=-1cm, xshift = 0.1cm] {\scriptsize Coverage Probability};
		\end{tikzpicture}
	\end{minipage}\\
	
	\begin{minipage}[t]{0.24\textwidth}	
		\begin{tikzpicture}
		\node (img1)  {\includegraphics[scale=0.16]{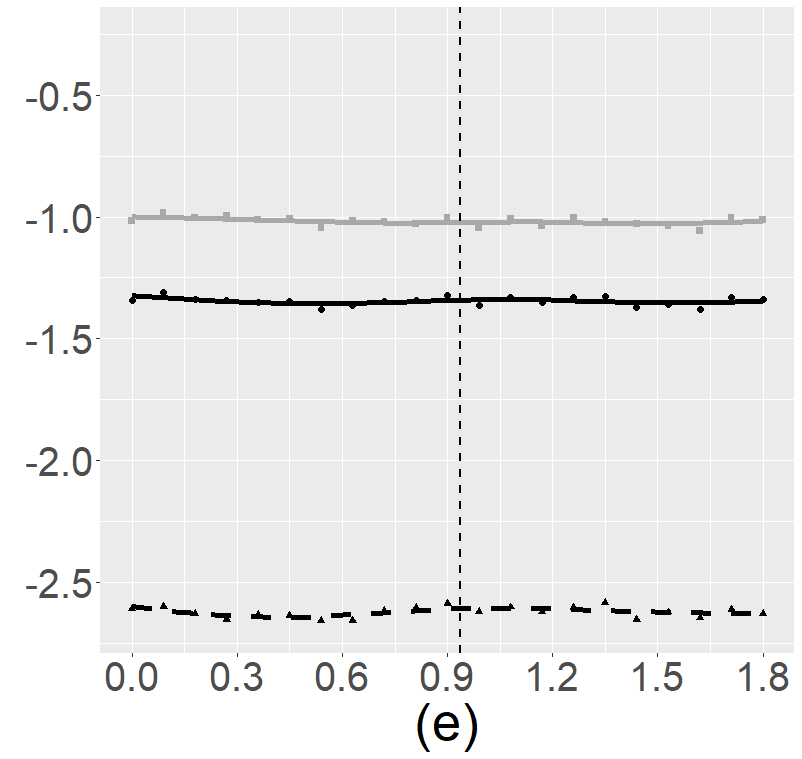}};
		\node[left=of img1, node distance=0cm, rotate=90, anchor=center,yshift=-1cm, xshift = 0.1cm, font=\color{black}] {\scriptsize Avg Log $\bm{\hat{\beta}}_{11}$ Variance};
		\node[left=of img1, node distance=0cm, rotate=90, anchor=center,yshift=-.5cm,font=\color{black}] { $\sigma_0 = 1, \text{ } \sigma_1 = 3$};
		\end{tikzpicture}
	\end{minipage}
	\hfill
	\begin{minipage}[t]{0.2\textwidth}
		\begin{tikzpicture}
		\node (img2)  {\includegraphics[scale=0.16]{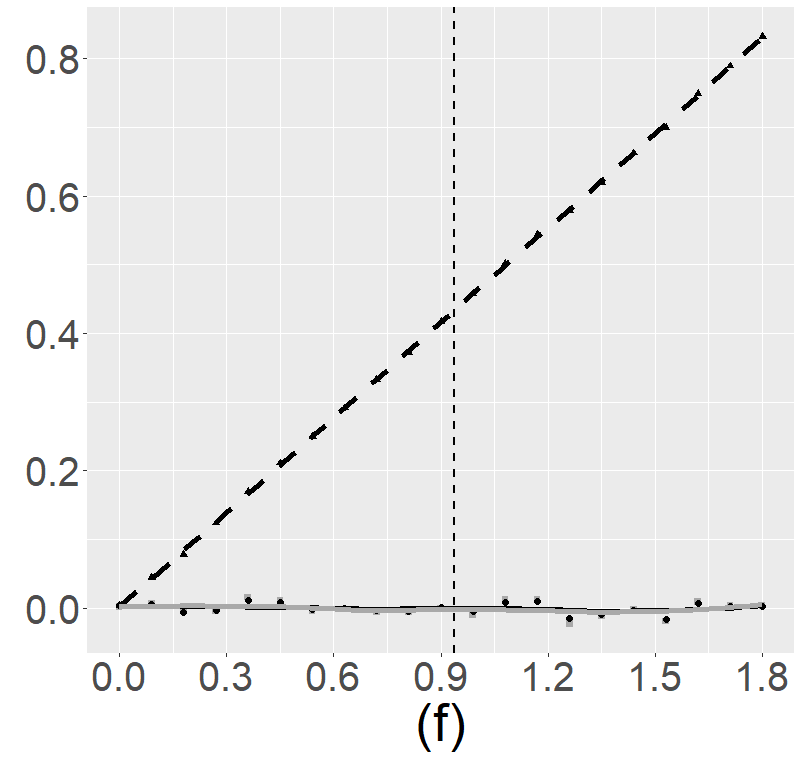}};
		\node[left=of img2, node distance=0cm, rotate=90, anchor=center,yshift=-1cm] {\scriptsize $\bm{\hat{\beta}}_{11}$ Bias};
		\end{tikzpicture}
	\end{minipage}
	\hfill
	\begin{minipage}[t]{0.2\textwidth}
		\begin{tikzpicture}
		\node (img3)  {\includegraphics[scale=0.16]{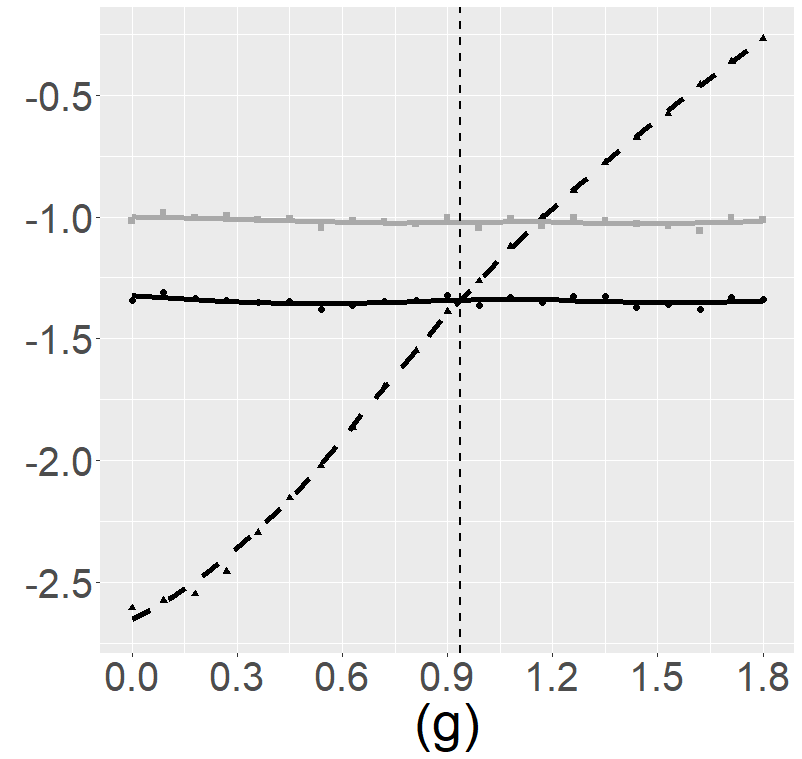}};
		\node[left=of img3, node distance=0cm, rotate=90, anchor=center,yshift=-1cm] {\scriptsize Log $\bm{\hat{\beta}}_{11}$ MSE};
		\end{tikzpicture}
	\end{minipage}
	\hfill
	\begin{minipage}[t]{0.2\textwidth}
		\begin{tikzpicture}
		\node (img4)  {\includegraphics[scale=0.16]{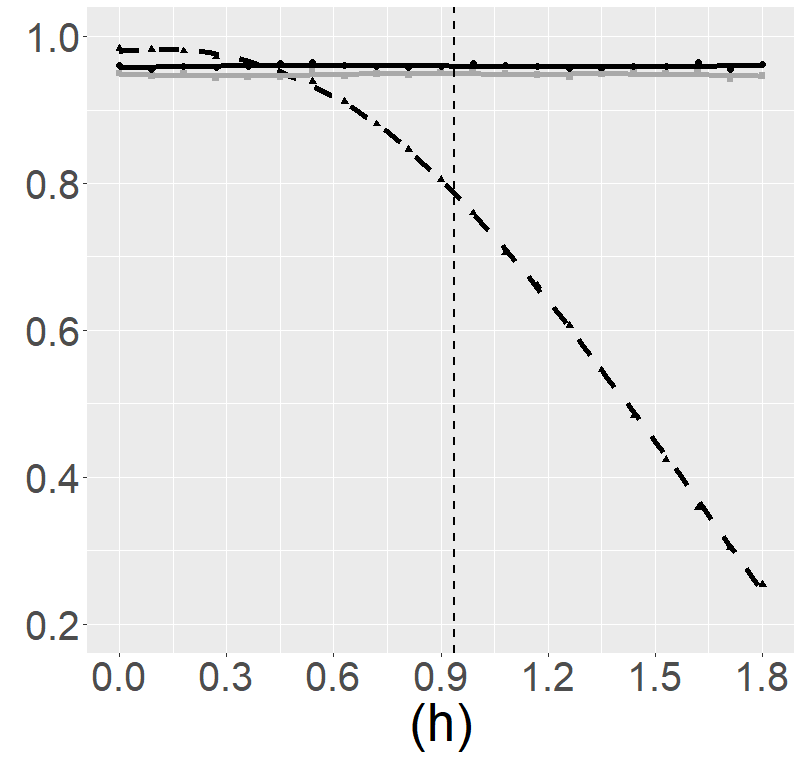}};
		\node[left=of img4, node distance=0cm, rotate=90, anchor=center,yshift=-1cm, xshift = 0.1cm] {\scriptsize Coverage Probability};
		\end{tikzpicture}
	\end{minipage}
	\caption{Panels (a)-(d) present the average log variance, bias, log MSE, and coverage probability for the posterior mean of $\beta_{11}$, respectively, as a function of the true value of $\beta_{11}$ plotted on the x-axis for the	case where $\sigma_0 =3, \sigma_1 =1$ for the scale transformed power prior, power prior, and uniform improper prior. Panels (e)-(h) present the same information for the case where $\sigma_0 =1, \sigma_1 =3$. 
	}\label{fig:NN sim}
\end{figure}

Figure~\ref{fig:NN sim} panels (a)-(d) present the results of the first set of the simulation studies and panels (e)-(h) present the results from the second set. For the first set of simulation studies, since the rescaling action of the straPP leads to more precise information about the parameter in the current data model, the theoretical threshold $\beta_{11}^*$ for MSE equivalence between the power prior and straPP is never crossed and it can be seen that the straPP has uniformly better performance than the power prior and the uniform improper prior. For the second set, where the variance of the historical data is less than that of the current data, the rescaling action of the straPP results in a prior that provides more accurate inference (e.g., less bias in the posterior mean) but also less precision on average (i.e., larger variability in the posterior mean). For this case, there is a tradeoff in terms of MSE with the straPP having superior MSE only for values of $\beta_{11}$ that exceed $\beta_{11}^*$ = 0.9364. Nonetheless, the straPP still provides more accurate information even if the MSE of the posterior mean is not always superior to that of the power prior. Furthermore, the coverage probability of the straPP remains approximately 0.95, while that of the power prior decreases greatly as the true value of $\beta_{11}$ increases.
Additional simulation studies were performed using other inputs. Importantly, the results presented here are indicative of the general behavior of the straPP with the only material difference between simulations being that the threshold for MSE equality between the straPP and the power prior varies depending on the particular inputs used for simulation in accordance with the formula given above. Note that the asymptotic power prior was not considered for the normal-normal case because, in this scenario, it is equivalent to the power prior.

\subsection{Simulation Studies for the Binary-Poisson Case \label{sim:BP}}

In the binary-Poisson simulation, we add a third covariate, standardized age, to the historical and current data. Standardized age was randomly generated using a standard normal distribution. Additionally, for these simulations we did not borrow on the intercept. We performed simulation studies using parameter values that obey the assumption of the partial-borriwng straPP transformation (e.g., $\bdeta_{1} = g(\beta_{00},\bbeta_1)$, in which $\bdeta_{1}^T = (\beta_{01}, \beta_{02})$). However, under the binary-Poisson simulation, both the historical and current information matrices involve the regression parameters, thus we must use a Metropolis-Hastings sampling algorithm in which proposed values of $\bbeta_0$ and $\bbeta_1$ satisfy the partial-borrowing straPP transformation. More details on this can be found in Appendix~\ref{alg}.

For the binary-Poisson simulation, the following were considered: $n_0$ = 150, $n_1$ =  150, $\beta_{10} = 0.2$, $\beta_{11} \in [0.05,0.35)$, $\beta_{12} = -0.1$, and $\beta_{00} = -0.6$. The value of $\bdeta_{1}$ was calculated as $\bdeta_{1} = g(\beta_{00},\bbeta_{1})$ via a non-linear problem solver. The values of $\bbeta_0$ and $\bbeta_1$ were chosen such that all probabilities from the historical response were between 0.15 and 0.85, and all means for the current response were between 1 and 3. For these simulations, we use $a_0 = 1$, so as to directly compare the Gen-straPP to the commensurate prior as described in Section~\ref{relation}. For the Gen-straPP, we considered $\omega_0 = 1, 2, 4$. For the commensurate prior, we specified the inverse scale hyperparameter for the gamma prior on the commensurability parameter to be one of $b_0$ = 1, 8, 16.

\begin{figure}[h]
	\centering
	\begin{minipage}[t]{0.1\textwidth}
		\text{ }
	\end{minipage}
	\hfill
	\begin{minipage}[t]{0.1\textwidth}
		\begin{tikzpicture}
		\node (ls)  {	\includegraphics[scale=0.05]{./figures/strapp_legend.png} };
		\node[right=of ls, node distance=0cm, anchor=center, xshift = -0.5cm, font=\color{black}] {\scriptsize straPP};
		\end{tikzpicture}
	\end{minipage}
	\hfill
	\begin{minipage}[t]{0.1\textwidth}
		\begin{tikzpicture}
		\node (lp)  {	\includegraphics[scale=0.05]{./figures/pp_legend.png} };
		\node[right=of lp, node distance=0cm, anchor=center, xshift = -0.5cm, font=\color{black}] {\scriptsize PP};
		\end{tikzpicture}
	\end{minipage}
	\hfill
	\begin{minipage}[t]{0.1\textwidth}
		\begin{tikzpicture}
		\node (lp)  {	\includegraphics[scale=0.05]{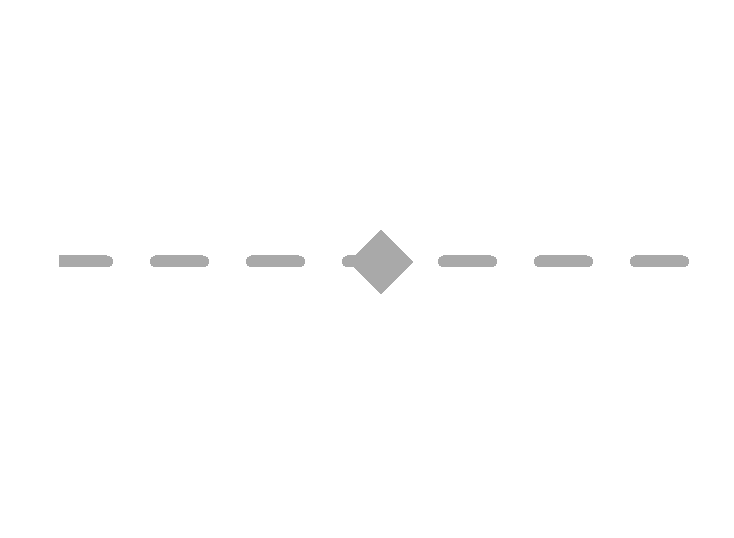} };
		\node[right=of lp, node distance=0cm, anchor=center, xshift = -0.5cm, font=\color{black}] {\scriptsize APP};
		\end{tikzpicture}
	\end{minipage}
	\hfill
	\begin{minipage}[t]{0.1\textwidth}
		\begin{tikzpicture}
		\node (lu)  {	\includegraphics[scale=0.05]{./figures/ui_legend.png} };
		\node[right=of lu, node distance=0cm, anchor=center, xshift = -0.5cm, font=\color{black}] {\scriptsize UIP};
		\end{tikzpicture}
	\end{minipage}
	\hfill
	\begin{minipage}[t]{0.1\textwidth}
		\text{ }
	\end{minipage}
	\vspace{-0.4cm}
	
	\begin{minipage}[t]{0.23\textwidth}
		\begin{tikzpicture}
		\node (img1)  {\includegraphics[scale=0.16]{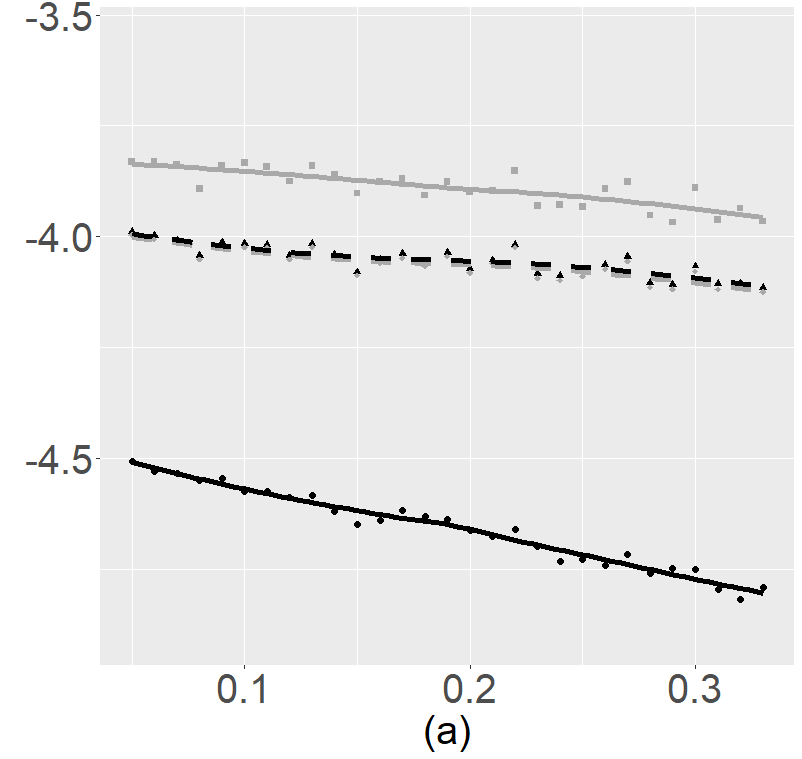}};
		\node[left=of img1, node distance=0cm, rotate=90, anchor=center,yshift=-1cm, xshift = 0.1cm, font=\color{black}] {\scriptsize Avg Log $\bm{\hat{\beta}}_{11}$ Variance};
		\end{tikzpicture}
	\end{minipage}
	\hfill
	\begin{minipage}[t]{0.23\textwidth}
		\begin{tikzpicture}
		\node (img2)  {\includegraphics[scale=0.16]{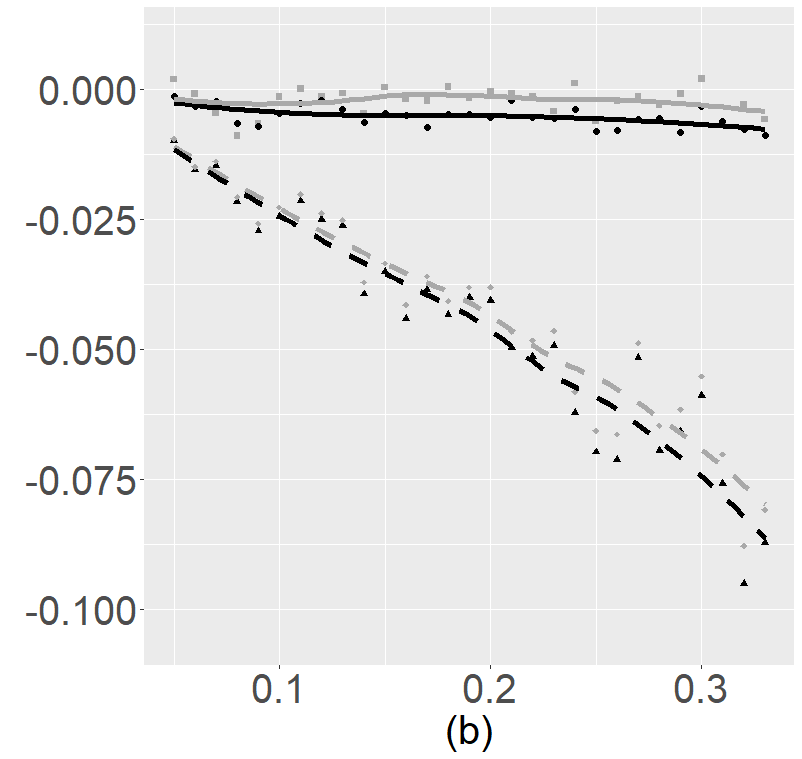}};
		\node[left=of img2, node distance=0cm, rotate=90, anchor=center,yshift=-1cm] {\scriptsize $\bm{\hat{\beta}}_{11}$ Bias};
		\end{tikzpicture}
	\end{minipage}
	\hfill
	\begin{minipage}[t]{0.23\textwidth}
		\begin{tikzpicture}
		\node (img3)  {\includegraphics[scale=0.16]{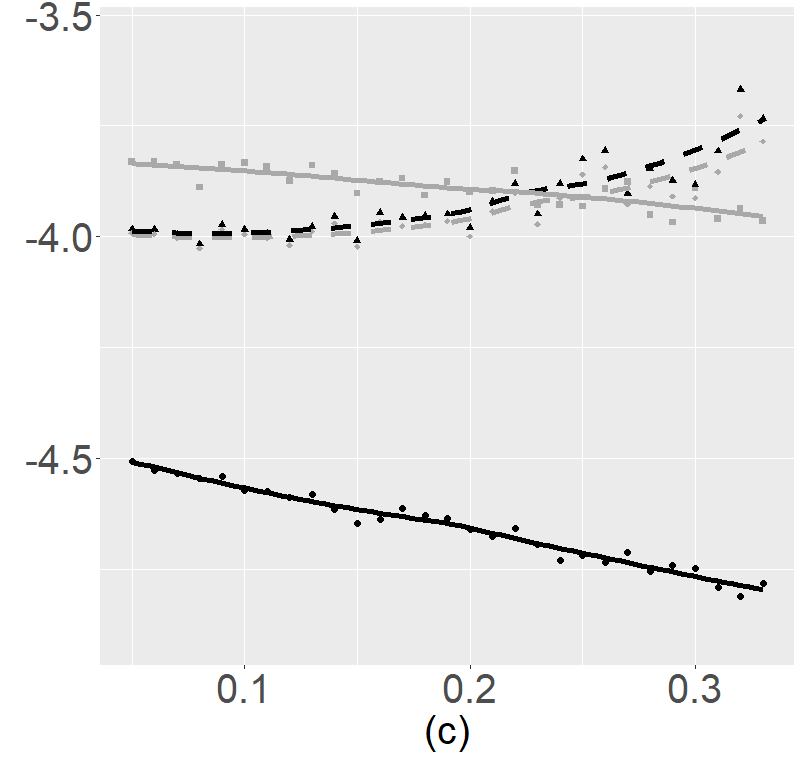}};
		\node[left=of img3, node distance=0cm, rotate=90, anchor=center,yshift=-1cm] {\scriptsize Log $\bm{\hat{\beta}}_{11}$ MSE};
		\end{tikzpicture}
	\end{minipage}
	\hfill
	\begin{minipage}[t]{0.23\textwidth}
		\begin{tikzpicture}
		\node (img4)  {\includegraphics[scale=0.16]{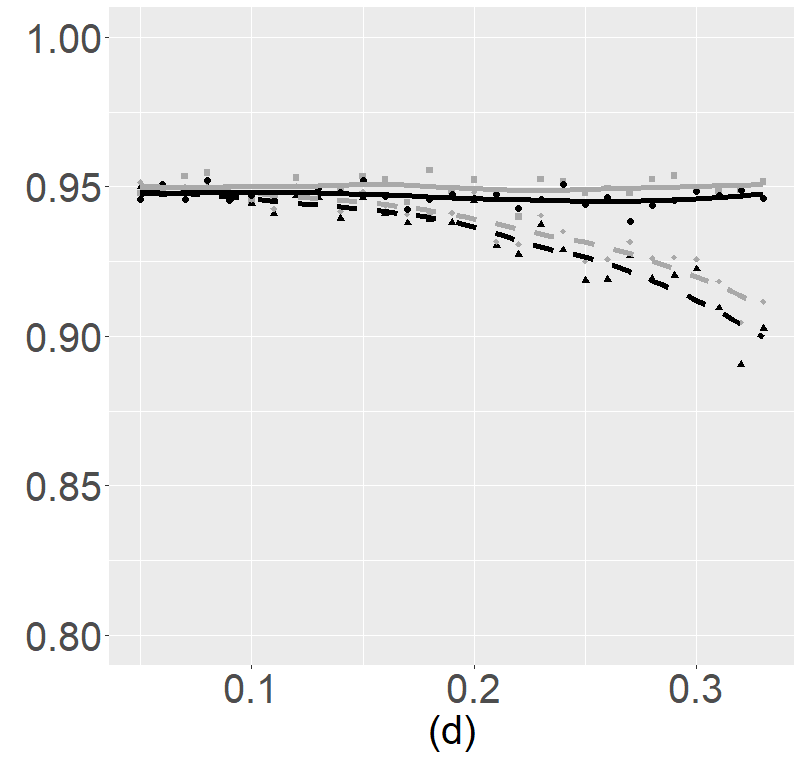}};
		\node[left=of img4, node distance=0cm, rotate=90, anchor=center,yshift=-1cm, xshift = 0.1cm] {\scriptsize Coverage Probability};
		\end{tikzpicture}
	\end{minipage}
	\vspace{-0.2cm}
	
	\begin{minipage}[t]{0.05\textwidth}
		\text{ }
	\end{minipage}
	\hfill
	\begin{minipage}[t]{0.1\textwidth}
		\begin{tikzpicture}
		\node (ls)  {	\includegraphics[scale=0.05]{./figures/strapp_legend.png} };
		\node[right=of ls, node distance=0cm, anchor=center, xshift = -0.5cm, font=\color{black}] {\scriptsize straPP};
		\end{tikzpicture}
	\end{minipage}
	\hfill
	\begin{minipage}[t]{0.15\textwidth}
		\begin{tikzpicture}
		\node (lp)  {	\includegraphics[scale=0.05]{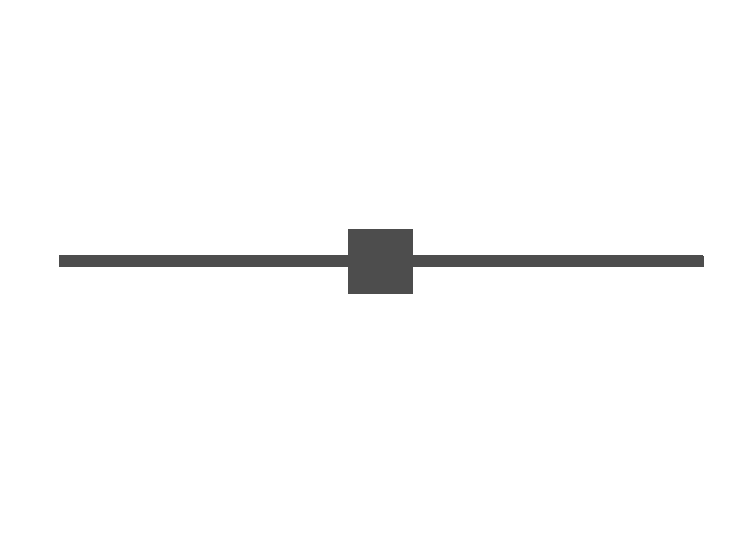} };
		\node[right=of lp, node distance=0cm, anchor=center, xshift = -0.3cm, font=\color{black}] {\scriptsize GS($\omega_0 = 1$)};
		\end{tikzpicture}
	\end{minipage}
	\hfill
	\begin{minipage}[t]{0.15\textwidth}
		\begin{tikzpicture}
		\node (lp)  {	\includegraphics[scale=0.05]{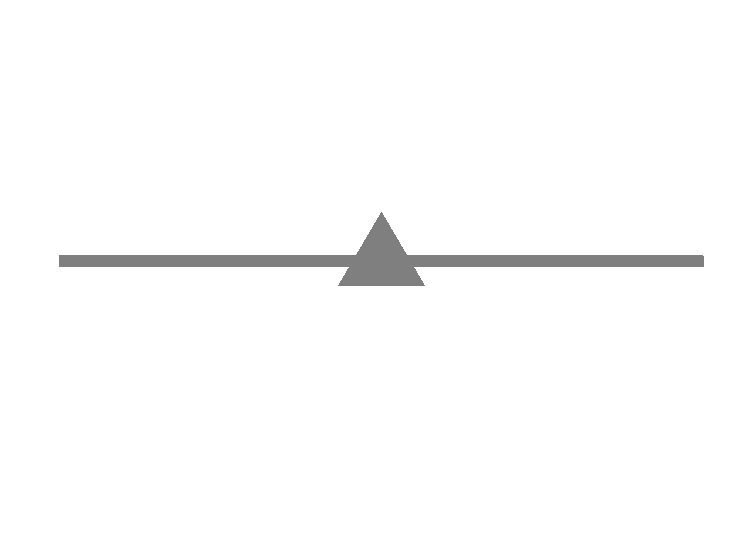} };
		\node[right=of lp, node distance=0cm, anchor=center, xshift = -0.3cm, font=\color{black}] {\scriptsize GS($\omega_0 = 2$)};
		\end{tikzpicture}
	\end{minipage}
	\hfill
	\begin{minipage}[t]{0.15\textwidth}
		\begin{tikzpicture}
		\node (lu)  {	\includegraphics[scale=0.05]{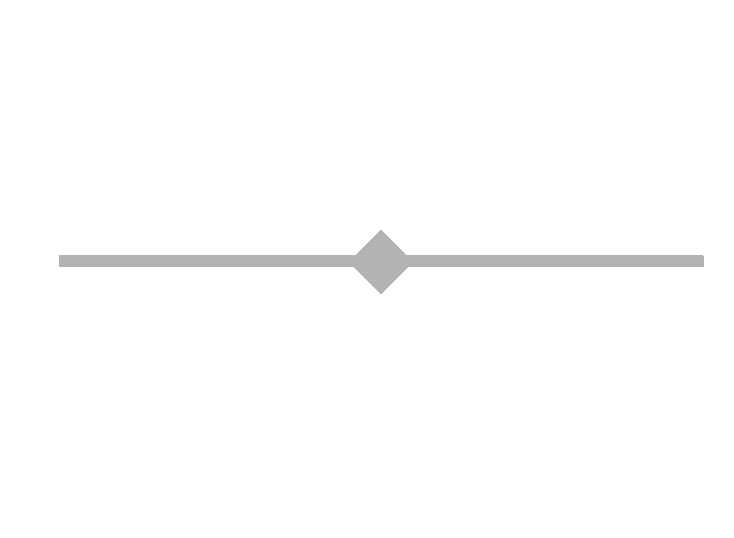} };
		\node[right=of lu, node distance=0cm, anchor=center, xshift = -0.3cm, font=\color{black}] {\scriptsize GS($\omega_0 = 4$)};
		\end{tikzpicture}
	\end{minipage}
	\hfill
	\begin{minipage}[t]{0.05\textwidth}
		\text{ }
	\end{minipage}\\
	\vspace{-0.6cm}
	\begin{minipage}[t]{0.05\textwidth}
		\text{ }
	\end{minipage}
	\hfill
	\begin{minipage}[t]{0.15\textwidth}
		\begin{tikzpicture}
		\node (lp)  {	\includegraphics[scale=0.05]{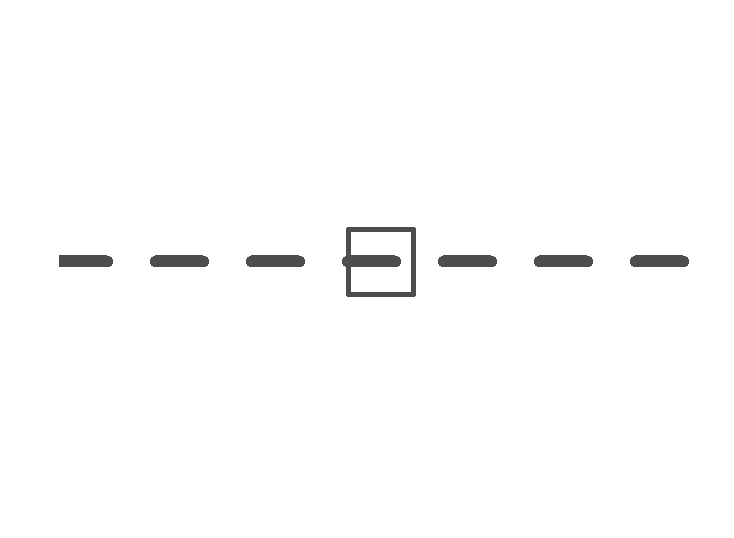} };
		\node[right=of lp, node distance=0cm, anchor=center, xshift = -0.3cm, font=\color{black}] {\scriptsize COM($b_0$ = 1)};
		\end{tikzpicture}
	\end{minipage}
	\hfill
	\begin{minipage}[t]{0.15\textwidth}
		\begin{tikzpicture}
		\node (lp)  {	\includegraphics[scale=0.05]{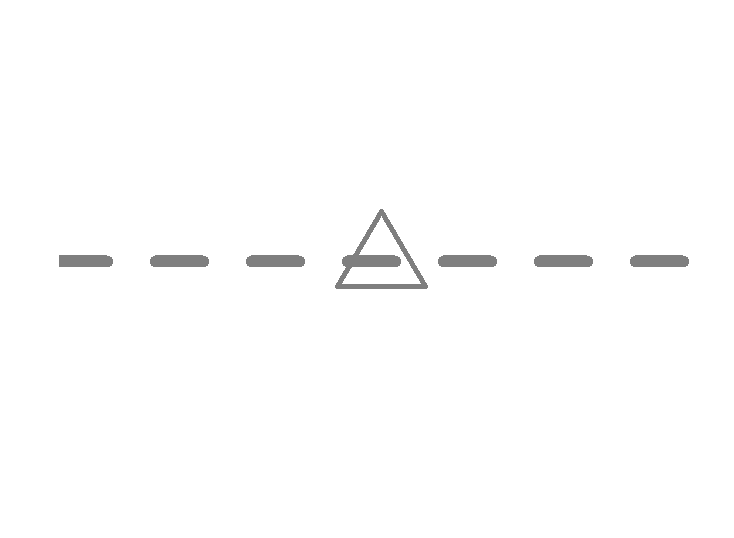} };
		\node[right=of lp, node distance=0cm, anchor=center, xshift = -0.3cm, font=\color{black}] {\scriptsize COM($b_0$ = 4)};
		\end{tikzpicture}
	\end{minipage}
	\hfill
	\begin{minipage}[t]{0.15\textwidth}
		\begin{tikzpicture}
		\node (lu)  {	\includegraphics[scale=0.05]{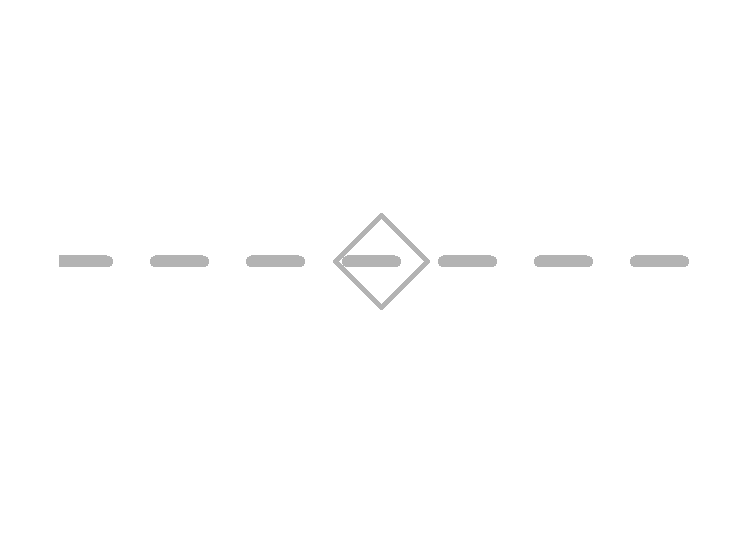} };
		\node[right=of lu, node distance=0cm, anchor=center, xshift = -0.3cm, font=\color{black}] {\scriptsize COM($b_0$ = 16)};
		\end{tikzpicture}
	\end{minipage}
	\hfill
	\begin{minipage}[t]{0.05\textwidth}
		\text{ }
	\end{minipage}

	\vspace{-0.4cm}
	
	\begin{minipage}[t]{0.23\textwidth}
		\begin{tikzpicture}
		\node (img5)  {\includegraphics[scale=0.16]{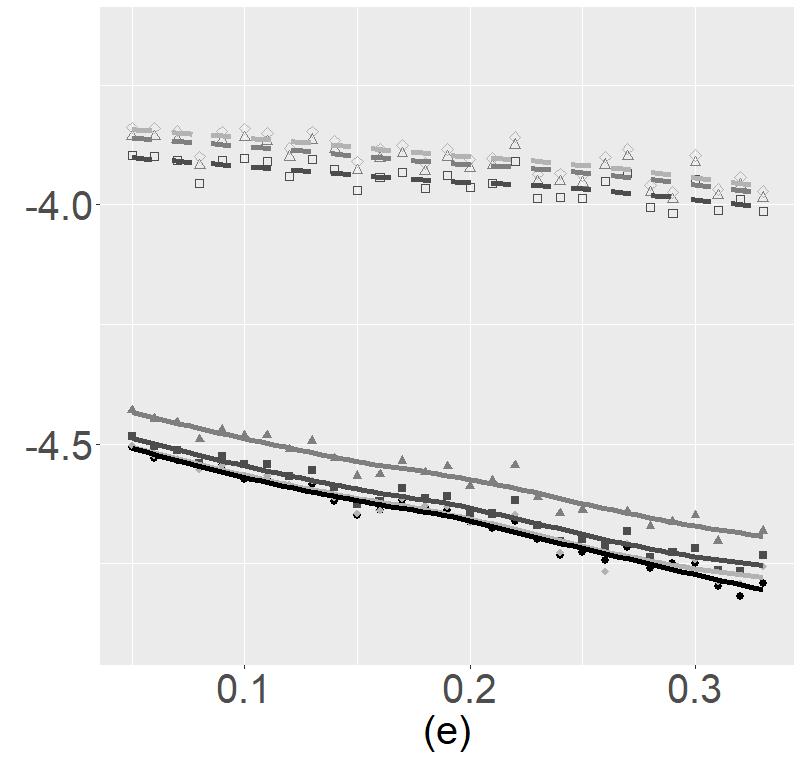}};
		\node[left=of img5, node distance=0cm, rotate=90, anchor=center,yshift=-1cm, xshift = 0.1cm, font=\color{black}] {\scriptsize Avg Log $\bm{\hat{\beta}}_{11}$ Variance};
		\end{tikzpicture}
	\end{minipage}
	\hfill
	\begin{minipage}[t]{0.23\textwidth}
		\begin{tikzpicture}
		\node (img6)  {\includegraphics[scale=0.16]{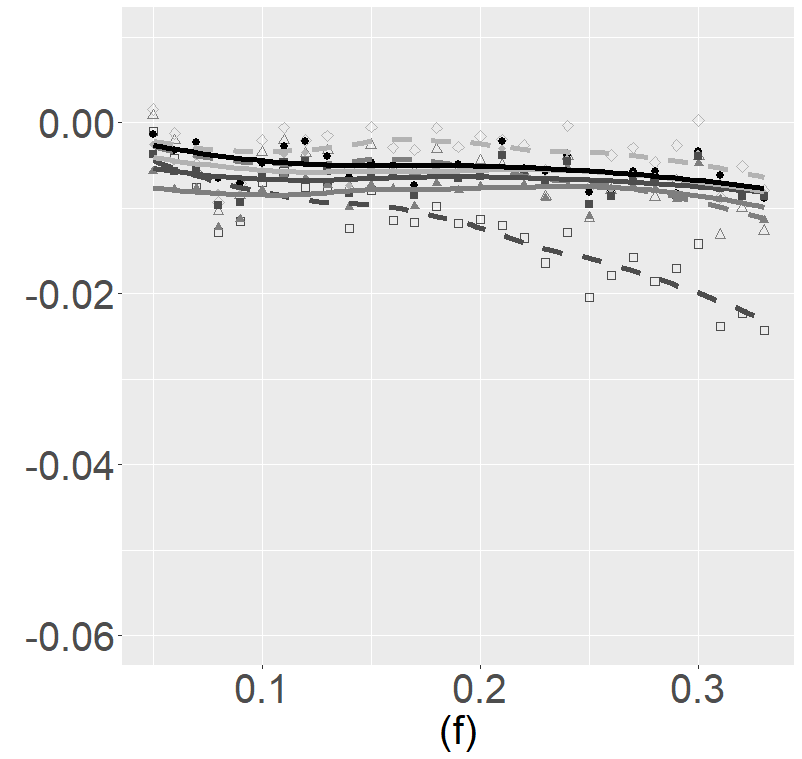}};
		\node[left=of img6, node distance=0cm, rotate=90, anchor=center,yshift=-1cm] {\scriptsize $\bm{\hat{\beta}}_{11}$ Bias};
		\end{tikzpicture}
	\end{minipage}
	\hfill
	\begin{minipage}[t]{0.23\textwidth}
		\begin{tikzpicture}
		\node (img7)  {\includegraphics[scale=0.16]{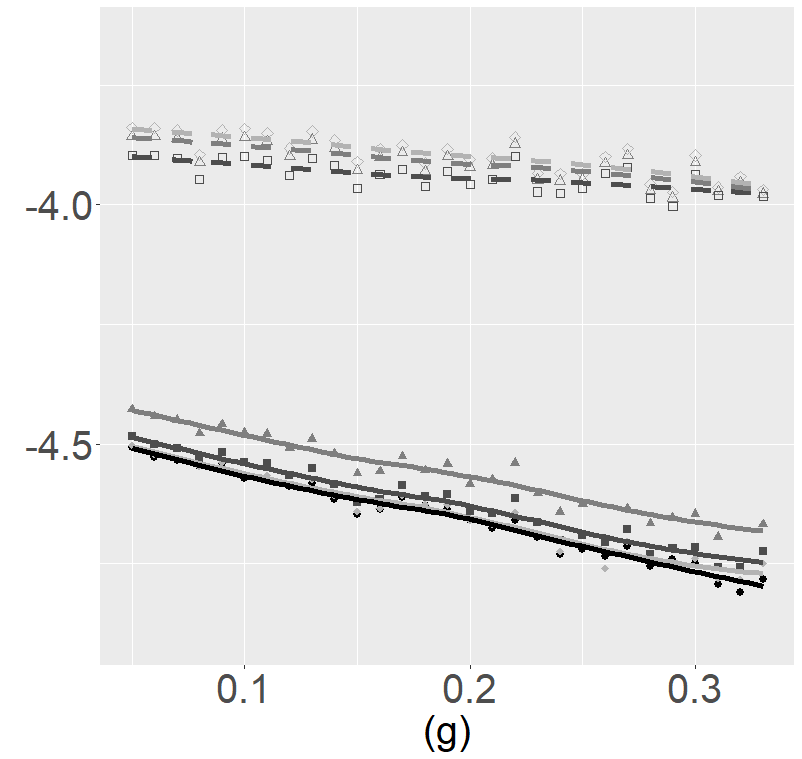}};
		\node[left=of img7, node distance=0cm, rotate=90, anchor=center,yshift=-1cm] {\scriptsize Log $\bm{\hat{\beta}}_{11}$ MSE};
		\end{tikzpicture}
	\end{minipage}
	\hfill
	\begin{minipage}[t]{0.23\textwidth}
		\begin{tikzpicture}
		\node (img8)  {\includegraphics[scale=0.16]{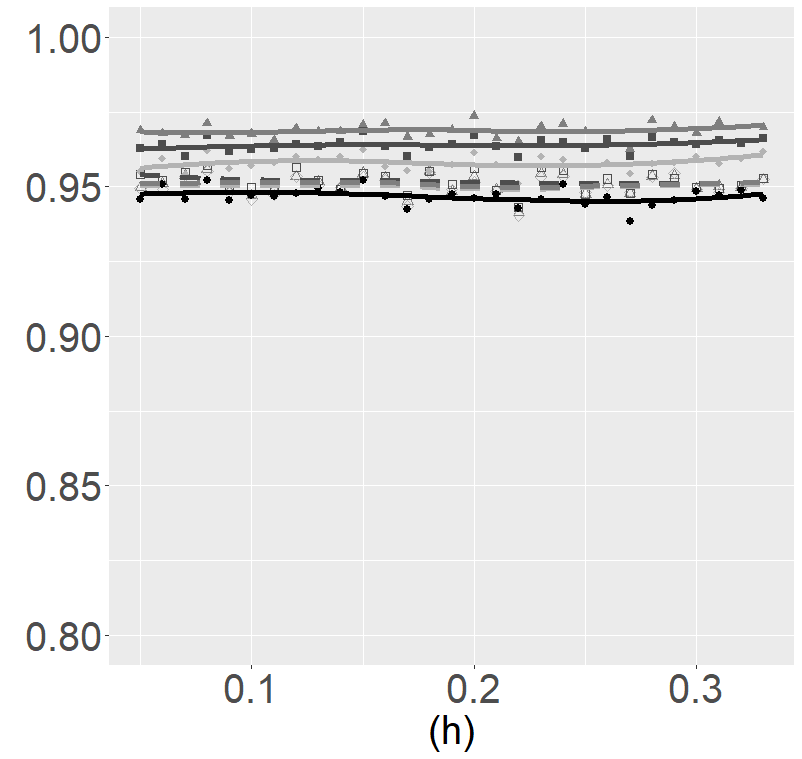}};
		\node[left=of img8, node distance=0cm, rotate=90, anchor=center,yshift=-1cm, xshift = 0.1cm] {\scriptsize Coverage Probability};
		\end{tikzpicture}
	\end{minipage}
	
	\vspace{-0.2cm}
	
	\caption{Panels (a)-(d) present the average log variance, bias, log MSE, and coverage probability for the posterior mean of $\beta_{11}$,	respectively, as a function of the true value of $\beta_{11}$ plotted on the x-axis for the straPP, power prior, asymptotic power prior and uniform improper prior. Panels (e)-(h) present the same information for the straPP, Gen-straPP with $\omega_0 = 1, 2, 4$, and commensurate prior with $b_0$ = 1, 4, 16. straPP, scale transformed power prior; PP, power prior; APP, asymptotic power prior; UIP, uniform improper prior; GS, generalized scale transformed power prior; COM, commensurate prior.}\label{bp-plot}
\end{figure}

Figure~\ref{bp-plot} panels (a)-(d) present results comparing performance characteristics of the straPP, power prior, asymptotic power prior and uniform improper prior. Figure~\ref{bp-plot} panels (e)-(h) present similar results for the straPP, the Gen-straPP  for several choices of $\omega_0$, and the commensurate prior for several choices of $b_0$.

The posterior mean based on the straPP has the smallest variance on average and smallest MSE in comparison to the power prior, asymptotic power prior, and uniform improper prior as seen in Figure \ref{bp-plot} (a) and (c). In Figure \ref{bp-plot} (b), the posterior mean based on the straPP has a bias that is smaller than the power prior and asymptotic power prior, and that is slightly larger than that of the uniform improper prior. The coverage probabilities based on analysis with the straPP is close to 0.95, as seen in Figure \ref{bp-plot} (d). In Figure \ref{bp-plot} (e)-(g), the posterior mean of the straPP and Gen-straPP have smaller variance on average, smaller MSE, and comparable bias to the commensurate prior. In Figure \ref{bp-plot} (h) the coverage probabilities based on analysis with the straPP and commensurate prior is essentially 0.95, while that of the Gen-straPP is slightly larger than 0.95. 

\subsection{Simulation Studies for the Binary-Normal Case -- straPP Transformation Violated \label{sim:BN violate}}

We now consider the binary-normal case where the parameters in the historical and current data models do not satisfy the assumption of the straPP transformation. The purpose of these simulations is to explore the robustness of the Gen-straPP to account for such violations. To operationalize this investigation, we assumed 
\begin{equation}
\beta_1 = I_1^{-1/2}I_0^{1/2}(\bbeta_0)\bbeta_0 - I_1^{-1/2}\bm{c}_0, \label{viol}
\end{equation}
where $\bm{c}_0^T = (0, c_{01})$ and $c_{01}$ was varied. 

In the binary-normal case, the historical data information matrix depends on the regression 
parameter, so we sample from the complementary posterior distribution of the straPP and Gen-straPP to obtain samples for $\bbeta_0$ and transform them to obtain samples for $\bbeta_1$, as described in Appendix~\ref{comp_straPP}. 

We considered the following inputs: $n_0 = 100$, $n_1 = 50$, $a_0 = 1.0$, $\beta_{00} = 0.5$, $\beta_{01} = 0.25$, $\sigma_1 = 2$, and $c_{01} \in [-1.5,1.5]$. The values of the current data model parameters were then identified by solving \eqref{viol}. As we are allowing the difference in standardized parameters to be as large as 1.5 in absolute value, we took $\omega_0 = 1, 4, \omega_{0,E}$, where $\omega_{0,E}$ is calculated as described in Section~\ref{methods: straPP: gen}, to recover the true value of $\beta_{11}$. For the commensurate prior, we considered the inverse scale hyperparameter for the gamma prior on the commensurability parameter as $b_0$ = 2, 4, 8.

\begin{figure}[h]
	\centering
	\begin{minipage}[t]{0.05\textwidth}
		\text{ }
	\end{minipage}
	\hfill
	\begin{minipage}[t]{0.1\textwidth}
		\begin{tikzpicture}
		\node (ls)  {	\includegraphics[scale=0.05]{./figures/strapp_legend.png} };
		\node[right=of ls, node distance=0cm, anchor=center, xshift = -0.5cm, font=\color{black}] {\scriptsize straPP};
		\end{tikzpicture}
	\end{minipage}
	\hfill
	\begin{minipage}[t]{0.15\textwidth}
		\begin{tikzpicture}
		\node (lp)  {	\includegraphics[scale=0.05]{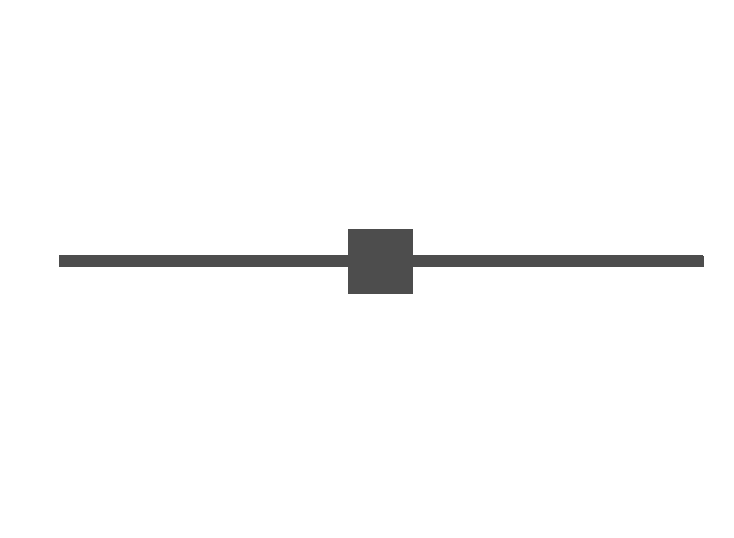} };
		\node[right=of lp, node distance=0cm, anchor=center, xshift = -0.3cm, font=\color{black}] {\scriptsize GS($\omega_0 = 1$)};
		\end{tikzpicture}
	\end{minipage}
	\hfill
	\begin{minipage}[t]{0.18\textwidth}
		\begin{tikzpicture}
		\node (lp)  {	\includegraphics[scale=0.05]{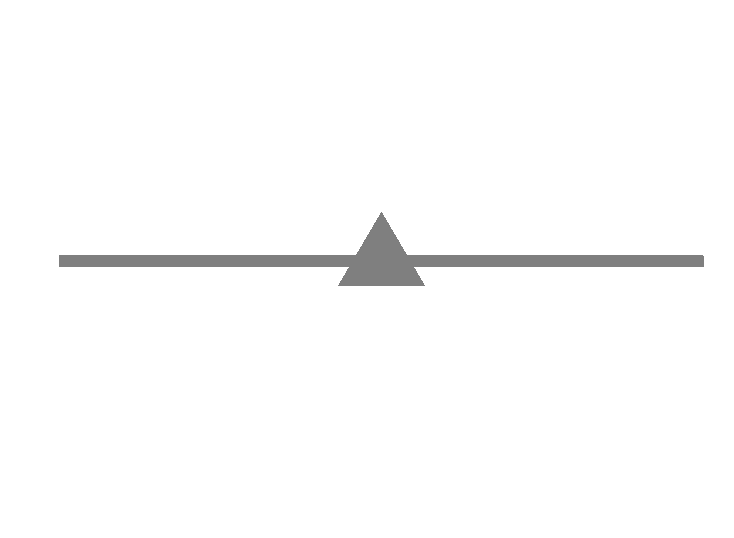} };
		\node[right=of lp, node distance=0cm, anchor=center, xshift = -0.1cm, font=\color{black}] {\scriptsize GS($\omega_0 = \omega_{0,E}$)};
		\end{tikzpicture}
	\end{minipage}
	\hfill
	\begin{minipage}[t]{0.15\textwidth}
		\begin{tikzpicture}
		\node (lu)  {	\includegraphics[scale=0.05]{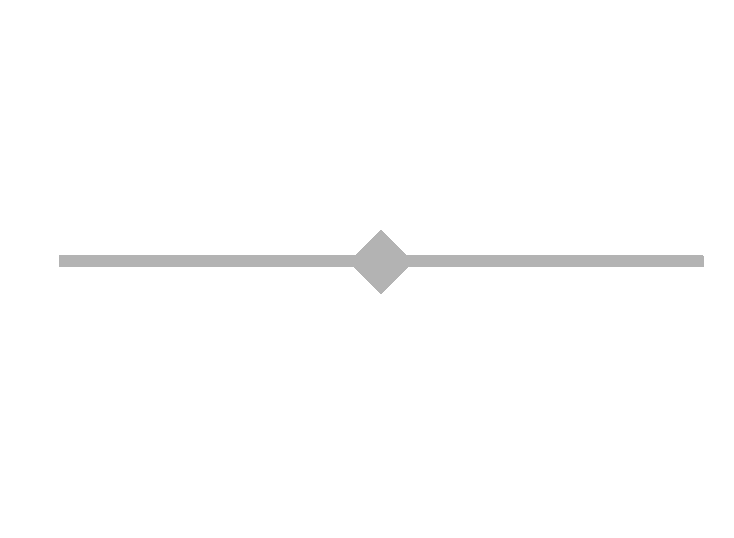} };
		\node[right=of lu, node distance=0cm, anchor=center, xshift = -0.3cm, font=\color{black}] {\scriptsize GS($\omega_0 = 4$)};
		\end{tikzpicture}
	\end{minipage}
	\hfill
	\begin{minipage}[t]{0.05\textwidth}
		\text{ }
	\end{minipage}\\
	\vspace{-0.6cm}
	\begin{minipage}[t]{0.05\textwidth}
		\text{ }
	\end{minipage}
	\hfill
	\begin{minipage}[t]{0.15\textwidth}
		\begin{tikzpicture}
		\node (lp)  {	\includegraphics[scale=0.05]{./figures/com_b2_legend.png} };
		\node[right=of lp, node distance=0cm, anchor=center, xshift = -0.3cm, font=\color{black}] {\scriptsize COM($b_0$ = 2)};
		\end{tikzpicture}
	\end{minipage}
	\hfill
	\begin{minipage}[t]{0.15\textwidth}
		\begin{tikzpicture}
		\node (lp)  {	\includegraphics[scale=0.05]{./figures/com_b4_legend.png} };
		\node[right=of lp, node distance=0cm, anchor=center, xshift = -0.3cm, font=\color{black}] {\scriptsize COM($b_0$ = 4)};
		\end{tikzpicture}
	\end{minipage}
	\hfill
	\begin{minipage}[t]{0.15\textwidth}
		\begin{tikzpicture}
		\node (lu)  {	\includegraphics[scale=0.05]{./figures/com_b8_legend.png} };
		\node[right=of lu, node distance=0cm, anchor=center, xshift = -0.3cm, font=\color{black}] {\scriptsize COM($b_0$ = 8)};
		\end{tikzpicture}
	\end{minipage}
	\hfill
	\begin{minipage}[t]{0.05\textwidth}
		\text{ }
	\end{minipage}

	\vspace{-0.4cm}
	
	\begin{minipage}[t]{0.23\textwidth}
		\begin{tikzpicture}
		\node (img1)  {\includegraphics[scale=0.16]{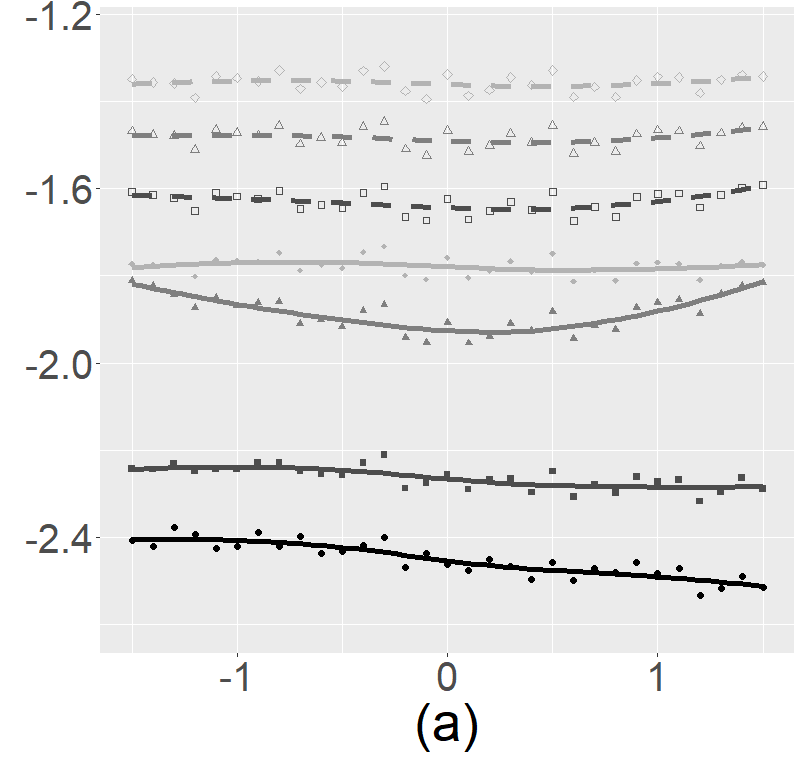}};
		\node[left=of img1, node distance=0cm, rotate=90, anchor=center,yshift=-1cm, xshift = 0.1cm, font=\color{black}] {\scriptsize Avg Log $\bm{\hat{\beta}}_{11}$ Variance};
		\end{tikzpicture}
		\label{fig:var_c01}
	\end{minipage}
	\hfill
	\begin{minipage}[t]{0.23\textwidth}
		\begin{tikzpicture}
		\node (img2)  {\includegraphics[scale=0.16]{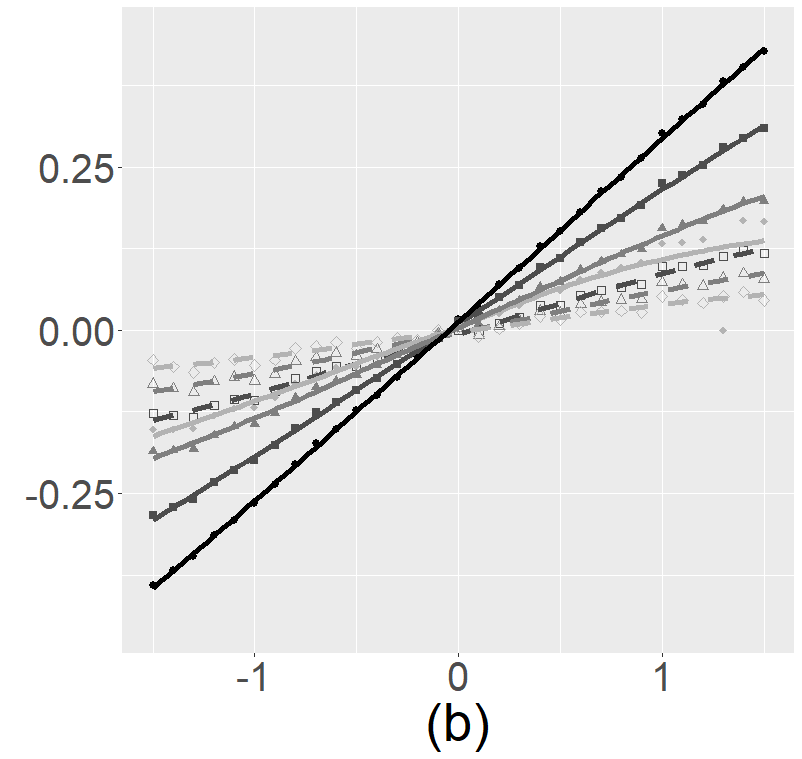}};
		\node[left=of img2, node distance=0cm, rotate=90, anchor=center,yshift=-1cm] {\scriptsize $\bm{\hat{\beta}}_{11}$ Bias};
		\end{tikzpicture}
		\label{fig:bias_c01}
	\end{minipage}
	\hfill
	\begin{minipage}[t]{0.23\textwidth}
		\begin{tikzpicture}
		\node (img3)  {\includegraphics[scale=0.16]{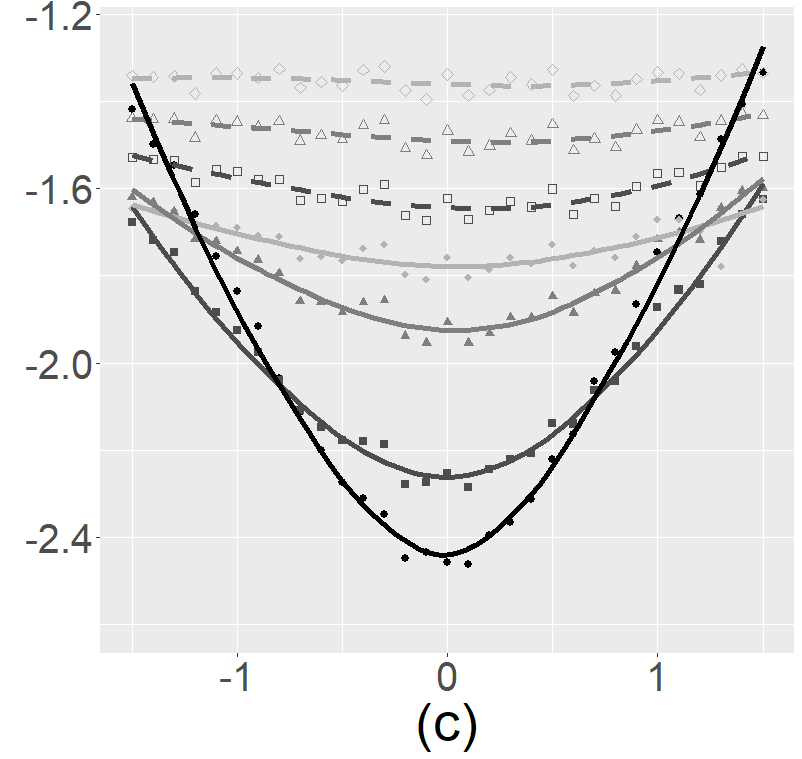}};
		\node[left=of img3, node distance=0cm, rotate=90, anchor=center,yshift=-1cm] {\scriptsize Log $\bm{\hat{\beta}}_{11}$ MSE};
		\end{tikzpicture}
		\label{fig:mse_c01}
	\end{minipage}
	\hfill
	\begin{minipage}[t]{0.23\textwidth}
		\begin{tikzpicture}
		\node (img4)  {\includegraphics[scale=0.16]{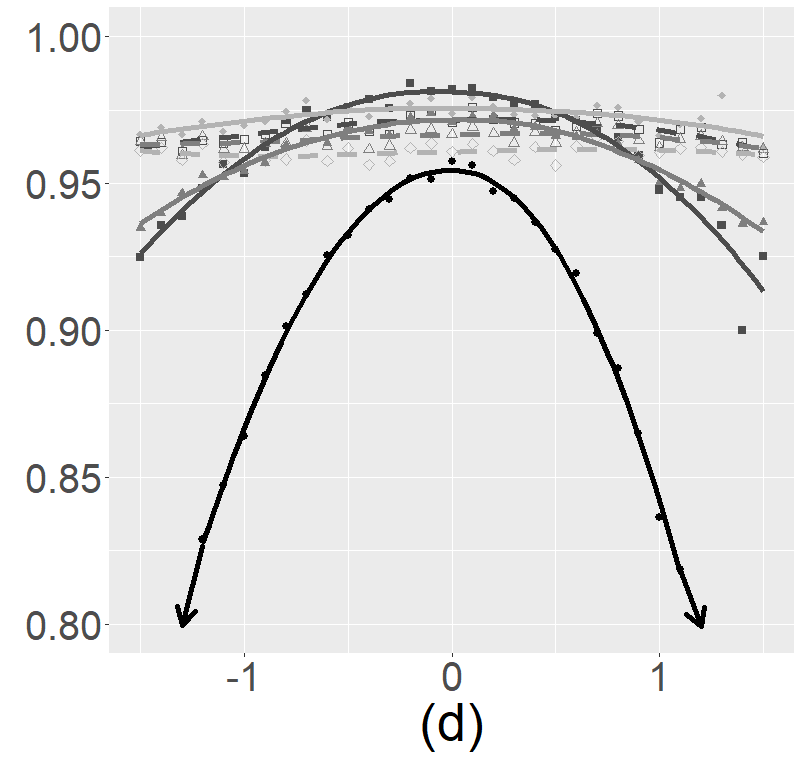}};
		\node[left=of img4, node distance=0cm, rotate=90, anchor=center,yshift=-1cm, xshift = 0.1cm] {\scriptsize Coverage Probability};
		\end{tikzpicture}
		\label{fig:cov_c01}
	\end{minipage}
	\vspace{-0.4cm}
	\caption{Panels (a)-(d) present the average log variance, bias, log MSE, and coverage probability for the posterior mean of $\beta_{11}$, respectively, as a function of $c_{01}$ plotted on the x-axis for the straPP, Gen-straPP with with $\omega_0 = 1, 4, \omega_{0,E}$ where $\omega_{0,E} = \max|I_0^{1/2}(\hat{\bbeta}_0)\hat{\bbeta}_0 - I_1^{1/2}(\hat{\bbeta}_1)\hat{\bbeta}_1|$, and commensurate prior with $b_0$ = 2, 4, 8. straPP, scale transformed power prior; GS, generalized scale transformed power prior; COM, commensurate prior.}\label{c01-plot}
\end{figure}

Figure~\ref{c01-plot} panels (a)-(d) present results comparing performance characteristics of the straPP, the Gen-straPP for several choices of $\omega_0$, and the commensurate prior for several choices of $b_0$ with varying value of $c_{01}$. 

The average log variance of the posterior mean based on the straPP is smaller than the Gen-straPP and commensurate prior, as seen in Figure \ref{c01-plot}(a).
Figure~\ref{c01-plot}(b) illustrates that the magnitude of the posterior mean estimator bias increases as $c_{01}$ increases and is greater for the straPP compared to the Gen-straPP for each of the choices for $\omega_0$. This illustrates the robustification provided by the Gen-straPP in terms of bias reduction when the assumption of the straPP does not hold.
In Figure~\ref{c01-plot}(c), the MSE is lower for the straPP when the value of $c_{01}$ is near zero but exceeds the Gen-straPP and commensurate prior when the quantity becomes sufficiently large in absolute value. 
The coverage probabilities based on analysis with the Gen-straPP gets closer to 0.95 as $\omega_0$ increases, while the coverage probabilities based on analysis with the commensurate prior gets closer to 0.95 as $b_0$ increases, as seen in Figure~\ref{c01-plot}(d).
Of note, the straPP and Gen-straPP priors emit point estimators that have lower MSE than that based on the commensurate prior. Thus, one can conclude that the Gen-straPP may give better performance compared to the commensurate prior in terms of MSE even when the assumption of the straPP are badly violated.

\section{\label{compass} Analysis of the COMPASS Study Data}

For the following analyses, the formulation of the partial-borrowing straPP utilized an incidence of falls indicator for the Phase 1 historical data
and the continuous PROMIS outcome for the Phase 2 data. We assume that the historical patient outcomes are independently distributed according to a logistic regression model and the current patient outcomes are independently distributed according to a linear regression model. 

As with any real data analysis, the true value of the dispersion parameter for the linear regression model is unknown, and thus we appropriately treat that parameter as random. The focus of this analysis is to borrow information on the covariate effects as there is no rationale for borrowing information on the intercept parameters. 
To complete specification of the partial-borrowing straPP and partial-borrowing Gen-straPP priors, we specified a uniform improper initial prior, a uniform improper prior on the historical and current data model intercepts and a gamma prior on the dispersion parameter with shape and inverse scale parameters equal to 0.01. We specified a normal prior for $\bm{c}_0$, as stated in Section~\ref{methods: straPP: gen}.

For the partial-borrowing power prior and partial-borrowing asymptotic power prior, the initial prior for the regression parameters and the prior on the historical and current data model intercepts were taken to be uniform improper. The prior for the dispersion parameter was taken to be gamma with shape and inverse scale parameters equal to 0.01. For the partial-borrowing commensurate prior, the prior on the historical and current data model intercepts were taken to be uniform improper. The prior for the dispersion parameter was taken to be gamma with shape and inverse scale hyperparameters equal to 0.01, and the prior for the commensurability parameter was taken to be gamma with shape and inverse scale hyperparameters equal to 2.

To compare the overall quality of performance based on the set of selected priors, we used the deviance 
information criterion (DIC) developed by \cite{dic}, where $\mbox{DIC}(a_0)= 2E\{\text{Dev}(\bm{\xi}_1)| D_1,D_0,a_0\}  -\mbox{Dev}( \overline{\bm{\xi}}_1)$,
where $\overline{\bm{\xi}_1}=E\{\bm{\xi}_1| D_1,D_0,a_0\}$ and $\mbox{Dev}(\bm{\xi}_1)= -2 \sum^n_{i=1} \log f(y_{1i}| \bm{x}_{1i},\bm{\xi}_1)$. Lower values of DIC indicate better performance of the associated prior. A similar approach was taken by \cite{IbrahimEtAl15}.

For our illustration, we perform an analysis on the COMPASS dataset to evaluate the covariate effect associated with receipt of the COMPASS eCare plan, a key component of the intervention. The regression model included an indicator for receipt of the COMPASS eCare plan, an indicator for having a history of stroke or TIA, NIHSS score,
and an indicator for having non-white race. After removing observations with missing values for the covariates of interest, the historical data sample size was 244 and the current data sample size was 385. 

Before comparing the posterior estimates from an analysis based on the set of priors considered, we determined an appropriate value of $(\omega_0,a_0)$ for the Gen-straPP by investigating which combination resulted in the smallest DIC. Table \ref{nu0} displays the DIC for the Gen-straPP with varying values of  $(\omega_0,a_0)$. The DIC values were calculated based on 20,000 posterior samples obtained from a Metropolis-Hastings MCMC sampling algorithm. 

\begin{table}
	\caption{DIC for the Gen-straPP with Various ($a_0, \omega_0$)\label{nu0}}
	\centering
	\footnotesize
	\fbox{%
		\begin{tabular}{*{8}{c}}
			& & \multicolumn{6}{c}{$a_0$}\\
			\cline{3-8}
			\multicolumn{1}{c}{$\omega_0$} & & \multicolumn{1}{c}{0.0**} & \multicolumn{1}{c}{0.1} & \multicolumn{1}{c}{0.25} & \multicolumn{1}{c}{0.5} & \multicolumn{1}{c}{0.75} & \multicolumn{1}{c}{1.0} \\ 
			\hline
			0.00* & & 2816.56 & 2815.34 & 2815.97 & 2817.47 & 2819.10 & 2820.68 \\ 
			0.10~ & & 2816.56 & 2816.43 & 2816.32 & 2817.32 & 2818.71 & 2819.70 \\ 
			0.25~ & & 2816.56 & 2816.41 & 2816.28 & 2816.96 & 2817.93 & 2818.91 \\ 
			0.50~ & & 2816.56 & 2817.02 & 2816.24 & 2816.68 & 2817.39 & 2818.03 \\   
			0.75~ & & 2816.56 & 2817.12 & 2816.25 & 2816.41 & 2816.91 & 2817.39 \\   
			1.00~ & & 2816.56 & 2817.12 & 2816.43 & 2816.43 & 2816.64 & 2816.98 \\
			\hline
			\multicolumn{8}{l}{Note: *$\omega_0 = 0$ is equivalent to the straPP and **$a_0 = 0$ is equivalent to}\\
			\multicolumn{8}{l}{ the uniform improper prior.}
	\end{tabular}}
\end{table}
The smallest DIC in Table \ref{nu0} is associated with $\omega_0 = 0$ and $a_0 = 0.1$, which corresponds to a straPP with $a_0 = 0.1$. We used a  Gen-straPP with $a_0 = 0.1$ and $\omega_0 = 0.1$ for comparison with other priors as it was the partial-borrowing Gen-straPP with the smallest DIC in which $\omega_0 \neq 0$. We note that we calculated the DIC for the partial-borrowing Gen-straPP for higher values of $\omega_0$ than those shown here (up to $\omega_0=1$) which resulted in similar patterns of DIC with varying $a_0$ for a given value of $\omega_0$. Note, we calculated the DIC for the Gen-straPP with $a_0 = 0.1$, and $\omega_0$ equal to the empirical estimator as described in Section~\ref{methods: straPP: gen}, and found that it was larger than the Gen-straPP with $\omega_0 = 0.1$, thus it was excluded from further consideration.

Table \ref{gen-res} displays the DIC, posterior estimates and 95\% highest posterior density (HPD) intervals for the partial-borrowing priors and uniform improper prior. The posterior estimates were calculated based on 20,000 posterior samples obtained from a Metropolis-Hastings MCMC sampling algorithm.  The posterior summary for the intercepts and variance for the regression model can be found in Appendix~\ref{int-var}.
\begin{table}
	\caption{Posterior Estimates for the COMPASS Data\label{gen-res}}
	\setlength{\tabcolsep}{3.5pt} 
	\renewcommand{\arraystretch}{1.2} 
	\resizebox{\textwidth}{!}{%
		\fbox{%
			\begin{tabular}{lrrrrrrrrrrrrrrr}
				&   &\multicolumn{2}{c}{eCare Plan} & & \multicolumn{2}{c}{History of Stroke}  & &\multicolumn{2}{c}{Minor NIHSS} & &\multicolumn{2}{c}{Moderate-Severe NIHSS} & &\multicolumn{2}{c}{Non-white}  \\
				\cline{3-4} \cline{6-7} \cline{9-10} \cline{12-13} \cline{15-16}
				~~Model  & DIC~~ &  Mean (SD) & 95\%HPD~~ & & Mean (SD) & 95\%HPD~~ & & Mean (SD) & 95\%HPD~~ & & Mean (SD)    & 95\%HPD~~  & & Mean (SD)    & 95\%HPD~~\\
				\hline
				straPP & 2815.34 & 0.84 (0.97) & (-1.13, 2.67) & & -0.84 (1.13) & (-3.04, 1.40) & & -1.24 (1.05) & (-3.25, 0.87) & & -3.51 (1.06) & (-5.61, -1.44) & & -1.45 (1.47) & (-3.96, 1.68)\\ 
				APP &  2815.85 & 0.28 (0.77) & (-1.21, 1.79) & & -0.37 (0.96) & (-2.23, 1.56) & & -0.73 (0.80) & (-2.31, 0.83) & & -3.07 (1.17) & (-5.42, -0.82) & & -2.17 (1.88) & (-5.92, 1.46)\\
				PP & 2815.97 & 0.29 (0.77) & (-1.16, 1.86) & & -0.36 (0.96) & (-2.27, 1.48) & & -0.72 (0.81) & (-2.30, 0.86) & & -3.04 (1.20) & (-5.47, -0.78) & & -2.19 (1.87) & (-5.86, 1.40)\\
				Gen-straPP &  2816.41 & 0.85 (0.98) & (-1.00, 2.80) & & -0.79 (1.16) & (-2.98, 1.56) & & -1.22 (1.09) & (-3.31, 0.96) & & -3.59 (1.21) & (-6.00, -1.30) & & -1.50 (1.65) & (-4.70, 1.85)\\
				UIP & 2816.56 & 0.85 (0.98) & (-1.08, 2.79) & & -1.18 (1.24) & (-3.60, 1.27) & & -1.69 (1.10) & (-3.81, 0.50) & & -4.69 (1.32) & (-7.19, -2.00) & & -2.23 (2.12) & (-6.40, 1.91)\\ 
				COM & 2817.45 &  0.45 (0.84) & (-1.16, 2.14) & & -0.47 (0.99) & (-2.43,  1.44) & & 
				-0.48 (0.93) & (-2.39, 1.21) & & -2.35 (1.25) & (-4.79, ~0.03) & & 
				-1.53 (1.37) & (-4.31, 1.07)\\
				\hline
				\multicolumn{16}{l}{ straPP, scale transformed 
					power prior; APP, asymptotic power prior; PP, power prior; Gen-straPP, generalized scale transformed power prior;   UIP, uniform}\\
				\multicolumn{16}{l}{ improper prior; COM, commensurate prior.}\\
	\end{tabular}}}
\end{table}
In Table \ref{gen-res}, analysis with the partial-borrowing straPP resulted in the smallest DIC when compared to analyses with all other priors. This suggests that, for this data set, the rescaling action of the partial-borrowing straPP is sufficient to 
effectively translate the information on covariate effects from the incidence of falls outcome to the effects on the continuous PROMIS outcome. In Table \ref{com_ests} in Appendix~\ref{compass_com}, we see that the commensurate prior has larger values of DIC than the straPP when using different values for the inverse scale hyperparameter for the gamma prior on the commensurability parameter.

Aside from the general performance of the priors as measured by DIC, we also investigated how posterior estimates compared to analyses based on the priors considered. 
The posterior mean of each regression parameter based on the partial-borrowing straPP and partial-borrowing Gen-straPP 
were larger in absolute value than the corresponding posterior mean based on the other three partial-borrowing priors. 
The posterior standard deviation of each regression parameter in Table \ref{gen-res} was smaller for the partial-borrowing straPP than the uniform improper prior and partial-borrowing Gen-straPP.
The modestly increased variability of the posterior distribution based on analyses using the Gen-straPP is a result of the robustification afforded by incorporating $\bm{c}_0$ into the prior construction. The partial-borrowing straPP has larger posterior standard deviations, for all regression parameters, than both the partial-borrowing power prior and partial-borrowing asymptotic power prior. This is due to the asymptotic power prior and power prior reflecting the variance of the binary historical outcome, which is smaller than the variance of the outcome for the current data. The posterior standard deviation of each regression parameter, except moderate-to-severe NIHSS, was larger for the partial-borrowing straPP than the partial-borrowing commensurate prior.

\section{Discussion}\label{disc}

In this paper, we developed the straPP to provide a mechanism for informative prior elicitation using historical data when the outcome model differs from that of the current data. The straPP is developed based on an assumption that parameter values between the two models are equivalent after appropriate rescaling. The Gen-straPP was developed to provide robustness to the underlying assumption of the straPP. However, the Gen-straPP can be computationally expensive, and thus its use is intended for sensitivity analysis. For any application where the historical data arise from a model that differs from the current data, translating the information from the historical dataset can be challenging. The straPP arises from assuming an intuitive relationship between the parameters in the two models. Our empirical analysis results based on the COMPASS data suggests that the straPP assumption sufficiently holds to provide improved inference over other commonly used priors and priors that do not make use of the historical data. 

In this paper the straPP and Gen-straPP were developed specifically for univariate GLMs. In future work the authors plan to extend development of the straPP and Gen-straPP to allow for historical and/or current data models with time-to-event outcomes (e.g., proportional hazards models). Developing the straPP for time-to-event data poses several new challenges not addressed in this paper, including dealing with right-censoring and modeling higher dimensional nuisance parameters (e.g., baseline hazard). 

\section*{Acknowledgments}

This research was partially supported by  NIH grants \#GM 70335 and P01CA142538, and NIEHS grant T32ES007018.

\bibliographystyle{unsrtnat}
\bibliography{References} 

\begin{appendices}
	\vspace{1cm}
\textbf{\Large Appendix}

	\section{Complementary Posterior Distribution of the Scale Transformed Power Prior \label{comp_straPP}}
	When the historical data matrix depends on the regression parameter, but the current data information matrix does not, one can perform analysis using the straPP by sampling from what we refer to as the complementary posterior distribution. Here we can write the transformation as $\btheta = g^{-1}(\bdeta) =  I_1^{-1/2}I_0^{1/2}(\bdeta) \bdeta$. In this case, the current and historical data have their roles reversed and one essentially draws samples from the posterior distribution for $\bdeta$ using the straPP while treating the current data as if it were historical for the purposes of the straPP construction. The complementary posterior distribution of the straPP can be written as,
	\begin{eqnarray}
	f_s(\bdeta \mid D_1, D_0) &\propto & \mathcal{L}(\bdeta \mid D_0)^{a_0}\pi_0(\bdeta)\mathcal{L}(g^{-1}(\bdeta)\mid D_1)\left|\frac{dg^{-1}(\bdeta)}{d\bdeta}\right|,\label{dual-straPP}
	\end{eqnarray} 
	where $D_0 = (n_0, \bm{Y}_0, \bm{X}_0)$  and $D_1 = (n_1, \bm{Y}_1, \bm{X}_1)$ denote the historical and current data respectively, and for $D_k$, $n_k$ denotes the sample size, $\bm{Y}_k$ denotes the $n_k\times 1$ response vector, and $\bm{X}_k$ denotes the $n_k\times p$ covariate matrix. The expression $\left|\{dg^{-1}(\bdeta)\}/d\bdeta\right|$ denotes the determinant of the Jacobian of the transformation. The expression for the Jacobian can be calculated analogously to what is described in Section~\ref{methods: straPP: jacobian}.
	
	The resulting samples from \eqref{dual-straPP} can then be transformed according to $\btheta = g^{-1}(\bdeta)$ to obtain samples from the posterior distribution for $\btheta$. The complementary posterior distribution of the Gen-straPP can be developed similarly.
	
	\subsection{Complementary Posterior Distribution for the straPP and Gen-straPP for the Binary-Normal Simulations}
	For the case where the historical data follow a logistic regression model and the current data follow a linear regression model with known variance, which we call the binary-normal case, we performed simulation studies using parameter values that obey the assumption of the straPP transformation. 
	However, in the binary-normal case, the historical data information matrix depends on the regression 
	parameter, so we sample from the complementary posterior to obtain samples for $\bbeta_0$ and transform them to obtain samples
	for $\bbeta_1$ (e.g., $\bbeta_1 = g^{-1}(\bbeta_0) = I_1^{-1/2}I_0^{1/2}(\bbeta_0)\bbeta_0$). 
	The complementary posterior distribution of the straPP and Gen-straPP are written in \eqref{dual-st-GLM} and \eqref{dual-GS}, respectively.
	\begin{eqnarray}
	f_s(\bbeta_0 \mid \sigma_1, D_1,  D_0) &\propto & \mathcal{L}(\bbeta_0 \mid D_0)^{a_0}\pi_0(\bbeta_0)\mathcal{L}(g^{-1}(\bbeta_0), \sigma_1 \mid D_1)\left|\frac{dg^{-1}(\bbeta_0)}{d\bbeta_0}\right|,\label{dual-st-GLM}\\
	f_g(\bbeta_0, c_0 \mid\sigma_1, D_1,  D_0) &\propto & \mathcal{L}(\bbeta_0 \mid D_0)^{a_0}\pi_0(\bbeta_0)\mathcal{L}(g_{c0}^{-1}(\bbeta_0), \sigma_1 \mid D_1)\left|\frac{dg_{c0}^{-1}(\bbeta_0)}{d\bbeta_0}\right|\pi_0(\bm{c}_0),\label{dual-GS}
	\end{eqnarray} 
	where $g_{c_0}^{-1}(\bbeta_0) \equiv I_1^{-1/2}I_0^{1/2}(\bbeta_0)\bbeta_0 + I_1^{-1/2}\bm{c}_0$.
	
	\section{Algorithm for Sampling Using the straPP when Neither Information Matrix is Parameter Free \label{alg}}
	
	For the case where neither information matrix is free of the parameter, it is still possible to analyze data using the straPP by considering a posterior representation involving both $\bdeta$ and $\btheta$, and using a Metropolis-Hastings sampling algorithm where proposed values of $\bdeta$ and $\btheta$ satisfy the constraint in \eqref{transf1}. Let $T$ denote the total number of samples and $\btheta^{(t)}$ denote the $t^{th}$ sample, for $t = 1, \ldots, T$.  We propose the following algorithm for sampling the posterior distribution of the straPP. 
	
	\begin{itemize}
		\item[1] Propose $\btheta^{(t)} \sim N\left(\btheta \mid \btheta^{(t-1)}, \widehat{\Sigma}\right)$, where $\widehat{\Sigma}$ is the maximum likelihood estimate (MLE) for the current data analysis.
		\item[2] Compute $h(\btheta^{(t)}) = I_1^{1/2}(\btheta^{(t)})\btheta^{(t)}$, where $h(\cdot)$ denotes the scaled parameter.
		\item[3] Solve for $\bdeta^{(t)}$  in $I_0^{1/2}(\bdeta^{(t)})\bdeta^{(t)} =h(\btheta^{(t)})$, via a nonlinear programming (NLP) solver (e.g., PROC NLP in SAS or optimx in R).
		\item[4] Perform a Metropolis-Hastings step based on the proposal value of 
		$\left(\bdeta^{(t)}, \btheta^{(t)}\right)$.
	\end{itemize}
	
	\section{Expression for the Jacobian for Generalized Linear Models with the Canonical Link \label{jac_canon}}
	In this section, we assume that outcomes for the historical and current data arise from the class of generalized linear models (GLMs), as described in Section~\ref{method: glm}. Further, we specify the canonical link for the historical and current data model. Then the historical information matrix for the regression parameter can be written as $I_{0}(\bbeta_0) = \phi_0X_0^TV_0(\bbeta_0)X_0$ and current data information matrix can be written as $I_{1}(\bbeta_1\mid X_0) = \phi_1X_0^TV_1(\bbeta_1)X_0,$ 
	where for $k = 0,1$, $V_{k}(\bbeta_k) = \text{diag}(v_{ki}(\bbeta_k))$, as stated in Section~\ref{glm}. Then following Section~\ref{methods: straPP: jacobian}, we can write the derivatives as
	\begin{eqnarray*}
		\left\{\frac{d}{d\bbeta_0}I_0^{1/2}(\bbeta_0)\right\}\bbeta_0 & = & \left(\left\{\frac{d}{d\beta_{0,0}}I_0^{1/2}(\bbeta_0)\right\}\bbeta_0, \ldots, \left\{\frac{d}{d\beta_{0,p-1}}I_0^{1/2}(\bbeta_0)\right\}\bbeta_0\right),\\
		\left\{\frac{d}{d\bbeta_1}I_1^{1/2}(\bbeta_1\mid X_0)\right\}\bbeta_1 & = & \left(\left\{\frac{d}{d\beta_{1,0}}I_1^{1/2}(\bbeta_1\mid X_0)\right\}\bbeta_1, \ldots, \left\{\frac{d}{d\beta_{1,p-1}}I_1^{1/2}(\bbeta_1\mid X_0)\right\}\bbeta_1\right).
	\end{eqnarray*}
	Using the results from equation \eqref{i1_sqrt} in Section~\ref{methods: straPP: jacobian}, for a given $j = 0, \ldots, p-1$, we find
	\begin{align*}
	\text{vec}\left(\frac{dI_0^{1/2}(\bbeta_0\mid X_0)}{d\beta_{0,j}}\right)
	&=\left\{I_0^{1/2}(\bbeta_0)\otimes \bm{I}_{p \times p} +\bm{I}_{p \times p}\otimes I_0^{1/2}(\bbeta_0\mid X_0)\right\}^{-1}\\
	& \tab \times \text{vec}\left(X_0^T\text{diag}\left(\frac{dv_{0i}(\bbeta_0)}{d\beta_{0,j}}\right)
	X_0\right),\\
	\text{vec}\left(\frac{dI_1^{1/2}(\bbeta_1\mid X_0)}{d\beta_{1,j}}\right)
	&=\left\{I_1^{1/2}(\bbeta_1\mid X_0)\otimes \bm{I}_{p \times p} +\bm{I}_{p \times p}\otimes I_1^{1/2}(\bbeta_1\mid X_0)\right\}^{-1}\\
	& \tab \times \text{vec}\left(X_0^T\text{diag}\left(\frac{dv_{1i}(\bbeta_1)}{d\beta_{1,j}}\right)
	X_0\right).
	\end{align*}

	\section{Additional Simulations \label{add_sims}}
	\renewcommand{\thefigure}{\Alph{section}.\arabic{figure}}
	\setcounter{figure}{0}
	
	In this section, we present and discuss results from a collection of simulation studies designed to evaluate the performance of the straPP and Gen-straPP compared to each other as well as to the power prior, asymptotic power prior, commensurate prior, and uniform improper prior. For the commensurate prior, the prior for the commensurability parameter was taken to be gamma with shape hyperparameter equal to 2 and inverse scale hyperparameter, which we denote $b_0$, to be 2, 4 or 8. For these simulations, we use $a_0 = 1$, so as to directly compare the Gen-straPP to the commensurate prior as described in Section~\ref{relation}. 
	
	In Apendix~\ref{bn-strapp-holds}, we present simulation studies for a case where the historical data follow a logistic regression model and the current data follow a linear regression model with known variance, which we call the binary-normal case. In Apendix~\ref{pe-sim} we present simulation studies for a case in which the historical data follow a loglinear regression model and the current data follow an exponential model with the log link, which we call the Poisson-exponential case.
	In both the binary-normal case and the Poisson-exponential case, the historical data information matrix depends on the regression 
	parameter, so we sample from the complementary posterior distribution to obtain samples for $\bbeta_0$ and transform them to obtain samples
	for $\bbeta_1$, as described in Appendix~\ref{comp_straPP}. 
	
	In all of the simulations, we simulated the historical and current data to have an intercept and a treatment indicator
	such that 50\% of simulated patients were treated. 
	
	For each simulation, we generated 5,000 historical and current data sets for each unique parameter combination, and 
	used Metropolis-Hastings Markov chain Monte Carlo (MCMC) methods to obtain 25,000 samples from the posterior using 
	various priors to analyze each dataset.

	\subsection{Binary-Normal Case Simulation -- straPP Transformation Holds \label{bn-strapp-holds}}
	
	For the binary-normal case, we first performed simulation studies using parameter values that obey the assumption of the straPP transformation. Results based on the straPP and Gen-straPP are compared to the power prior and asymptotic power prior as given in. We analyze the Gen-straPP with $\omega_0 = 0.1, 0.5,$ and $1.0.$
	We consider the binary-normal case where the parameters in the historical and current data models satisfy the
	assumption of the straPP transformation. 
	
	Unlike the normal-normal case, the form of $I_1^{-1/2}I_0^{1/2}(\bbeta_0)$ 
	does not simplify in this setting and thus no elegant relationship between the parameters can be seen. 
	$p_1 = \exp(\beta_{00} + \beta_{10})/(1+\exp(\beta_{00} + \beta_{01}))$ be the probability of response for the 
	control and treated patients.  
	Then we calculate $A^{-1}(\bbeta_0)$ as
	\begin{equation}
	A^{-1}(\bm{\beta}_0) = \left\{\frac{n_0}{2\sigma_1^2}\begin{pmatrix} 2 & 1 \\ 1 & 1\end{pmatrix}\right\}^{-1/2} \left\{\frac{n_0}{2}\begin{pmatrix}	p_0(1-p_0)  + p_1(1-p_1) & p_1(1-p_1) \\ p_1(1-p_1) & p_1(1-p_1) \end{pmatrix}\right\}^{1/2}. \label{bn-g}
	\end{equation}
	
	We considered the following inputs: $n_0 = 100$, $n_1 = 50$, $a_0 = 1$, $\beta_{00} = 0.5$, $\sigma_1 = 2$, and $\beta_{01} \in [0,2]$. The values of the current data model parameters were then identified by solving $\bbeta_1 = g^{-1}(\bbeta_0)$. 
	Note that the value of the parameters for the historical data model were chosen so that 
	$0.05 \le p_0, p_1 \le  0.95$. This was done so as 
	to ensure the simulated datasets from the logistic model had sufficient variability in the outcome across treatment groups 
	to avoid instability issues in model fitting.
	
\begin{figure}[H]
		\centering
		\begin{minipage}[t]{0.1\textwidth}
			\text{ }
		\end{minipage}
		\hfill
		\begin{minipage}[t]{0.1\textwidth}
			\begin{tikzpicture}
			\node (ls)  {	\includegraphics[scale=0.05]{./figures/strapp_legend.png} };
			\node[right=of ls, node distance=0cm, anchor=center, xshift = -0.5cm, font=\color{black}] {\scriptsize straPP};
			\end{tikzpicture}
		\end{minipage}
		\hfill
		\begin{minipage}[t]{0.1\textwidth}
			\begin{tikzpicture}
			\node (lp)  {	\includegraphics[scale=0.05]{./figures/pp_legend.png} };
			\node[right=of lp, node distance=0cm, anchor=center, xshift = -0.5cm, font=\color{black}] {\scriptsize PP};
			\end{tikzpicture}
		\end{minipage}
		\hfill
		\begin{minipage}[t]{0.1\textwidth}
			\begin{tikzpicture}
			\node (lp)  {	\includegraphics[scale=0.05]{./figures/app_legend.png} };
			\node[right=of lp, node distance=0cm, anchor=center, xshift = -0.5cm, font=\color{black}] {\scriptsize APP};
			\end{tikzpicture}
		\end{minipage}
		\hfill
		\begin{minipage}[t]{0.1\textwidth}
			\begin{tikzpicture}
			\node (lu)  {	\includegraphics[scale=0.05]{./figures/ui_legend.png} };
			\node[right=of lu, node distance=0cm, anchor=center, xshift = -0.5cm, font=\color{black}] {\scriptsize UIP};
			\end{tikzpicture}
		\end{minipage}
		\hfill
		\begin{minipage}[t]{0.1\textwidth}
			\text{ }
		\end{minipage}
		\vspace{-0.4cm}
		
		\begin{minipage}[t]{0.23\textwidth}
			\begin{tikzpicture}
			\node (img1)  {\includegraphics[scale=0.16]{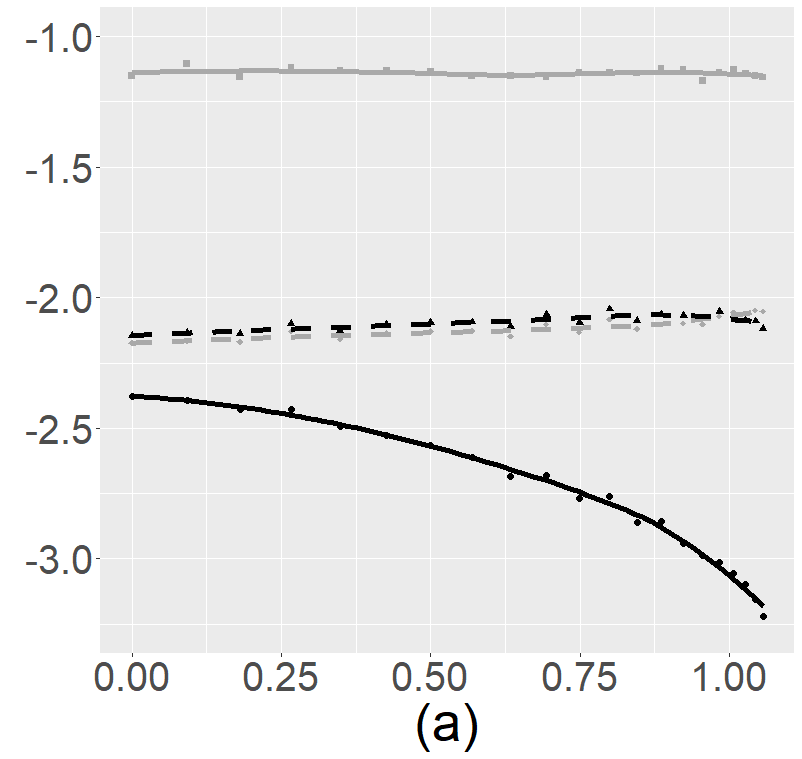}};
			\node[left=of img1, node distance=0cm, rotate=90, anchor=center,yshift=-1cm, xshift = 0.1cm, font=\color{black}] {\scriptsize Avg Log $\bm{\hat{\beta}}_{11}$ Variance};
			\end{tikzpicture}
			\label{fig:var_bn}
		\end{minipage}
		\hfill
		\begin{minipage}[t]{0.23\textwidth}
			\begin{tikzpicture}
			\node (img2)  {\includegraphics[scale=0.16]{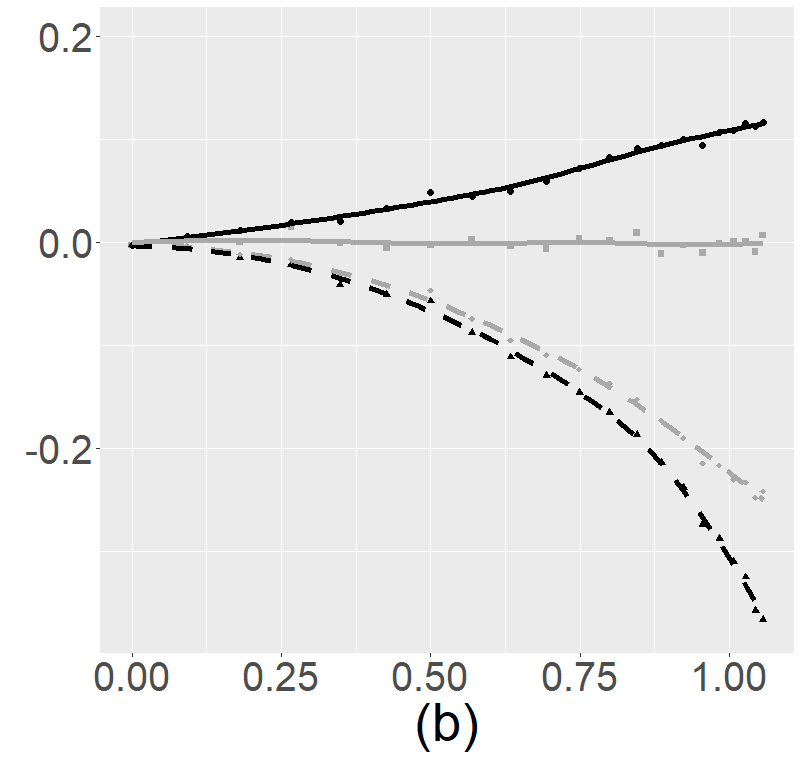}};
			\node[left=of img2, node distance=0cm, rotate=90, anchor=center,yshift=-1cm] {\scriptsize $\bm{\hat{\beta}}_{11}$ Bias};
			\end{tikzpicture}
			\label{fig:bias_bn}
		\end{minipage}
		\hfill
		\begin{minipage}[t]{0.23\textwidth}
			\begin{tikzpicture}
			\node (img3)  {\includegraphics[scale=0.16]{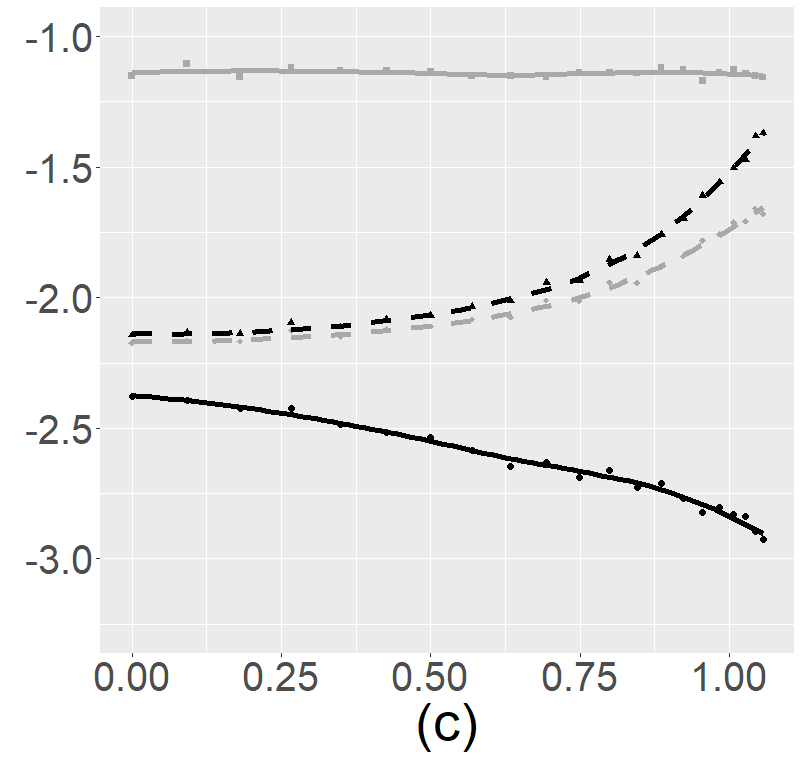}};
			\node[left=of img3, node distance=0cm, rotate=90, anchor=center,yshift=-1cm] {\scriptsize Log $\bm{\hat{\beta}}_{11}$ MSE};
			\end{tikzpicture}
			\label{fig:mse_bn}
		\end{minipage}
		\hfill
		\begin{minipage}[t]{0.23\textwidth}
			\begin{tikzpicture}
			\node (img4)  {\includegraphics[scale=0.16]{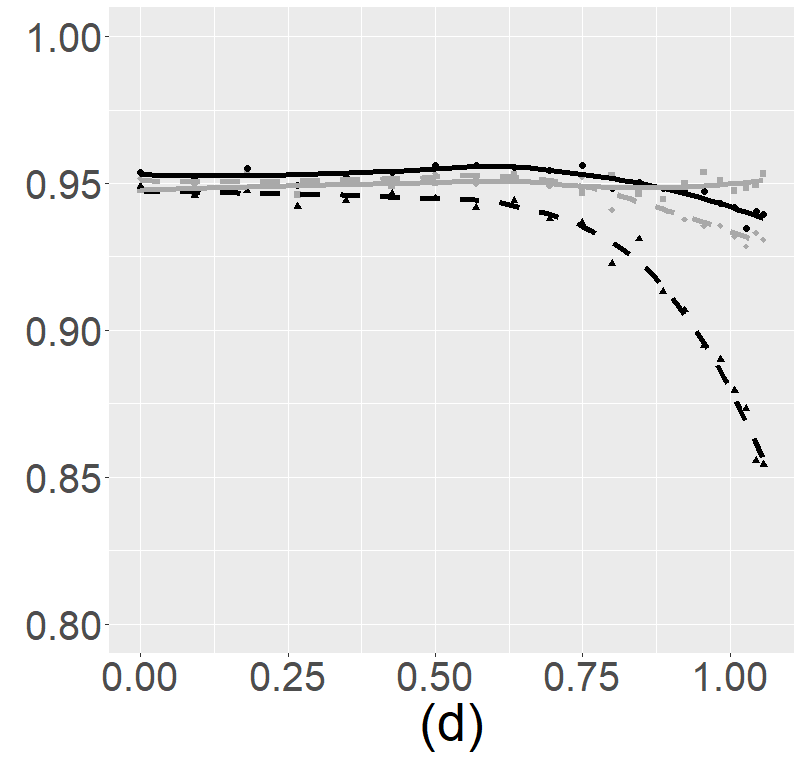}};
			\node[left=of img4, node distance=0cm, rotate=90, anchor=center,yshift=-1cm, xshift = 0.1cm] {\scriptsize Coverage Probability};
			\end{tikzpicture}
			\label{fig:cov_bn}
		\end{minipage}
		\vspace{-0.5cm}
		
		\begin{minipage}[t]{0.05\textwidth}
			\text{ }
		\end{minipage}
		\hfill
		\begin{minipage}[t]{0.1\textwidth}
			\begin{tikzpicture}
			\node (ls)  {	\includegraphics[scale=0.05]{./figures/strapp_legend.png} };
			\node[right=of ls, node distance=0cm, anchor=center, xshift = -0.5cm, font=\color{black}] {\scriptsize straPP};
			\end{tikzpicture}
		\end{minipage}
		\hfill
		\begin{minipage}[t]{0.15\textwidth}
			\begin{tikzpicture}
			\node (lp)  {	\includegraphics[scale=0.05]{./figures/gs01_legend.png} };
			\node[right=of lp, node distance=0cm, anchor=center, xshift = -0.3cm, font=\color{black}] {\scriptsize GS($\omega_0 = 0.1$)};
			\end{tikzpicture}
		\end{minipage}
		\hfill
		\begin{minipage}[t]{0.15\textwidth}
			\begin{tikzpicture}
			\node (lp)  {	\includegraphics[scale=0.05]{./figures/gs05_legend.png} };
			\node[right=of lp, node distance=0cm, anchor=center, xshift = -0.3cm, font=\color{black}] {\scriptsize GS($\omega_0 = 0.5$)};
			\end{tikzpicture}
		\end{minipage}
		\hfill
		\begin{minipage}[t]{0.15\textwidth}
			\begin{tikzpicture}
			\node (lu)  {	\includegraphics[scale=0.05]{./figures/gs1_legend.png} };
			\node[right=of lu, node distance=0cm, anchor=center, xshift = -0.3cm, font=\color{black}] {\scriptsize GS($\omega_0 = 1.0$)};
			\end{tikzpicture}
		\end{minipage}
		\hfill
		\begin{minipage}[t]{0.05\textwidth}
			\text{ }
		\end{minipage}\\
		\vspace{-0.6cm}
		\begin{minipage}[t]{0.05\textwidth}
			\text{ }
		\end{minipage}
		\hfill
		\begin{minipage}[t]{0.15\textwidth}
			\begin{tikzpicture}
			\node (lp)  {	\includegraphics[scale=0.05]{./figures/com_b2_legend.png} };
			\node[right=of lp, node distance=0cm, anchor=center, xshift = -0.3cm, font=\color{black}] {\scriptsize COM($b_0$ = 2)};
			\end{tikzpicture}
		\end{minipage}
		\hfill
		\begin{minipage}[t]{0.15\textwidth}
			\begin{tikzpicture}
			\node (lp)  {	\includegraphics[scale=0.05]{./figures/com_b4_legend.png} };
			\node[right=of lp, node distance=0cm, anchor=center, xshift = -0.3cm, font=\color{black}] {\scriptsize COM($b_0$ = 4)};
			\end{tikzpicture}
		\end{minipage}
		\hfill
		\begin{minipage}[t]{0.15\textwidth}
			\begin{tikzpicture}
			\node (lu)  {	\includegraphics[scale=0.05]{./figures/com_b8_legend.png} };
			\node[right=of lu, node distance=0cm, anchor=center, xshift = -0.3cm, font=\color{black}] {\scriptsize COM($b_0$ = 8)};
			\end{tikzpicture}
		\end{minipage}
		\hfill
		\begin{minipage}[t]{0.05\textwidth}
			\text{ }
		\end{minipage}

		\vspace{-0.4cm}
		
		\begin{minipage}[t]{0.23\textwidth}
			\begin{tikzpicture}
			\node (img5)  {\includegraphics[scale=0.16]{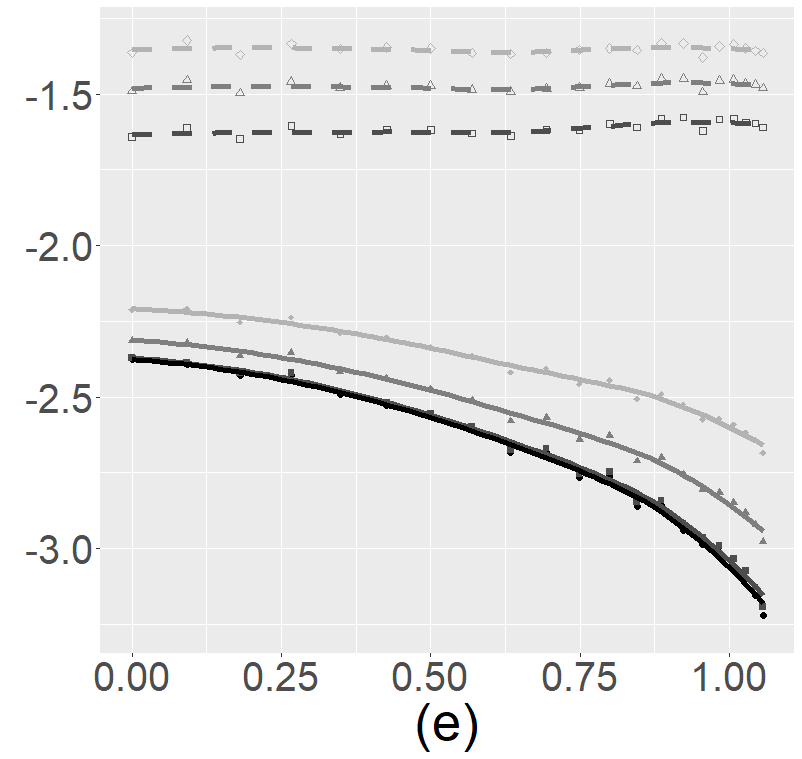}};
			\node[left=of img5, node distance=0cm, rotate=90, anchor=center,yshift=-1cm, xshift = 0.1cm,font=\color{black}] {\scriptsize Avg Log $\bm{\hat{\beta}}_{11}$ Variance};
			\end{tikzpicture}
			\label{fig:var_gs}
		\end{minipage}
		\hfill
		\begin{minipage}[t]{0.23\textwidth}
			\begin{tikzpicture}
			\node (img6)  {\includegraphics[scale=0.16]{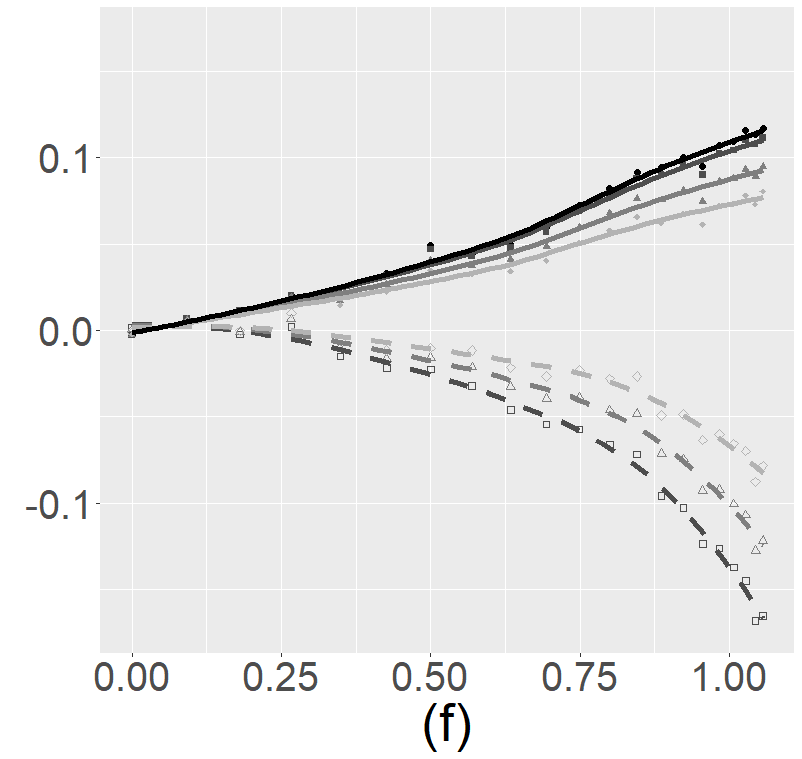}};
			\node[left=of img6, node distance=0cm, rotate=90, anchor=center,yshift=-1cm] {\scriptsize $\bm{\hat{\beta}}_{11}$ Bias};
			\end{tikzpicture}
			\label{fig:bias_gs}
		\end{minipage}
		\hfill
		\begin{minipage}[t]{0.23\textwidth}
			\begin{tikzpicture}
			\node (img7)  {\includegraphics[scale=0.16]{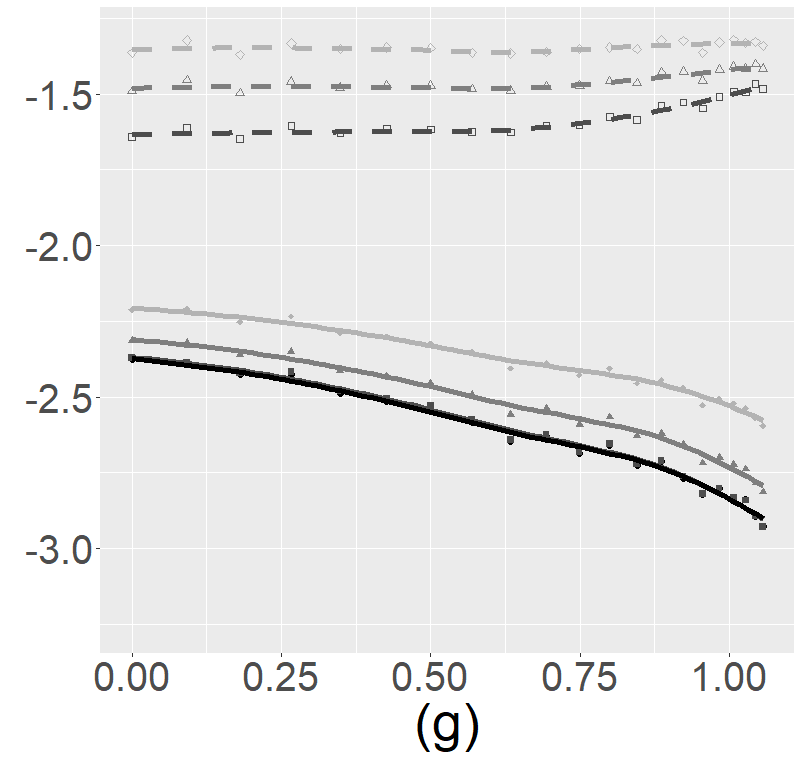}};
			\node[left=of img7, node distance=0cm, rotate=90, anchor=center,yshift=-1cm] {\scriptsize Log $\bm{\hat{\beta}}_{11}$ MSE};
			\end{tikzpicture}
			\label{fig:mse_gs}
		\end{minipage}
		\hfill
		\begin{minipage}[t]{0.23\textwidth}
			\begin{tikzpicture}
			\node (img8)  {\includegraphics[scale=0.16]{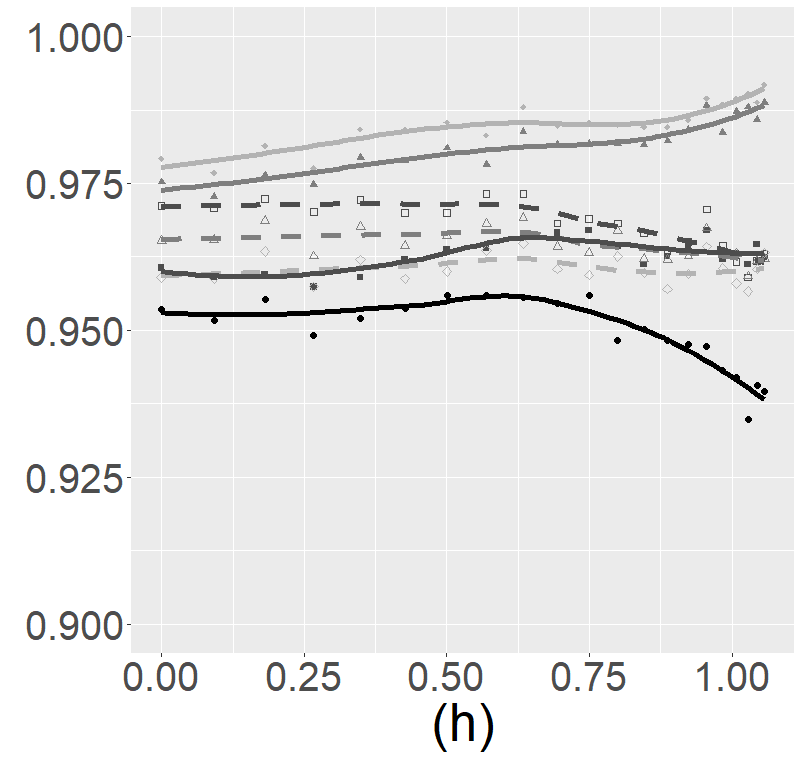}};
			\node[left=of img8, node distance=0cm, rotate=90, anchor=center,yshift=-1cm, xshift = 0.1cm] {\scriptsize Coverage Probability};
			\end{tikzpicture}
			\label{fig:cov_gs}
		\end{minipage}
		\vspace{-0.6cm}
		\caption{Panels (a)-(d) present the average log variance, bias, log MSE, and coverage probability for the posterior mean of $\beta_{11}$,	respectively, as a function of the true value of $\beta_{11}$ plotted on the x-axis for the straPP, power prior, asymptotic power prior and uniform improper prior. Panels (e)-(h) present the same information for the straPP, Gen-straPP with $\omega_0 = 0.1, 0.5, 1.0$, and commensurate prior with $b_0$ = 2, 4, 8. straPP, scale transformed power prior; PP, power prior; APP, asymptotic power prior; UIP, uniform improper prior; GS, generalized scale transformed power prior; COM, commensurate prior.}\label{bn-plot}
	\end{figure}

	Figure~\ref{bn-plot} panels (a)-(d) present results comparing performance characteristics of the straPP, power prior, asymptotic power prior and uniform improper prior. Figure~\ref{bn-plot} panels (e)-(h) present similar results for the straPP, the Gen-straPP  for several choices of $\omega_0$, and the commensurate prior for several choices of $b_0$.
	
	As shown in Figure~\ref{bn-plot} (a) and (e), the posterior mean based on the straPP has the smallest variance on average and the smallest MSE (Figure~\ref{bn-plot} (c) and (g)), though the Gen-straPP with $\omega_0 = 0.1$ has very similar results. In  Figure~\ref{bn-plot}(b), the straPP posterior mean no longer retains the unbiasedness property of the normal-normal case but still has bias that is less in absolute value than the power prior, and as good or better in absolute value than the asymptotic power prior. In Figure~\ref{bn-plot} panels (e)-(h), one can see that the straPP generally outperforms the Gen-straPP in this situation where the relationship between the historical and current data models satisfy the straPP transformation assumption. Additionally, the straPP and the Gen-straPP generally outperform the commensurate prior, except the Gen-straPP with $\omega_0 > 0.1$ in the case of the coverage probability.
	
	\subsection{Poisson-Exponential Case Simulation} \label{pe-sim}
	
	In the Poisson-exponential simulation, we include standardized age as an additional covariate for the historical and current data. Standardized age was randomly generated using a standard normal distribution. Additionally, for these simulations we did not wish to borrow on the intercept. We performed simulation studies using parameter values that obey the assumption of the partial-borrowing complimentary straPP transformation (e.g., $\btheta_{1} = g^{-1}(\beta_{10},\bbeta_0)$, in which $\btheta_{1}^T = (\beta_{11}, \beta_{12})$).
	
	We considered the following inputs: $n_0 = 100$, $n_1 = 100$, $a_0 = 1$, $\beta_{00} = 0.55$, $\beta_{01} \in [-0.35,0.5]$, $\beta_{02} = 0.1$, and $\beta_{10} = 0.25$. The values of the current data model parameters were then identified by solving $\btheta_{1} = g^{-1}(\beta_{10},\bbeta_0)$. The value of the parameters for the historical data model were chosen so that 
	all means for the historical response were between 1 and 3, to ensure the simulated datasets from the loglinear model had sufficient variability in the outcome across treatment groups to avoid instability issues in model fitting.
	
\begin{figure}[h]
		\centering
		\begin{minipage}[t]{0.1\textwidth}
			\text{ }
		\end{minipage}
		\hfill
		\begin{minipage}[t]{0.1\textwidth}
			\begin{tikzpicture}
			\node (ls)  {	\includegraphics[scale=0.05]{./figures/strapp_legend.png} };
			\node[right=of ls, node distance=0cm, anchor=center, xshift = -0.5cm, font=\color{black}] {\scriptsize straPP};
			\end{tikzpicture}
		\end{minipage}
		\hfill
		\begin{minipage}[t]{0.1\textwidth}
			\begin{tikzpicture}
			\node (lp)  {	\includegraphics[scale=0.05]{./figures/pp_legend.png} };
			\node[right=of lp, node distance=0cm, anchor=center, xshift = -0.5cm, font=\color{black}] {\scriptsize PP};
			\end{tikzpicture}
		\end{minipage}
		\hfill
		\begin{minipage}[t]{0.1\textwidth}
			\begin{tikzpicture}
			\node (lp)  {	\includegraphics[scale=0.05]{./figures/app_legend.png} };
			\node[right=of lp, node distance=0cm, anchor=center, xshift = -0.5cm, font=\color{black}] {\scriptsize APP};
			\end{tikzpicture}
		\end{minipage}
		\hfill
		\begin{minipage}[t]{0.1\textwidth}
			\begin{tikzpicture}
			\node (lu)  {	\includegraphics[scale=0.05]{./figures/ui_legend.png} };
			\node[right=of lu, node distance=0cm, anchor=center, xshift = -0.5cm, font=\color{black}] {\scriptsize UIP};
			\end{tikzpicture}
		\end{minipage}
		\hfill
		\begin{minipage}[t]{0.1\textwidth}
			\text{ }
		\end{minipage}
		\vspace{-0.4cm}
		
		\begin{minipage}[t]{0.23\textwidth}
			\begin{tikzpicture}
			\node (img1)  {\includegraphics[scale=0.16]{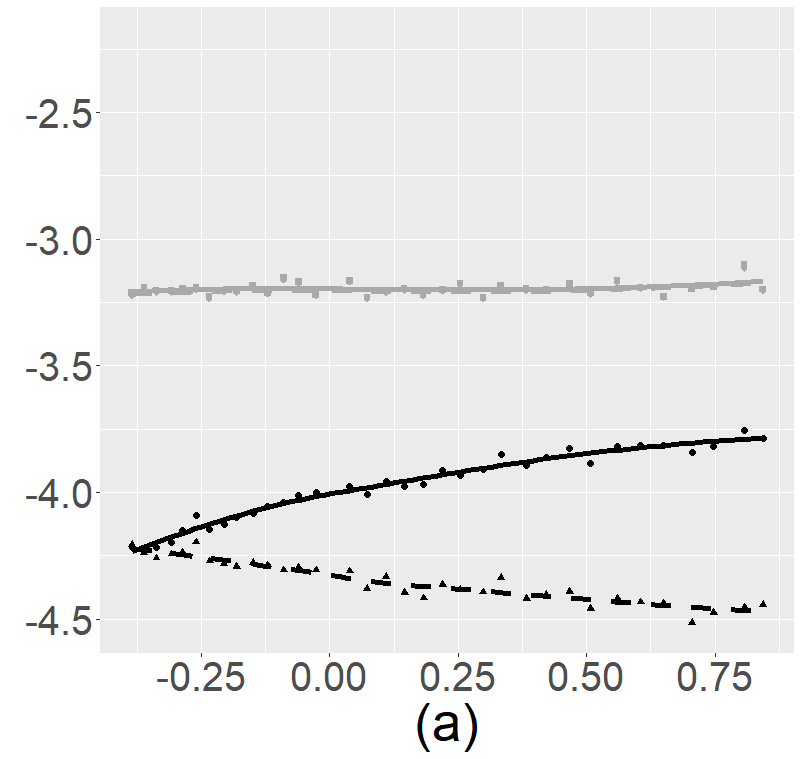}};
			\node[left=of img1, node distance=0cm, rotate=90, anchor=center,yshift=-1cm, xshift = 0.1cm, font=\color{black}] {\scriptsize Avg Log $\bm{\hat{\beta}}_{11}$ Variance};
			\end{tikzpicture}
		\end{minipage}
		\hfill
		\begin{minipage}[t]{0.23\textwidth}
			\begin{tikzpicture}
			\node (img2)  {\includegraphics[scale=0.16]{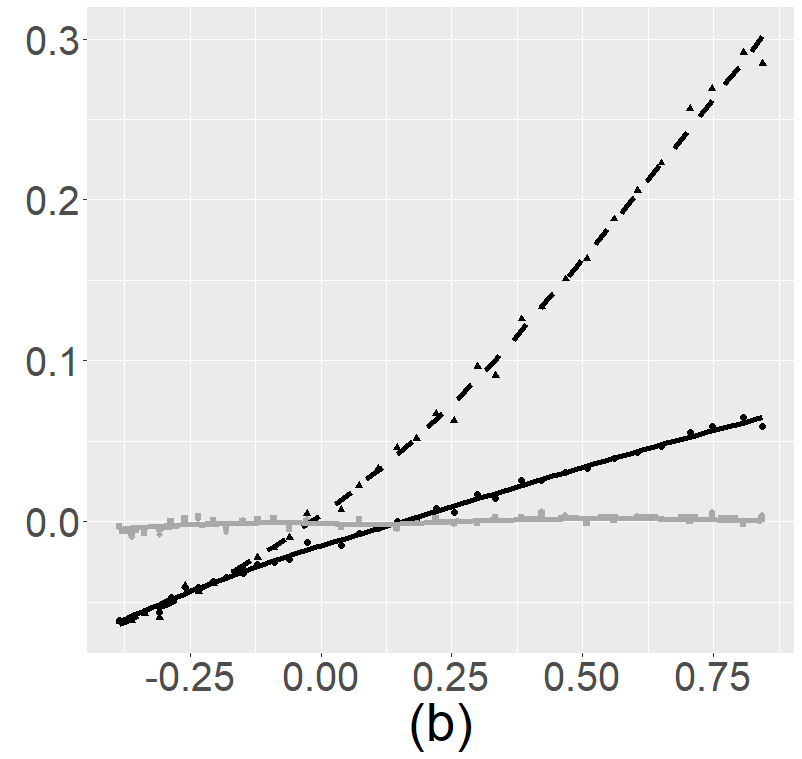}};
			\node[left=of img2, node distance=0cm, rotate=90, anchor=center,yshift=-1cm] {\scriptsize $\bm{\hat{\beta}}_{11}$ Bias};
			\end{tikzpicture}
		\end{minipage}
		\hfill
		\begin{minipage}[t]{0.23\textwidth}
			\begin{tikzpicture}
			\node (img3)  {\includegraphics[scale=0.16]{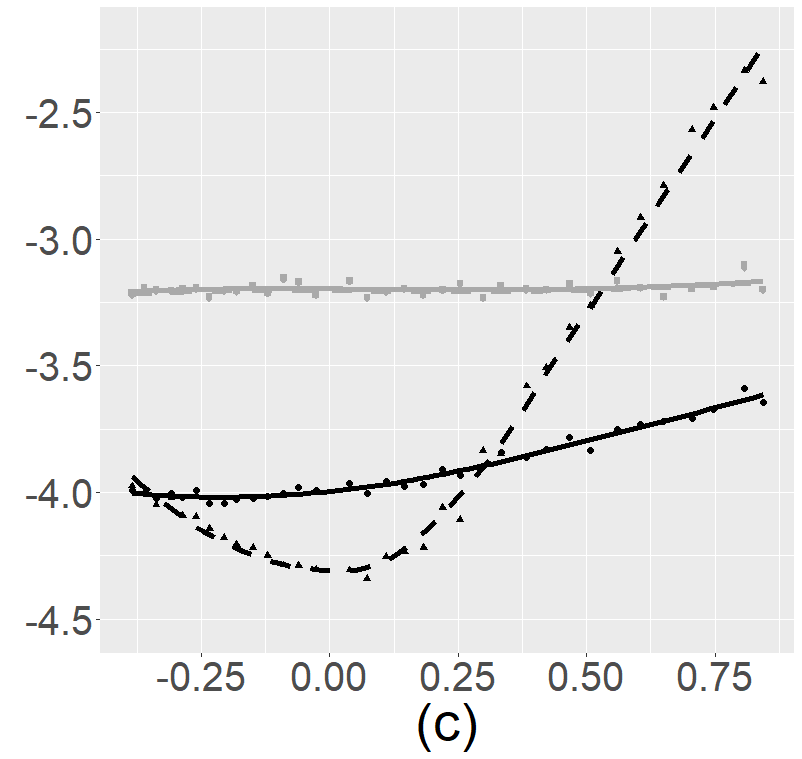}};
			\node[left=of img3, node distance=0cm, rotate=90, anchor=center,yshift=-1cm] {\scriptsize Log $\bm{\hat{\beta}}_{11}$ MSE};
			\end{tikzpicture}
		\end{minipage}
		\hfill
		\begin{minipage}[t]{0.23\textwidth}
			\begin{tikzpicture}
			\node (img4)  {\includegraphics[scale=0.16]{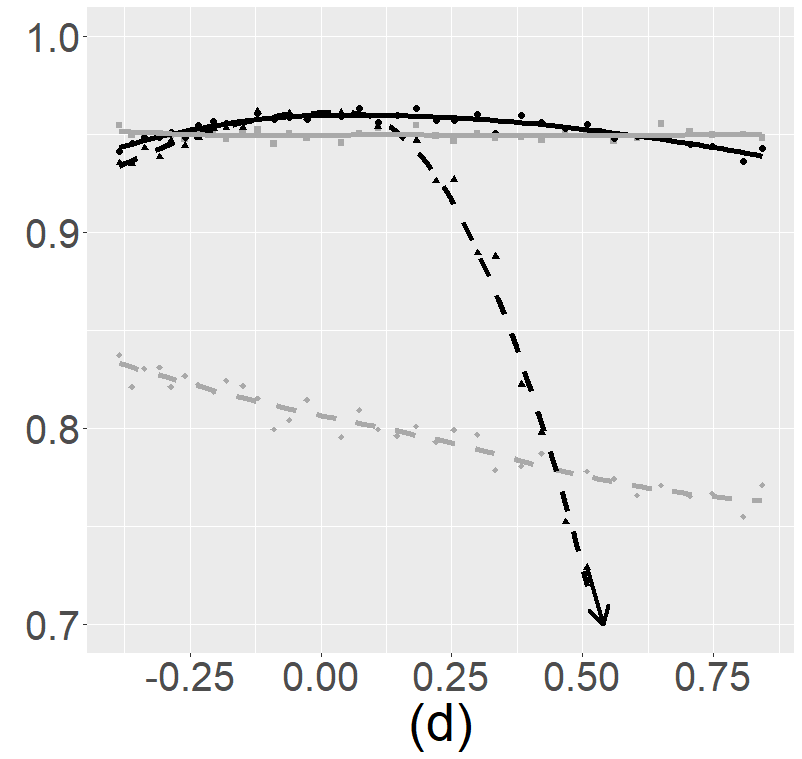}};
			\node[left=of img4, node distance=0cm, rotate=90, anchor=center,yshift=-1cm, xshift = 0.1cm] {\scriptsize Coverage Probability};
			\end{tikzpicture}
		\end{minipage}
		\vspace{-0.2cm}
		
		\begin{minipage}[t]{0.05\textwidth}
			\text{ }
		\end{minipage}
		\hfill
		\begin{minipage}[t]{0.1\textwidth}
			\begin{tikzpicture}
			\node (ls)  {	\includegraphics[scale=0.05]{./figures/strapp_legend.png} };
			\node[right=of ls, node distance=0cm, anchor=center, xshift = -0.5cm, font=\color{black}] {\scriptsize straPP};
			\end{tikzpicture}
		\end{minipage}
		\hfill
		\begin{minipage}[t]{0.15\textwidth}
			\begin{tikzpicture}
			\node (lp)  {	\includegraphics[scale=0.05]{./figures/gs01_legend.png} };
			\node[right=of lp, node distance=0cm, anchor=center, xshift = -0.3cm, font=\color{black}] {\scriptsize GS($\omega_0 = 0.1$)};
			\end{tikzpicture}
		\end{minipage}
		\hfill
		\begin{minipage}[t]{0.15\textwidth}
			\begin{tikzpicture}
			\node (lp)  {	\includegraphics[scale=0.05]{./figures/gs05_legend.png} };
			\node[right=of lp, node distance=0cm, anchor=center, xshift = -0.3cm, font=\color{black}] {\scriptsize GS($\omega_0 = 0.5$)};
			\end{tikzpicture}
		\end{minipage}
		\hfill
		\begin{minipage}[t]{0.15\textwidth}
			\begin{tikzpicture}
			\node (lu)  {	\includegraphics[scale=0.05]{./figures/gs1_legend.png} };
			\node[right=of lu, node distance=0cm, anchor=center, xshift = -0.3cm, font=\color{black}] {\scriptsize GS($\omega_0 = 1.0$)};
			\end{tikzpicture}
		\end{minipage}
		\hfill
		\begin{minipage}[t]{0.05\textwidth}
			\text{ }
		\end{minipage}\\
		\vspace{-0.6cm}
		\begin{minipage}[t]{0.05\textwidth}
			\text{ }
		\end{minipage}
		\hfill
		\begin{minipage}[t]{0.15\textwidth}
			\begin{tikzpicture}
			\node (lp)  {	\includegraphics[scale=0.05]{./figures/com_b2_legend.png} };
			\node[right=of lp, node distance=0cm, anchor=center, xshift = -0.3cm, font=\color{black}] {\scriptsize COM($b_0$ = 2)};
			\end{tikzpicture}
		\end{minipage}
		\hfill
		\begin{minipage}[t]{0.15\textwidth}
			\begin{tikzpicture}
			\node (lp)  {	\includegraphics[scale=0.05]{./figures/com_b4_legend.png} };
			\node[right=of lp, node distance=0cm, anchor=center, xshift = -0.3cm, font=\color{black}] {\scriptsize COM($b_0$ = 4)};
			\end{tikzpicture}
		\end{minipage}
		\hfill
		\begin{minipage}[t]{0.15\textwidth}
			\begin{tikzpicture}
			\node (lu)  {	\includegraphics[scale=0.05]{./figures/com_b8_legend.png} };
			\node[right=of lu, node distance=0cm, anchor=center, xshift = -0.3cm, font=\color{black}] {\scriptsize COM($b_0$ = 8)};
			\end{tikzpicture}
		\end{minipage}
		\hfill
		\begin{minipage}[t]{0.05\textwidth}
			\text{ }
		\end{minipage}

		\vspace{-0.4cm}
		
		\begin{minipage}[t]{0.23\textwidth}
			\begin{tikzpicture}
			\node (img5)  {\includegraphics[scale=0.16]{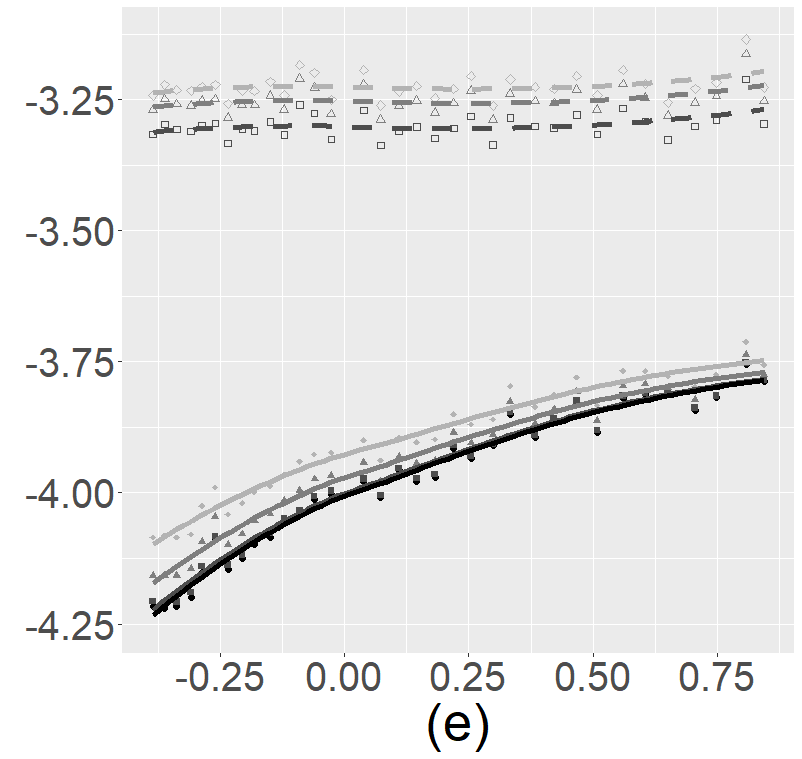}};
			\node[left=of img5, node distance=0cm, rotate=90, anchor=center,yshift=-1cm, xshift = 0.1cm, font=\color{black}] {\scriptsize Avg Log $\bm{\hat{\beta}}_{11}$ Variance};
			\end{tikzpicture}
		\end{minipage}
		\hfill
		\begin{minipage}[t]{0.23\textwidth}
			\begin{tikzpicture}
			\node (img6)  {\includegraphics[scale=0.16]{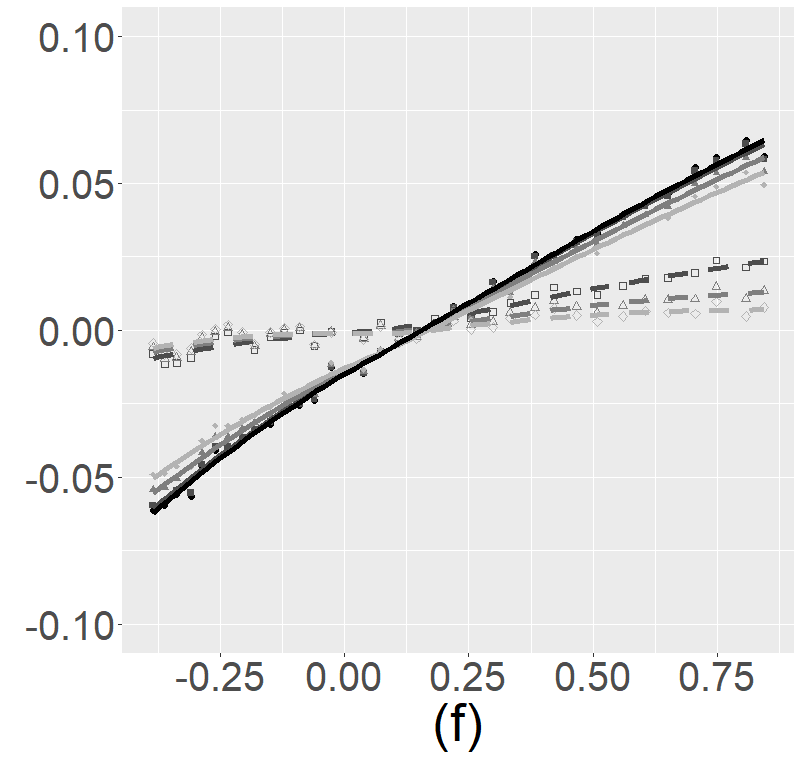}};
			\node[left=of img6, node distance=0cm, rotate=90, anchor=center,yshift=-1cm] {\scriptsize $\bm{\hat{\beta}}_{11}$ Bias};
			\end{tikzpicture}
		\end{minipage}
		\hfill
		\begin{minipage}[t]{0.23\textwidth}
			\begin{tikzpicture}
			\node (img7)  {\includegraphics[scale=0.16]{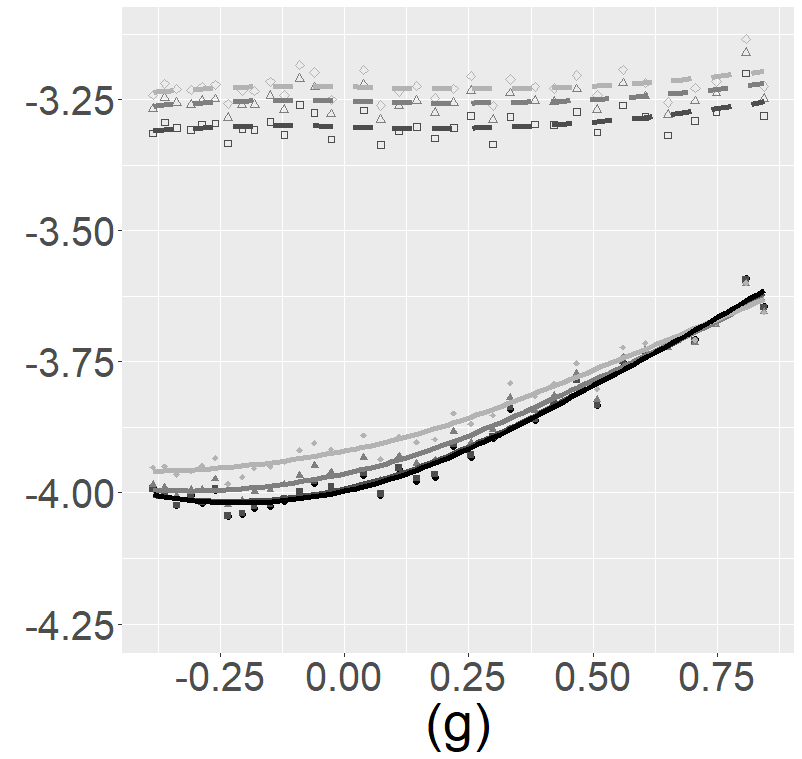}};
			\node[left=of img7, node distance=0cm, rotate=90, anchor=center,yshift=-1cm] {\scriptsize Log $\bm{\hat{\beta}}_{11}$ MSE};
			\end{tikzpicture}
		\end{minipage}
		\hfill
		\begin{minipage}[t]{0.23\textwidth}
			\begin{tikzpicture}
			\node (img8)  {\includegraphics[scale=0.16]{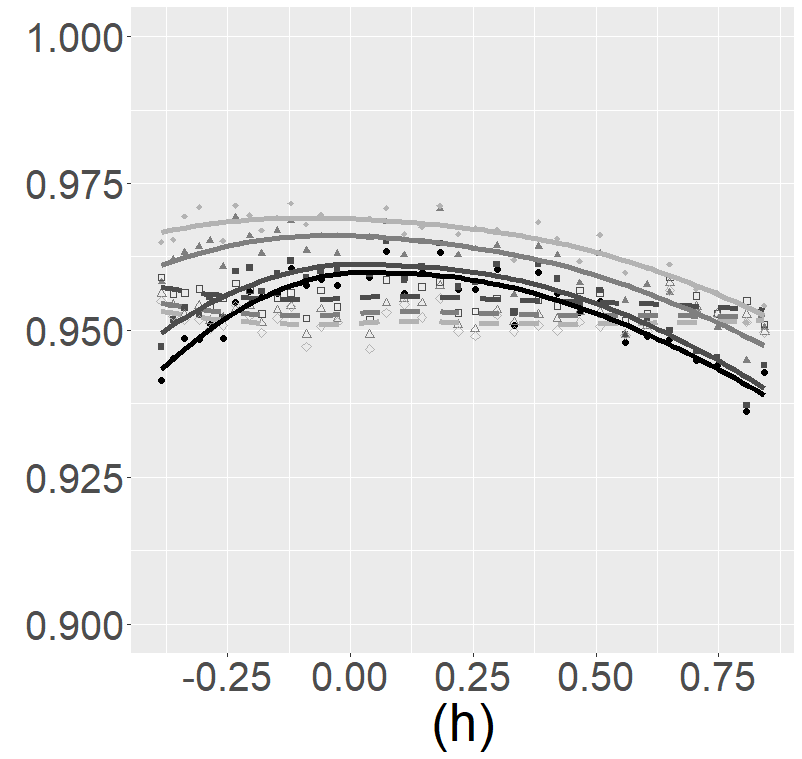}};
			\node[left=of img8, node distance=0cm, rotate=90, anchor=center,yshift=-1cm, xshift = 0.1cm] {\scriptsize Coverage Probability};
			\end{tikzpicture}
		\end{minipage}
		
		\vspace{-0.2cm}
		
		\caption{Panels (a)-(d) present the average log variance, bias, log MSE, and coverage probability for the posterior mean of $\beta_{11}$,	respectively, as a function of the true value of $\beta_{11}$ plotted on the x-axis for the straPP, power prior, asymptotic power prior and uniform improper prior. Panels (e)-(h) present the same information for the straPP, Gen-straPP with $\omega_0 = 0.1, 0.5, 1.0$, and commensurate prior with $b_0$ = 2, 4, 8. straPP, scale transformed power prior; PP, power prior; APP, asymptotic power prior; UIP, uniform improper prior; GS, generalized scale transformed power prior; COM, commensurate prior.}\label{pe-plot}
	\end{figure}

	Figure~\ref{pe-plot} panels (a)-(d) present results comparing performance characteristics of the straPP, power prior, asymptotic power prior and uniform improper prior. Figure~\ref{pe-plot} panels (e)-(h) present similar results for the straPP, the Gen-straPP  for  $\omega_0 = 0.1, 0.5, 1.0$, and the commensurate prior for several choices of $b_0$.
	
	The posterior mean based on the straPP has the smallest variance on average (Figure ~\ref{pe-plot}(a), ~\ref{pe-plot}(e)) and the smallest MSE (Figure ~\ref{pe-plot}(c), ~\ref{pe-plot}(g)), when compared to all other analyzed models, except the power prior. However, the posterior mean based on the straPP has lower bias in absolute value, and more accurate coverage probability, as well as a smaller MSE when $\beta_{11} > 0.3$, when compared to the power prior. Additionally, as seen in Figure ~\ref{pe-plot}(f), the posterior mean based on the straPP and Gen-straPP have a larger bias than the commensurate prior, though the bias is still relatively small at less than 0.1.
	
	\section{Posterior Estimate for the Intercepts and Linear Regression Variance \label{int-var}}
	\renewcommand{\thetable}{\Alph{section}.\arabic{table}}
	\setcounter{table}{0}
	
	In this section, we conducted additional analyses of the Comprehensive Post-Acute Stroke Services (COMPASS) Study \citep{compass}. We assume that the historical patient outcomes are independently distributed according to a logistic regression model and the current patient outcomes are independently distributed according to a linear regression model. To complete specification of the partial-borrowing straPP and partial-borrowing Gen-straPP priors, we specified a uniform improper prior on the initial prior for the regression parameters, a uniform improper prior on the current data model intercept and a gamma prior on the dispersion parameter with shape and inverse scale parameters equal to 0.01. We specified a normal prior for $\bm{c}_0$, as stated in Section~\ref{methods: straPP: gen}.
	
	For the partial-borrowing power prior and partial-borrowing asymptotic power prior, the initial prior for the regression parameters was taken to be uniform improper and the initial prior for the dispersion parameter was taken to be gamma with shape and inverse scale parameters equal to 0.01. For the partial-borrowing commensurate prior, the prior on the historical and current data model intercepts were taken to be uniform improper. The prior for the dispersion parameter was taken to be gamma with shape and inverse scale parameters equal to 0.01, and the prior for the commensurability parameter was taken to be gamma with shape and inverse scale parameters equal to 2.
	
	Table \ref{complete-int-var} displays the posterior estimates for the intercept parameter of the logistic regression model, denoted historical intercept, and the intercept and variance parameter of the current linear regression model when $a_0 = 0.1$. The partial-borrowing Gen-straPP with $\omega_0 = 0.1$ was used. The posterior mean, standard deviation (SD) and 95\% highest posterior density (HPD) intervals were calculated based on 20,000 posterior samples obtained from a Metropolis-Hastings MCMC sampling algorithm.  
	
	\begin{table}
		\caption{Posterior Estimates for Intercepts and Current Linear Regression Variance \label{complete-int-var}}
		\setlength{\tabcolsep}{3.5pt} 
		\renewcommand{\arraystretch}{1.2} 
		\resizebox{\textwidth}{!}{%
			\fbox{%
				\begin{tabular}{lrrrrrrrrr}
					\hline
					\hline
					&   &\multicolumn{2}{c}{Historical Intercept} & & \multicolumn{2}{c}{Current Intercept}  & &\multicolumn{2}{c}{Current Variance}\\
					\cline{3-4} \cline{6-7} \cline{9-10}
					~~Model  & DIC~~ &  Mean (SD) & 95\%HPD~~ & & Mean (SD) & 95\%HPD~~ & & Mean (SD) & 95\%HPD~~ \\
					\hline
					straPP &  2815.34 & -0.83 (0.51) & (-1.86, 0.14) & & 47.10 (1.03) & (45.05, 49.11) & & 88.07 (6.44) & (76.34, 101.30)\\
					APP & 2815.85 & -1.17 (0.88) & (-2.91, ~0.54) & & 47.06 (0.89) & (45.34, 48.80) & & 87.44 (6.35) & (74.98, ~~99.97)\\
					PP & 2815.97 & -1.18 (0.89) & (-2.94, ~0.54) & & 47.04 (0.89) & (45.29, 48.76) & & 87.42 (6.40) & (75.14, 100.30)\\ 
					Gen-straPP &  2816.41 & -0.91 (0.66) & (-2.18, 0.39) & & 47.10 (1.07) & (44.98, 49.22) & & 87.99 (6.45) & (75.74, 100.85)\\ 
					UIP & 2816.56 & -----------~~ & ------------~ & & 47.67 (1.08) & (45.53, 49.78) & & 87.16 (6.38) & (75.10, 100.00)\\ 
					COM & 2817.45 & -1.47 (0.36) & (-2.18, -0.79) & & 46.67 (0.94) & (44.82, 48.50) & & 
					87.15 (6.41) & (75.03, ~99.93)\\
					\hline
					\multicolumn{10}{l}{ straPP, scale transformed 
						power prior; APP, asymptotic power prior; PP, power  prior; Gen-straPP, }\\
					\multicolumn{10}{l}{generalized scale transformed power prior;  UIP, uniform improper prior; COM, commensurate prior.}\\
		\end{tabular}}}
	\end{table}
	In Table \ref{complete-int-var}, the posterior mean for the parameters are similar, with the partial-borrowing straPP and partial-borrowing Gen-straPP having a slightly smaller posterior mean for the historical intercept and current variance than the partial-borrowing power prior and partial-borrowing asymptotic power prior, but a larger posterior mean for the current intercept. The partial-borrowing straPP and Gen-straPP have a larger posterior mean and standard deviation for all parameters than the partial-borrowing commensurate prior. The partial-borrowing straPP and partial-borrowing Gen-straPP resulted in a smaller posterior standard deviation for the historical intercept and current variance than the partial-borrowing power prior and partial-borrowing asymptotic power prior, while they had larger posterior standard deviations for the current intercept.
	
	\subsection{Commensurate Prior: Sensitivity Analysis \label{compass_com}}
	
	For analysis using the commensurate prior, we investigated additional values for the hyperparameter of the gamma prior on the commensurability parameter. Here we retain the scale value of 2, but investigate various values of the inverse scale hyperparameter, denoted $b_0$.
	
	Table \ref{com_ests} displays the posterior estimates for the current covariates using the commensurate prior with varying values of $b_0$. All models analyzed have a larger DIC than the straPP in Table \ref{complete-int-var}, and when $b_0$ is less than 4, the commensurate prior has a larger DIC than the uniform improper prior. The posterior mean for eCare Plan increases as $b_0$ increases, while the posterior mean decreases as $b_0$ increases for all other covariates. The posterior standard deviation for all covariates increases as $b_0$ increases.

	\begin{table}
		\caption{Commensurate Prior: Posterior Estimates\label{com_ests}}
		\setlength{\tabcolsep}{3.5pt} 
		\renewcommand{\arraystretch}{1.2} 
		\resizebox{\textwidth}{!}{%
			\fbox{%
				\begin{tabular}{lrrrrrrrrrrrrrrr}
					\hline
					\hline
					&   &\multicolumn{2}{c}{eCare Plan} & & \multicolumn{2}{c}{History of Stroke}  & &\multicolumn{2}{c}{Minor NIHSS} & &\multicolumn{2}{c}{Moderate-Severe NIHSS} & &\multicolumn{2}{c}{Non-white}  \\
					\cline{3-4} \cline{6-7} \cline{9-10} \cline{12-13} \cline{15-16}
					~~$b_0$  & DIC~~ &  Mean (SD) & 95\%HPD~~ & & Mean (SD) & 95\%HPD~~ & & Mean (SD) & 95\%HPD~~ & & Mean (SD)    & 95\%HPD~~  & & Mean (SD)    & 95\%HPD~~\\
					\hline
					~0.25 & 2821.77 & -0.06 (0.64) & (-1.21, 1.31) & &  0.15 (0.73) & (-1.34,  1.53) & & 
					0.04 (0.66) & (-1.33, 1.28) & & -0.89 (1.05) & (-3.14, ~0.87) & & 
					-1.20 (1.06) & (-3.34, 0.80)\\
					~0.50 & 2820.21 &  0.13 (0.73) & (-1.22, 1.68) & & -0.06 (0.84) & (-1.78,  1.49) & & 
					-0.10 (0.77) & (-1.71, 1.33) & & -1.37 (1.19) & (-3.92, ~0.60) & &
					-1.30 (1.16) & (-3.65, 0.95)\\
					~1.00 & 2818.64 &  0.30 (0.78) & (-1.21, 1.85) & & -0.29 (0.93) & (-2.21,  1.45) & & 
					-0.28 (0.85) & (-1.97, 1.41) & & -1.87 (1.24) & (-4.37, ~0.35) & & 
					-1.41 (1.27) & (-3.92, 1.04)\\
					~2.00 & 2817.45 &  0.45 (0.84) & (-1.16, 2.14) & & -0.47 (0.99) & (-2.43,  1.44) & & 
					-0.48 (0.93) & (-2.39, 1.21) & & -2.35 (1.25) & (-4.79, ~0.03) & & 
					-1.53 (1.37) & (-4.31, 1.07)\\
					~4.00 & 2816.50 &  0.56 (0.87) & (-1.13, 2.26) & & -0.64 (1.04) & (-2.62,  1.47) & & 
					-0.67 (0.97) & (-2.57, 1.19) & & -2.80 (1.25) & (-5.27, -0.41) & & 
					-1.63 (1.48) & (-4.69, 1.12)\\
					~8.00 & 2815.95 &  0.64 (0.89) & (-1.12, 2.38) & & -0.78 (1.08) & (-2.96,  1.27) & & 
					-0.88 (1.00) & (-2.78, 1.12) & & -3.20 (1.25) & (-5.61, -0.77) & & 
					-1.73 (1.60) & (-4.87, 1.47)\\
					16.00 & 2815.69 &  0.70 (0.92) & (-1.09, 2.51) & & -0.91 (1.13) & (-3.14,  1.30) & & 
					-1.08 (1.02) & (-3.05, 0.96) & & -3.60 (1.25) & (-6.07, -1.20) & & 
					-1.82 (1.70) & (-5.12, 1.58)\\
					24.00 & 2815.62 &  0.73 (0.94) & (-1.03, 2.63) & & -0.96 (1.14) & (-3.21,  1.23) & & 
					-1.20 (1.03) & (-3.14, 0.85) & & -3.81 (1.25) & (-6.38, -1.45) & & 
					-1.89 (1.77) & (-5.41, 1.56)\\
					32.00 & 2815.66 &  0.76 (0.94) & (-1.07, 2.60) & & -1.01 (1.16) & (-3.22,  1.31) & & 
					-1.25 (1.04) & (-3.29, 0.79) & & -3.93 (1.24) & (-6.32, -1.48) & & 
					-1.96 (1.81) & (-5.54, 1.57)\\
					\hline
		\end{tabular}}}
	\end{table}
	
	\ref{com_else} displays the posterior estimates for historical and current intercepts, and the current variance. The posterior mean for the historical intercept and current variance decreases as $b_0$ increases, while the posterior mean for the current intercept increases as $b_0$ increases. The posterior standard deviation for the historical intercept and current variance increases as $b_0$ increases, while the posterior standard deviation for the current intercept decreases as $b_0$ increases.
	
	\begin{table}
		\caption{Commensurate Prior: Posterior Estimates for Intercepts and Current Variance\label{com_else}}
		\setlength{\tabcolsep}{3.5pt} 
		\renewcommand{\arraystretch}{1.2} 
		\resizebox{\textwidth}{!}{%
			\fbox{%
				\begin{tabular}{lrrrrrrrrr}
					\hline
					\hline
					&   &\multicolumn{2}{c}{Historical Intercept} & & \multicolumn{2}{c}{Current Intercept}  & &\multicolumn{2}{c}{Current Variance}\\
					\cline{3-4} \cline{6-7} \cline{9-10}
					~~$b_0$  & DIC~~ &  Mean (SD) & 95\%HPD~~ & & Mean (SD) & 95\%HPD~~ & & Mean (SD) & 95\%HPD~~ \\
					\hline
					~0.25 & 2821.77 & -1.43 (0.35) & (-2.12, -0.76) & & 46.27 (0.73) & (44.85, 47.76) & & 
					88.51 (6.55) & (76.08, 101.26)\\
					~0.50 & 2820.21 & -1.44 (0.35) & (-2.11, -0.74) & & 46.38 (0.82) & (44.79, 48.03) & & 
					88.03 (6.48) & (75.86, 101.12)\\
					~1.00 & 2818.64 & -1.46 (0.35) & (-2.19, -0.80) & & 46.51 (0.88) & (44.79, 48.26) & & 
					87.62 (6.47) & (75.22, 100.55)\\
					~2.00 & 2817.45 & -1.47 (0.36) & (-2.18, -0.79) & & 46.67 (0.94) & (44.82, 48.50) & & 
					87.15 (6.41) & (75.03, ~99.93)\\
					~4.00 & 2816.50 & -1.50 (0.36) & (-2.20, -0.81) & & 46.82 (0.97) & (44.92, 48.73) & & 
					86.86 (6.36) & (74.98, ~99.84)\\
					~8.00 & 2815.95 & -1.52 (0.36) & (-2.22, -0.81) & & 46.99 (1.00) & (44.96, 48.86) & & 
					86.76 (6.33) & (75.24, ~99.71)\\
					16.00 & 2815.69 & -1.54 (0.36) & (-2.25, -0.83) & & 47.17 (1.02) & (45.20, 49.17) & & 
					86.70 (6.38) & (74.51, ~99.33)\\
					24.00 & 2815.62 & -1.55 (0.37) & (-2.28, -0.85) & & 47.27 (1.03) & (45.22, 49.26) & & 
					86.60 (6.32) & (74.74, ~99.35)\\
					32.00 & 2815.66 & -1.56 (0.37) & (-2.30, -0.86) & & 47.31 (1.03) & (45.26, 49.33) & &
					86.63 (6.33) & (74.72, ~99.31)\\	
					\hline
		\end{tabular}}}
	\end{table}
	
\end{appendices}

\end{document}